\documentclass{article}
\usepackage[a4paper, total={6in, 8in}]{geometry}
\usepackage{tikz}
\usepackage{url} 
\usepackage{enumitem}
\usepackage{amsmath}
\usepackage{overpic}
\usepackage{pict2e}
\usepackage{picture}
\usepackage{authblk}
\usepackage[super,numbers,sort&compress]{natbib}
\title{Diurnal Self-Aggregation}

\author[1]{Jan O. Haerter} 
\author[1]{Bettina Meyer}
\author[1]{Silas Boye Nissen}
\affil[1]{Niels Bohr Institute, Copenhagen University, Blegdamsvej 17, 2100 Copenhagen, Denmark}

\def\Var{{\textrm{Var}}\,}

\begin{document}

\maketitle

\noindent
{\bf
Convective self-aggregation is a modelling paradigm for thunderstorm organisation over a constant-temperature tropical sea surface. 
This setup can give rise to cloud clusters over timescales of weeks. 
In reality, sea surface temperatures do oscillate diurnally, affecting the atmospheric state.
Over land, surface temperatures vary more strongly, and rain rate is significantly influenced.
Here, we carry out a substantial suite of cloud-resolving numerical experiments, and find that even weak surface temperature oscillations enable qualitatively different dynamics to emerge:
the spatial distribution of rainfall is only homogeneous during the first day. 
Already on the second day, the rain field is firmly structured.
In later days, the clustering becomes stronger and alternates from day-to-day.
We show that these features are robust to changes in resolution, domain size, and surface temperature, but can be removed by a reduction of the amplitude of oscillation, suggesting a transition to a clustered state. 
Maximal clustering occurs at a scale of $\mathbf{l_{max}\approx 180\;km}$, a scale we relate to the emergence of mesoscale convective systems. 
At $\mathbf{l_{max}}$ rainfall is strongly enhanced and far exceeds the rainfall expected at random. 
We explain the transition to clustering using simple conceptual modelling.
Our results may help clarify how continental extremes build up and how cloud clustering over the tropical ocean could emerge much faster than through conventional self-aggregation alone.
}


\noindent
Currently, general circulation models cannot simulate organised deep convection as these models describe convection as a collection of non-interacting convective cells. 
Yet, the increase in tropical rainfall was stated to predominantly stem from organised deep convection. \cite{tan2015increases} 
Similarly, in mid-latitudes, the majority of flood-producing rainfall was attributed to mesoscale convective systems (MCSs),\cite{stevenson201410,feng2018structure} that is, long-lived complexes of thunderstorms spanning $\sim$ 100 km in diameter.\cite{moeng1995atmospheric} 
In self-aggregation studies pronounced clustering occurs on the timescale of several weeks. \cite{bretherton2005energy,khairoutdinov2010aggregation,muller2012detailed} 
There, the radiative convective equilibrium scheme (RCE) \cite{held1993radiative,tompkins1998radiative} is usually employed, assuming spatially and temporally constant surface temperature ($\sim 300\;K$). 
In self-aggregation, radiation feedbacks have emerged as the "smoking gun" for sustaining and increasing clustering. \cite{bretherton2005energy} 
Still, factors such as sea surface feedbacks, \cite{hohenegger2016coupled} domain size, geometry, and resolution, \cite{muller2015favors} as well as cold pool effects, \cite{jeevanjee2013convective,haerter2019convective} all contribute.

Prescribing constant boundary conditions is an elegant model simplification, but not always realistic. 
Especially under weak surface wind conditions and strong insolation, sea surface temperature amplitudes were observed to be as large as two \cite{weller1996surface,johnson1999trimodal} to five kelvin \cite{kawai2007diurnal} and suggested to modify atmospheric properties. \cite{kawai2007diurnal}
Such temperature variations may even affect the Madden-Julian Oscillation (MJO) and the El Ni\~no phenomenon.
Diurnally varying precipitation intensity also emerges from numerical simulations that account for differences in insolation, \cite{liu1998numerical} and satellite observations show a diurnal cycle in cloud height for the MJO. \cite{tian2006modulation,suzuki2009diurnal}
Further, in contrast to self-aggregation simulations, observed tropical cloud clusters typically are more transient, as they hardly live as long as two days. \cite{chen1996multiscale}

Mechanistically, cold pools (CPs), that is, denser air formed by rain evaporation under thunderstorm clouds, were long implicated in the organisation of convection, \cite{droegemeier1985three,rotunno1988theory} both by thermodynamic and mechanical effects, \cite{tompkins2001organizationCold,boing2012influence,schlemmer2015modifications,feng2015mechanisms,moseley2016intensification} and were suggested to lead to clustering. \cite{haerter2019circling,lochbihler2019response}
CPs can be long-lived when they arise in MCSs.
Over ocean, \citeauthor{chen1997diurnal} (\citeyear{chen1997diurnal}) observed MCSs to undergo bi-diurnal oscillations of local cloudiness and referred to this dynamics as "diurnal dancing."
Over continental regions, the clustering of convection in MCS poses a risk to humans as severe storms can lead to intense downdraughts and flash flooding.
\cite{smith2001extreme,cintineo2013predictability,stevenson201410,feng2018structure}
The continental diurnal cycle of deep convection is typically driven by a build-up of Convective Available Potential Energy (CAPE) during the morning hours and its release in terms of precipitation during the afternoon or evening. \cite{petch2002impact,guichard2004modelling,schlemmer2011diurnal,moseley2016intensification,haerter2018intensified} 
Simulated domains were often chosen to be relatively small.
State-of-the-art climate models now simulate much larger areas ($>$ 1,000 $km$ horizontally at kilometre resolution).\cite{ban2015heavy,prein2017simulating,rasp2018variability,satoh2019global} 
Encompassing features such as terrain variation and inhomogeneous external forcing adds unique regional insight but makes mechanistic analysis of self-organised clustering more cumbersome.

To entirely focus on the spontaneous emergence of clustering, we analyse multi-day organisation in the thunderstorm rain field over land and sea through an idealised setup. 
We show that a transition from a homogeneous to a strongly-clustered state occurs spontaneously when the amplitude of surface temperature is increased sufficiently. 
We refer to these clusters as MCS due to their typical scale of $\sim 100\;km$. \cite{houze2004mesoscale}
Clustering oscillates and continues to strengthen from day-to-day --- findings we explore and explain using conceptual modelling.
Finally, we discuss the relevance for the emergence of extreme continental rainfall and the implications for organised convection over the tropical ocean.

\section*{Results}\label{sec:results}
\noindent
We carry out a suite of numerical experiments on square mesoscale model domains ($L\times L$) of up to $L=960\;km$ horizontal length and horizontal grid resolutions of one kilometre and finer. 
We provide sensitivity experiments exploring grid resolution, domain size, rain evaporation, and surface conditions ({\it Details:} Methods).
We contrast simulations with different amplitudes $T_a$ but equal average surface temperature and refer to experiments with $T_a=2\,K$, $3.5\,K$, and $5\,K$ as $A2$, $A3.5$, and $A5$, respectively.
Modifiers, such as $A5b$ or $A5sea$ ({\it see} Tab.~\ref{tab:experiments}), label sensitivity studies. 
Unless explicitly stated, our key results, which regard the value of $T_a$, are qualitatively robust under the sensitivity experiments.
Each numerical experiment is run for several days, allowing for a spin-up and quasi-steady-state period (Tab.~\ref{tab:experiments} and Fig.~\ref{fig:multi-day_timeseries}). 

\begin{figure*}[ht]
\centering
\includegraphics[width=1\linewidth]{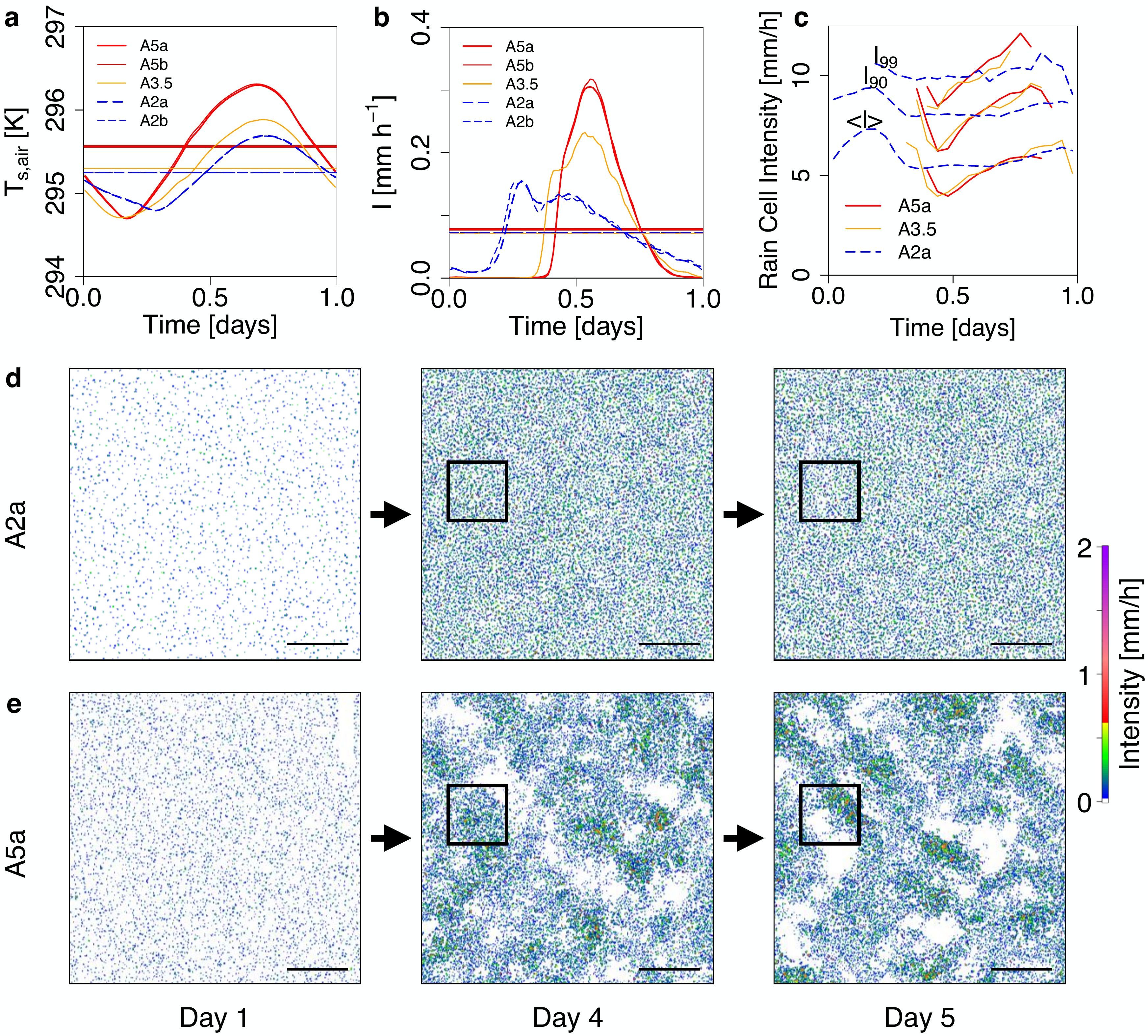}

\caption{{\bf Transition to a clustered rainfall state.}
{\bf a---c}, Diurnal cycles of domain averaged quantities.
Each quantity was horizontally averaged. 
Time-series represent a compound diurnal cycle, where equal times of day were averaged over all available model days.
{\bf a}, Near-surface temperature for simulations with different imposed surface temperature amplitudes, as labelled in legend. 
Horizontal lines of corresponding colours represent the time average of each simulation.
{\bf b}, Analogous to (a), but for rain intensity.
{\bf c}, Mean, 90'th, and 99'th percentiles of event rain intensity ($I>I_0=.5\;mm\;h^{-1}$) for $A5a$ and $A2a$, as labelled ({\it compare}: Tab.~\ref{tab:parameters} for experiment label details).
{\bf d}, Surface rainfall average during day 1 (spin up), day 4, and day 5 for $A2a$.
{\bf e}, Similar to (d), but for $A5a$.
Boxes of side length $200\;km$ highlight the spatial and temporal variation in panels (d) and (e). The scale bars in panels (d) and (e) also have a length of $200\;km$
({\it Details}: Fig.~\ref{fig:allData}
).
}
\label{fig:daily_mean}
\end{figure*}

\noindent
{\bf Domain-mean timeseries.}
Unsurprisingly, the differences in surface temperature amplitudes are reflected in larger amplitudes of atmospheric near-surface temperatures (Fig.~\ref{fig:daily_mean}a and Fig.~\ref{fig:multi-day_timeseries}).
Changes in the time-series of domain-mean rain rate are more profound (Fig.~\ref{fig:daily_mean}b): whereas $A5$ yields a relatively sharp mid-afternoon single-peak structure, the curve transitions to a broader and double-peaked structure for $A2$, 
where they approximate the diurnal cycle typical of oceanic convection \cite{yang2001diurnal}.
Again, the differences in the temporal mean (horizontal lines) are minimal, reflecting radiation constraints on rainfall \cite{held2006robust}.
Approximately proportional curves are found for rain area fraction, which, by contrast, differ from those of rain rate immediately before the central peak (Fig.~\ref{fig:daily_mean}b).
These differences are made more transparent when inspecting rain rates conditional on a threshold (Fig.~\ref{fig:daily_mean}c). 
Both mean and heavy precipitation show a pronounced evening peak for $A5$ and an early-morning peak for $A2$.
In summary, whereas time averages of rain rate are nearly identical for numerical experiments with varying forcing amplitude, the time-series 
differ markedly.

\noindent
{\bf Quantifying clustering.}
Now consider the spatial pattern formed by precipitation cells from day-to-day (Fig.~\ref{fig:daily_mean}d,e). 
During the spin-up from the initial condition (first day), both the $A2$ and $A5$ show modest and relatively homogeneous, convective activity throughout the domain. 
During the subsequent model days, convection intensifies for both simulations because near-surface temperatures gradually increase. 
In $A2$, the spatial pattern of events remains rather homogeneous.
In contrast, for $A5$, an inhomogeneous pattern self-organises, with several locations receiving pronounced average rainfall, whereas others are left all but dry. 
Besides, for $A5$, temporal alternations in surface rainfall rate are apparent when comparing one day to the next ({\it compare:} Fig.~\ref{fig:daily_mean}e):
a cluster on one day leaves an almost rain-free area the next day.

\begin{figure}[ht]
\centering
\begin{overpic}[width=0.4\textwidth]{dummy.pdf}
\put(-50,37){\includegraphics[trim={0cm 0cm 0cm 0cm}, clip, height=0.36\linewidth]{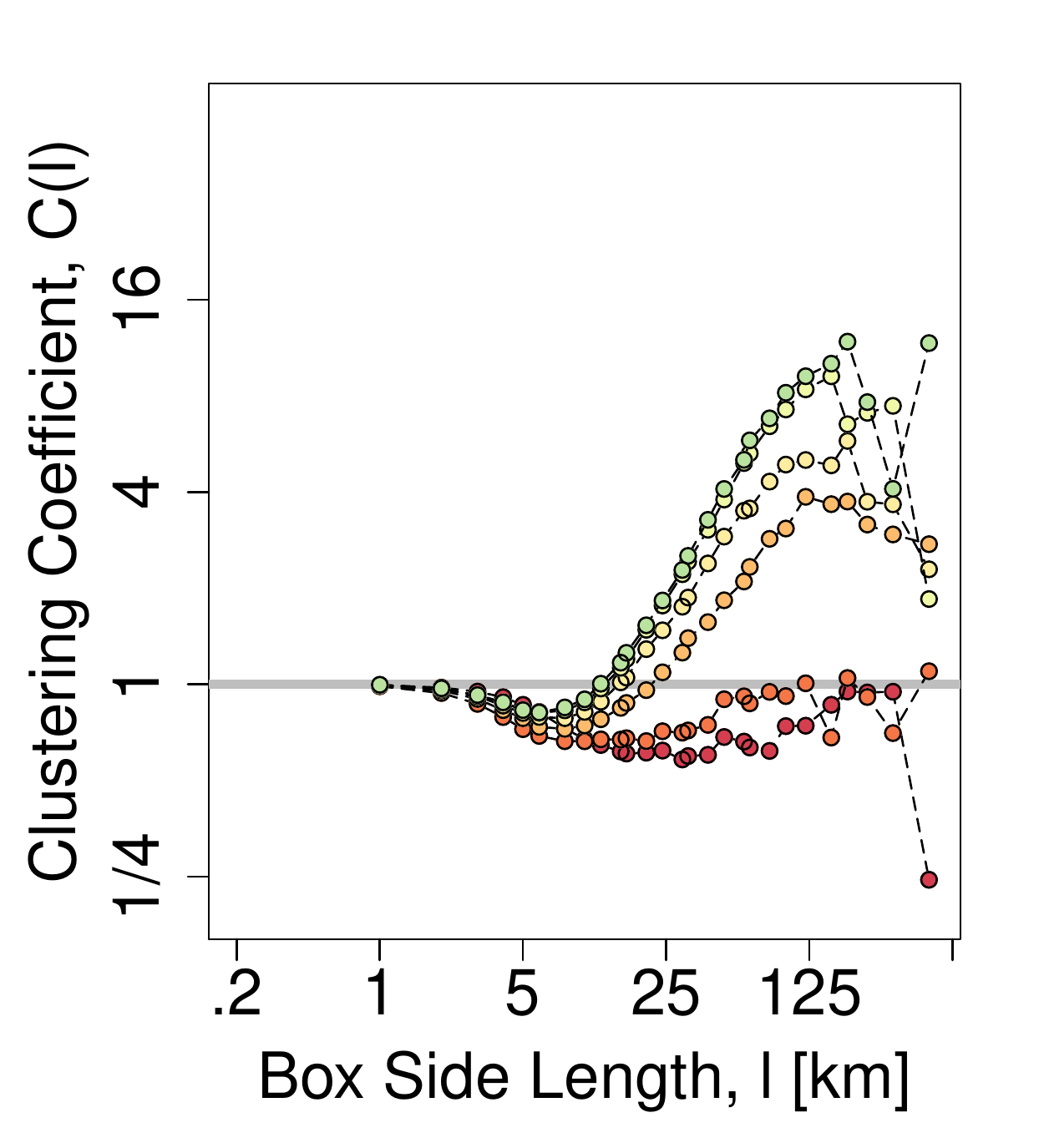}}
\put(5,37){\includegraphics[trim={0cm 0cm 0cm 0cm}, clip, height=0.36\linewidth]{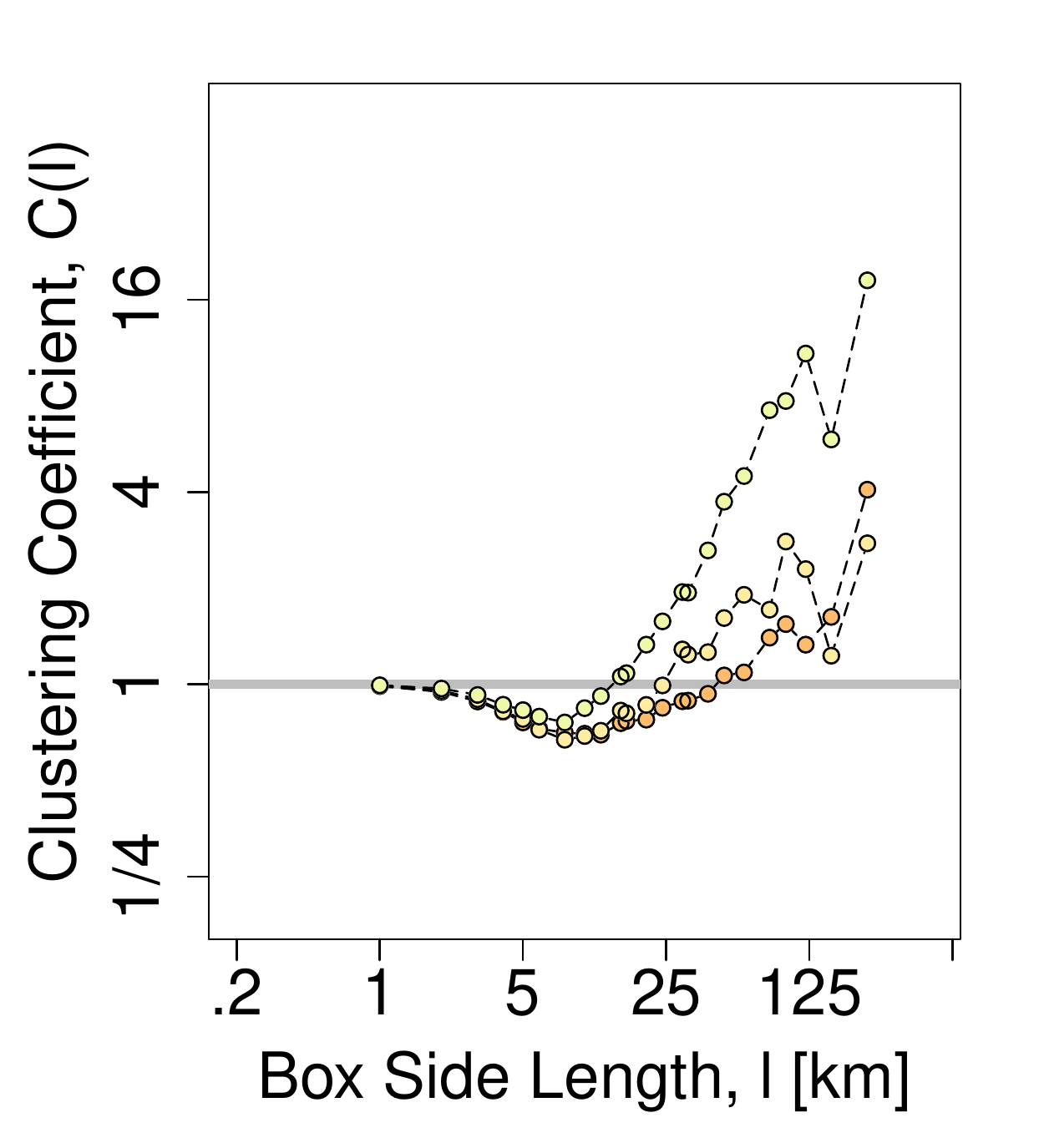}}
\put(60,37){\includegraphics[trim={0cm 0cm 0cm 0cm}, clip, height=0.36\linewidth]{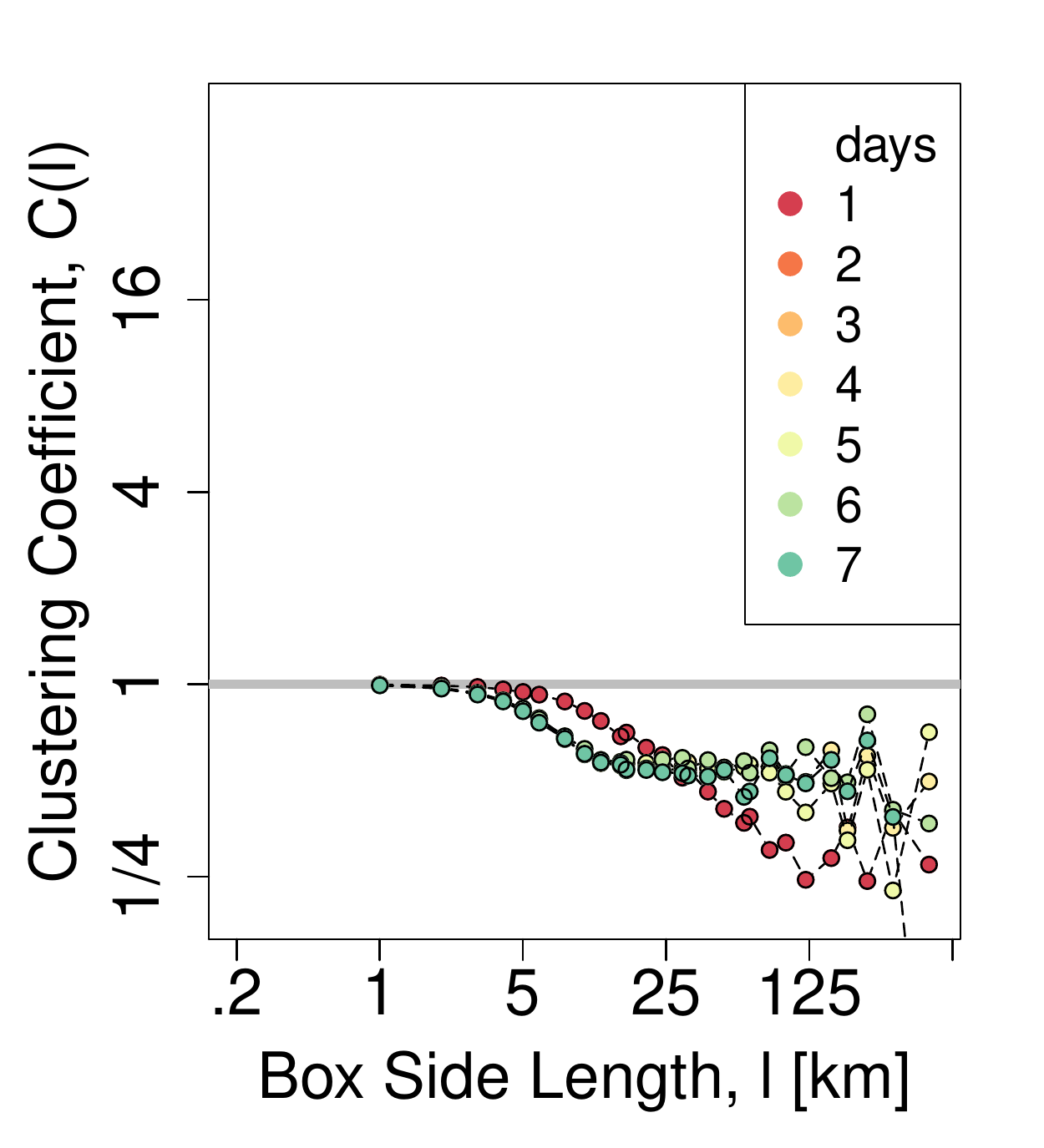}}
\put(-50,-25){\includegraphics[trim={0cm 0cm 0cm 0cm}, clip, height=0.36\linewidth]{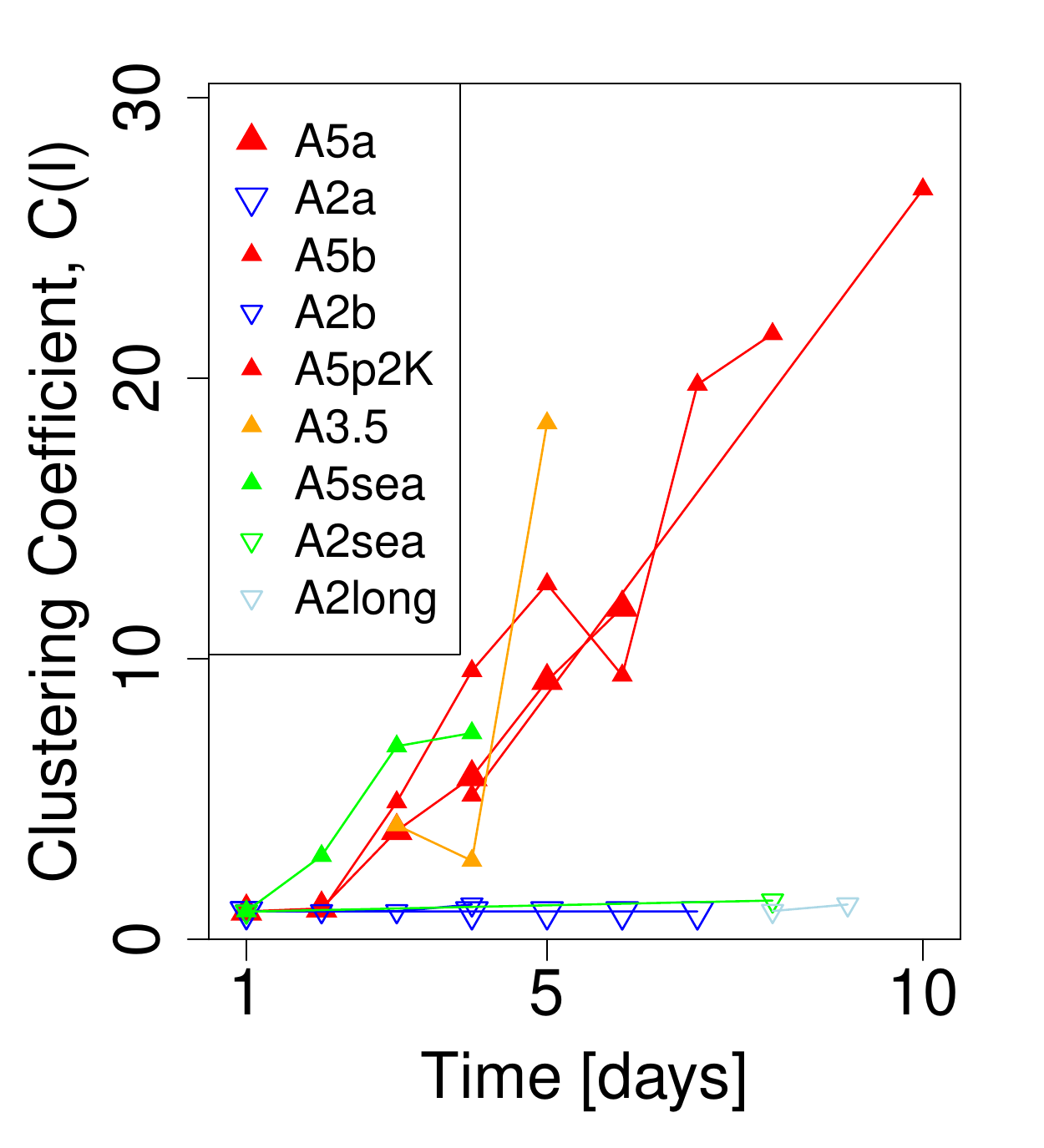}}
\put(5,-25){\includegraphics[trim={0cm 0cm 0cm 0cm}, clip, height=0.36\linewidth]{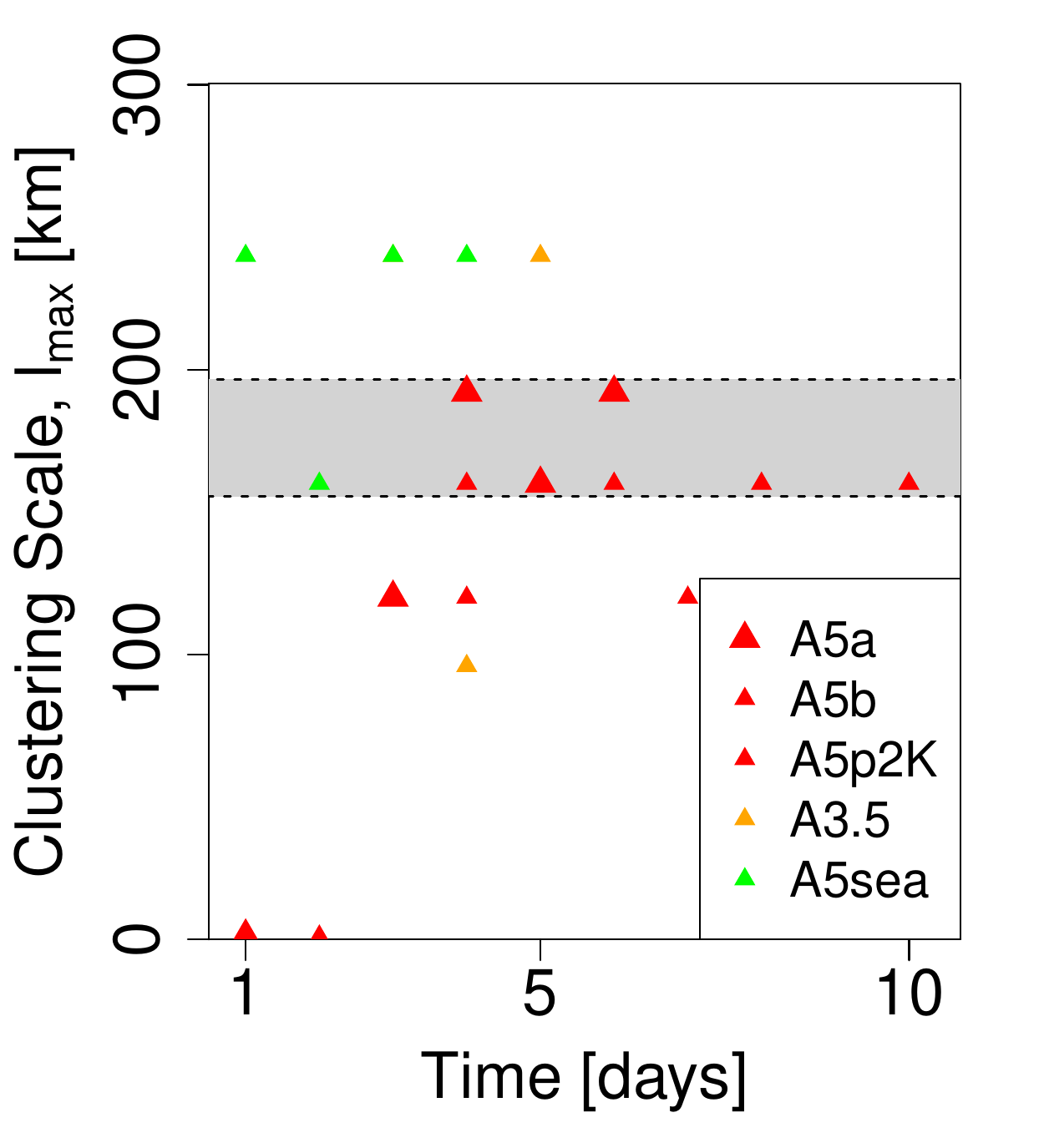}}
\put(60,-25){\includegraphics[trim={0cm 0cm 0cm 0cm}, clip, height=0.36\linewidth]{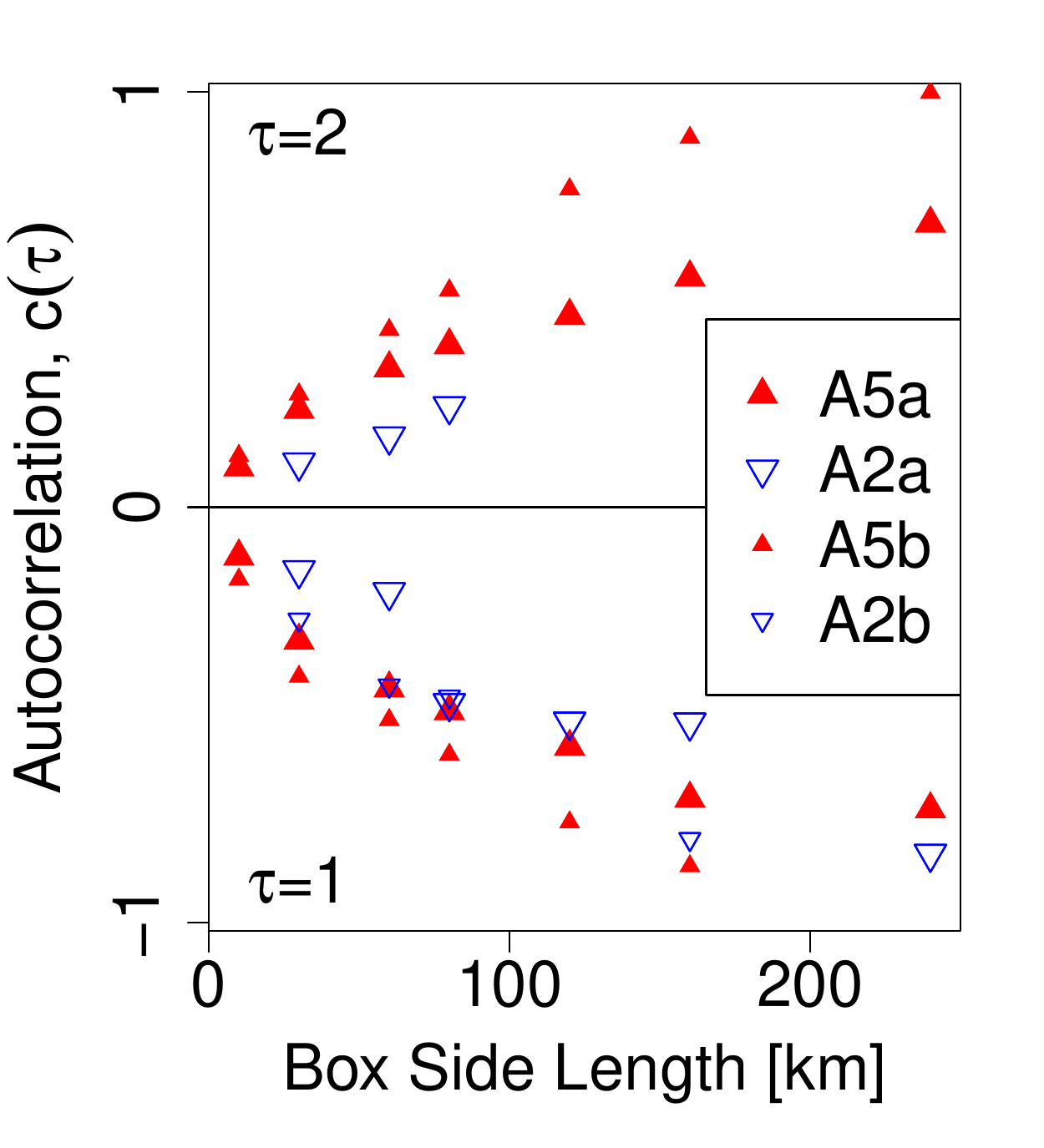}}
\put(-40,97){\large \bf a}
\put(15,97){\large \bf b}
\put(70,97){\large \bf c}
\put(-40,35){\large \bf d}
\put(15,35){\large \bf e}
\put(70,35){\large \bf f}
\put(-38,90){ $A5a$}
\put( 17,90){ $A3.5$}
\put( 72,90){ $A2a$}


\put(-36,53){regular}
\put(-36,75){clustered}

\end{overpic}
\vspace{1.8cm}
\caption{{\bf Quantifying clustering using spatial variance.}
{\bf a}, Spatial variance at different box sizes for $A5$.
Curves of colours ranging from red (day one) to green (day six) correspond to increasing time.
Note that at small scales ($\sim 10\;km$) or early times ($t<2\;d$), rain events are distributed regularly, while at larger scales ($\sim 180\;km$) and later times, events are clustered.
Note the double-logarithmic axis scaling.
{\bf b}, Analogous to (a), but for $A3.5$.
{\bf c}, Analogous to (a), but for $A2$. 
Note that the normalized variance remains in the normal range and does not exceed unity.
{\bf d}, Maximum of variance anomalies vs. time for different simulations ({\it see}: legend and Tab.~\ref{tab:experiments}). 
Note the general increase for large $\Delta T$ but flat behaviour for small $\Delta T$.
{\bf e}, Scales of clustering, i.e., the position of the variance maxima in $A3.5$ and $A5$. 
The grey shaded area marks the standard error of $l_{max}$, averaged over all times $t\geq 3d$. 
{\bf f}, Autocorrelation $c(\tau)$ for $\tau=1$ and $\tau=2$ show increasing $c(\tau)$ with scale for both A2 and A5.
Several points for $A2$ are not shown due to lack of statistical significance at the one percent confidence level.
}
\label{fig:quantifying_clustering}
\end{figure}

To quantify spatiotemporal inhomogeneities, we determine all surface rain event tracks and compute their center-of-mass positions ({\it Details:} Materials and Methods).
We then break the horizontal domain area down into square boxes of side length $l$, yielding $n(l)\equiv (L/l)^2$ such boxes, and determine the number of tracks located in each of the boxes.
The probability $p_l$ of a track occurring within one of the boxes at random would be $p_l=n(l)^{-1}$
and the binomial
\begin{equation}
    P_l(m)\equiv \binom{N}{m} p_l^m\left( 1-p_l \right)^{N-m}
    \label{eq:binomial}
\end{equation}
hence describes the probability of $m$ of $N$ randomly distributed tracks lying in one of these boxes during the model day.
The variance of counts $m$ at side length $l$ is \cite{feller1957introduction} 
\begin{equation}
    \Var_{ran}(l;N) = N\;p_l(1-p_l)\;,
    \label{eq:var_ran}
\end{equation}
which we compare to the variance of the empirical data 
\begin{equation}
    \Var_{emp}(l;N) = \sum_{i=1}^{n(l)}(m_i-\langle m\rangle)^2\;,
    \label{eq:var_emp}
\end{equation}
where $\langle m\rangle\equiv N/n(l)$ is the average number of tracks per box, and the sum is over all boxes $i$.
We now define a clustering coefficient $\mathcal{C}(l)\equiv \Var_{emp}(l;N)/\Var_{ran}(l;N)$, which is below or above unity, when tracks are regularly spaced or clustered, respectively.
Results show that, for low amplitudes ($A2$, Fig.~\ref{fig:quantifying_clustering}), $\mathcal{C}(l)<1$ for all box sizes $l$, hence, the spacing of tracks is generally more regular than expected at random.
For larger amplitudes ($A3.5$ and $A5$) regular spacing \cite{tompkins2017organization} with $\mathcal{C}(l)<1$ is found only for relatively small box sizes of $l\approx 20\;km$, whereas spacing at larger box sizes is strongly clustered, that is, $\mathcal{C}(l)\gg 1$.
Besides, this clustering increases over time (Fig.~\ref{fig:quantifying_clustering}d).

We also define a box size $l_{max}$, at which $\mathcal{C}(l)$ is maximal.
Despite some variation, $l_{max}\approx 180\;km$ can be identified from A5 (Fig.~\ref{fig:quantifying_clustering}e).
Additionally, we measure the autocorrelation $c(\tau)$ between day $d$ and $d+\tau$ by computing the Pearson correlation coefficient of daily mean precipitation rates from all grid boxes (Fig.~\ref{fig:quantifying_clustering}f). 
We find rainfall to be anticorrelated from one day to the next for all box sizes ($c(1)<0$), suggesting a local inhibitory effect of rain and positively correlated two days into the future ($c(2)>0$).
The magnitudes of $½½c(1)$ and $c(2)$ both increase with box size $l$, but appear to level off near $l=200\;km$.

\begin{figure}[ht]
\centering
\begin{overpic}[width=0.4\textwidth]{dummy.pdf}

\put(-50,15){\includegraphics[trim={0cm 2.15cm 1.8cm 0cm}, clip, height=0.205\linewidth]{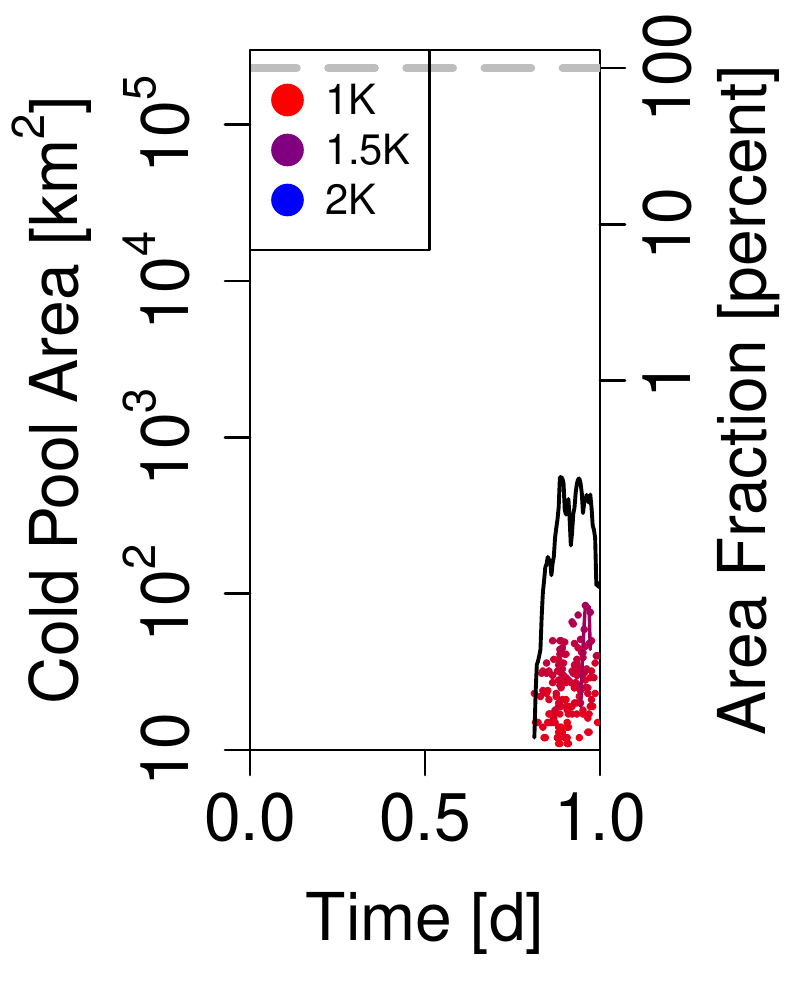}}
\put(-23,15){\includegraphics[trim={1.95cm 2.15cm 1.8cm 0cm}, clip, height=.205\linewidth]{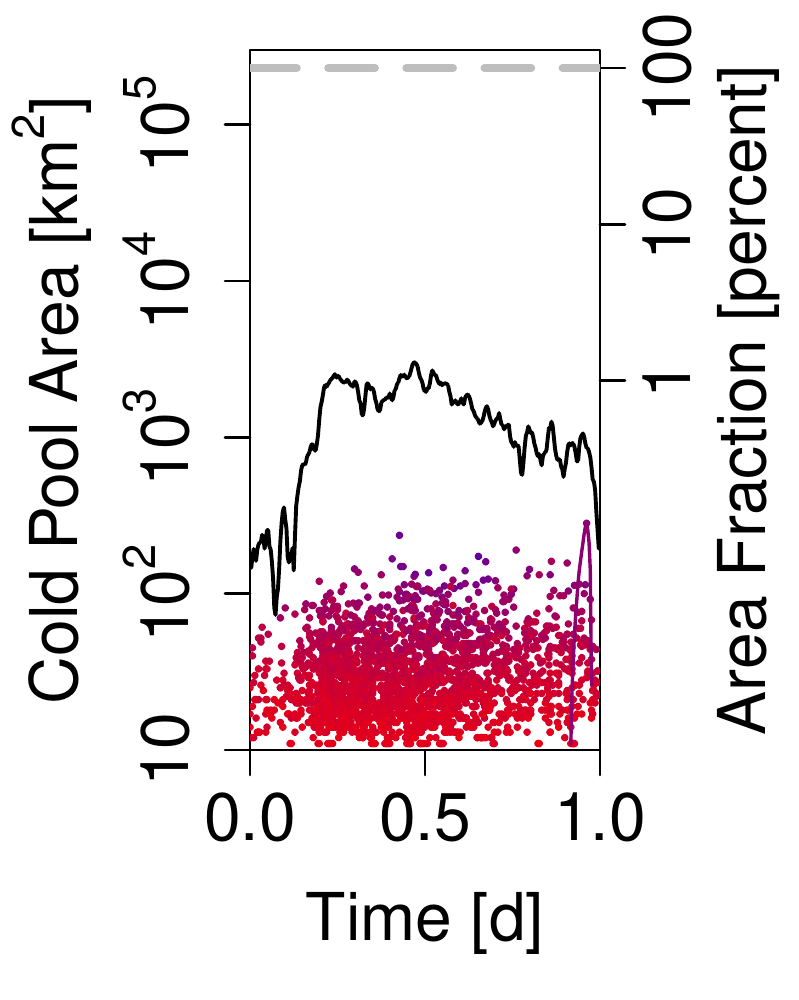}}
\put(-5,15){\includegraphics[trim={1.95cm 2.15cm 0cm 0cm}, clip, height=0.205\linewidth]{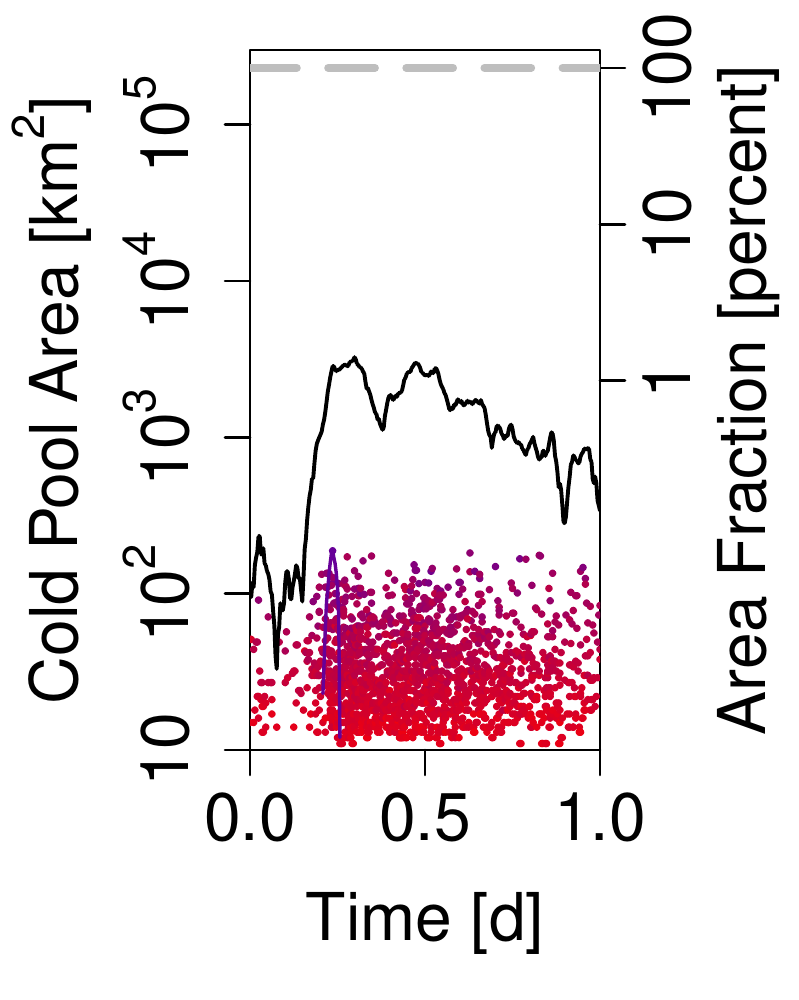}}

\put(-50,-31){\includegraphics[trim={0cm 0cm 1.8cm 0cm}, clip, height=0.255\linewidth]{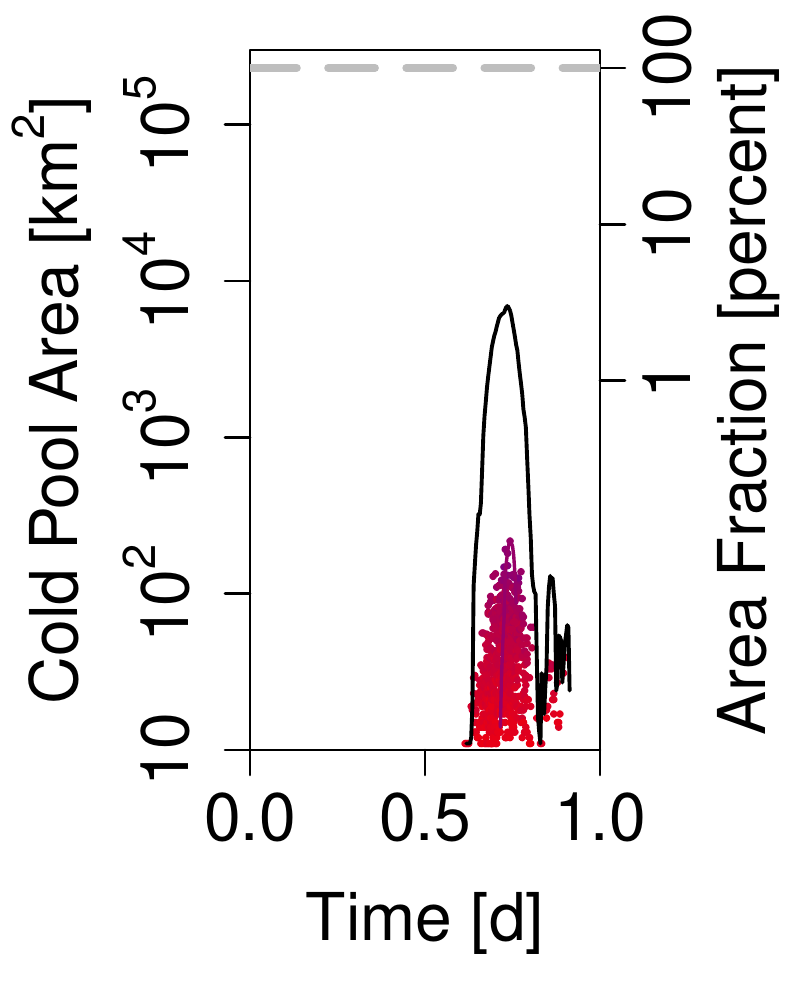}}
\put(-23,-31){\includegraphics[trim={1.95cm 0cm 1.8cm 0cm}, clip, height=0.255\linewidth]{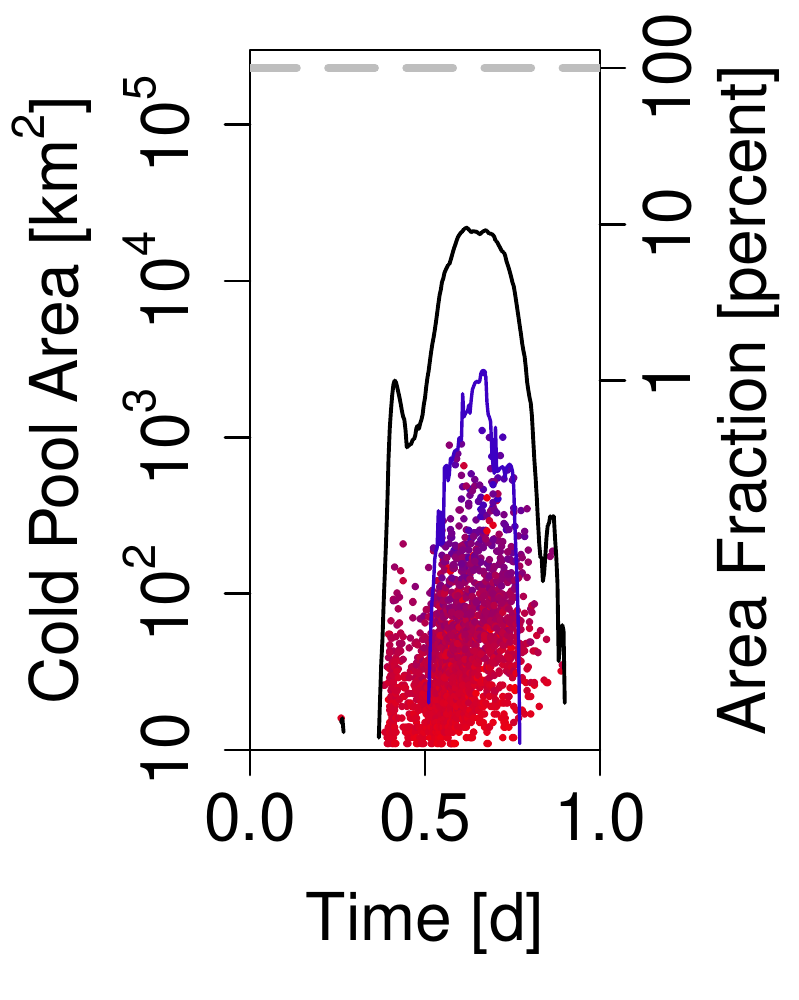}}
\put(-5,-31){\includegraphics[trim={1.95cm 0cm 0cm 0cm}, clip, height=0.255\linewidth]{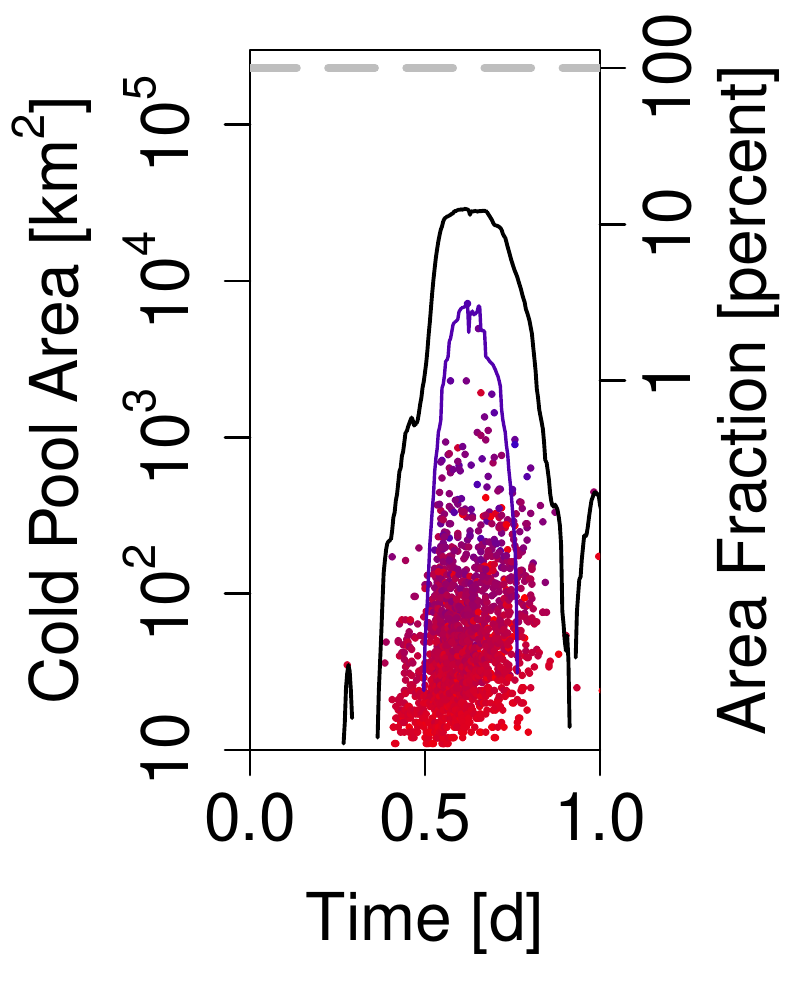}}

\put(18,16){\includegraphics[trim={.0cm 2.25cm 0cm 0cm}, clip, height=0.20\linewidth]{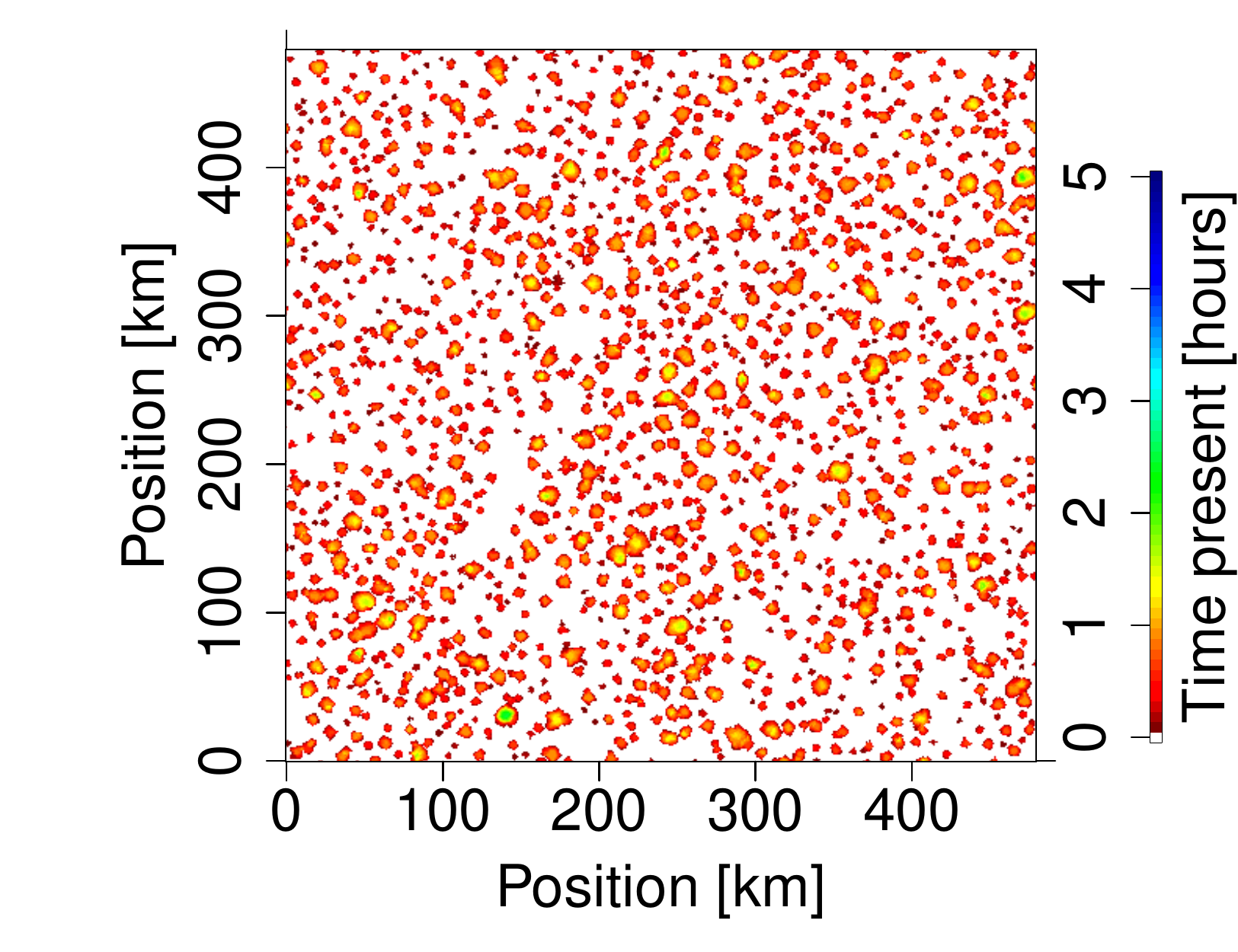}}
\put(18,-28){\includegraphics[trim={.0cm 0cm 0cm 0cm}, clip,height=0.24\linewidth]{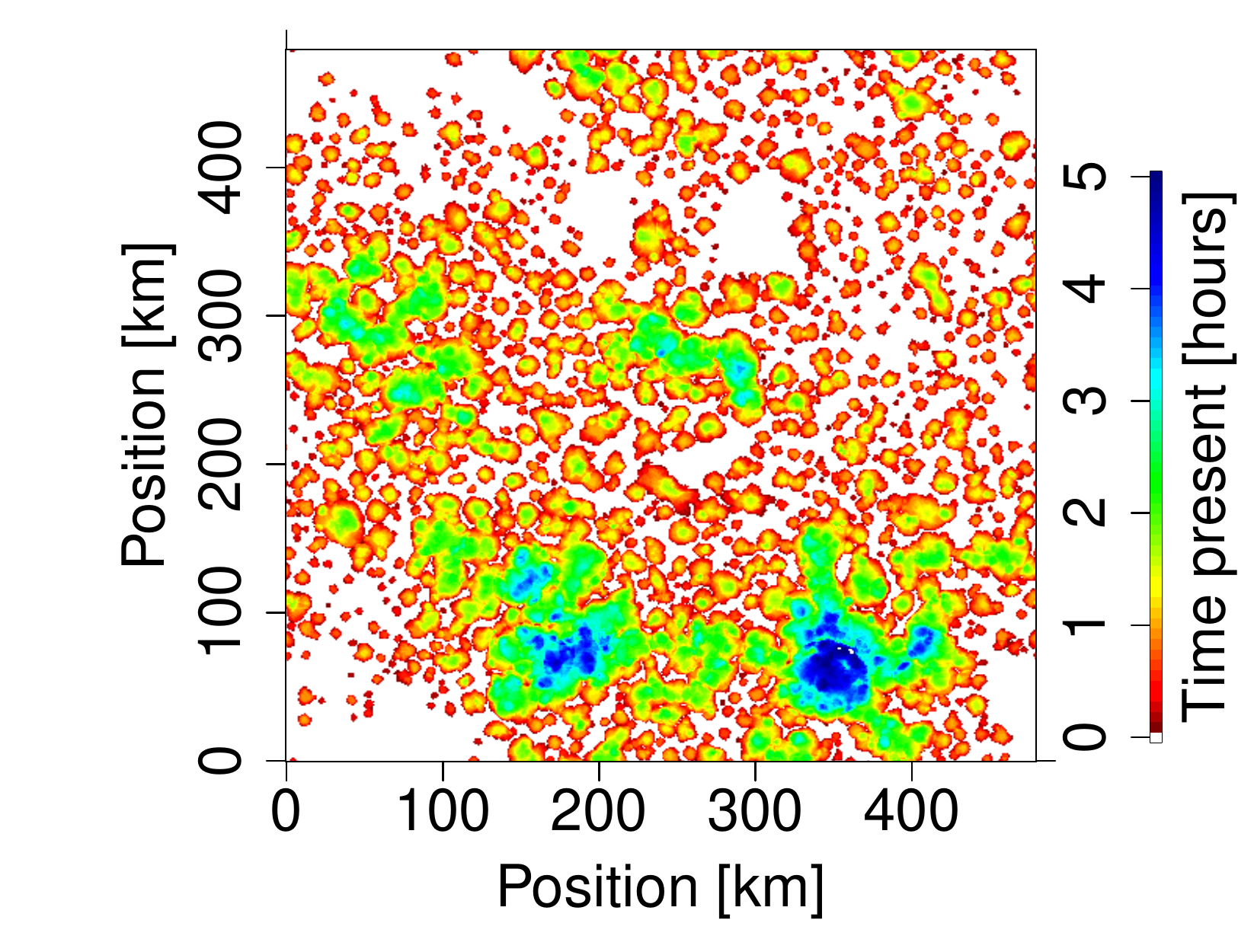}}

\put(75,16){\includegraphics[trim={.0cm 2.15cm 0cm 0cm}, clip, width=0.27\linewidth]{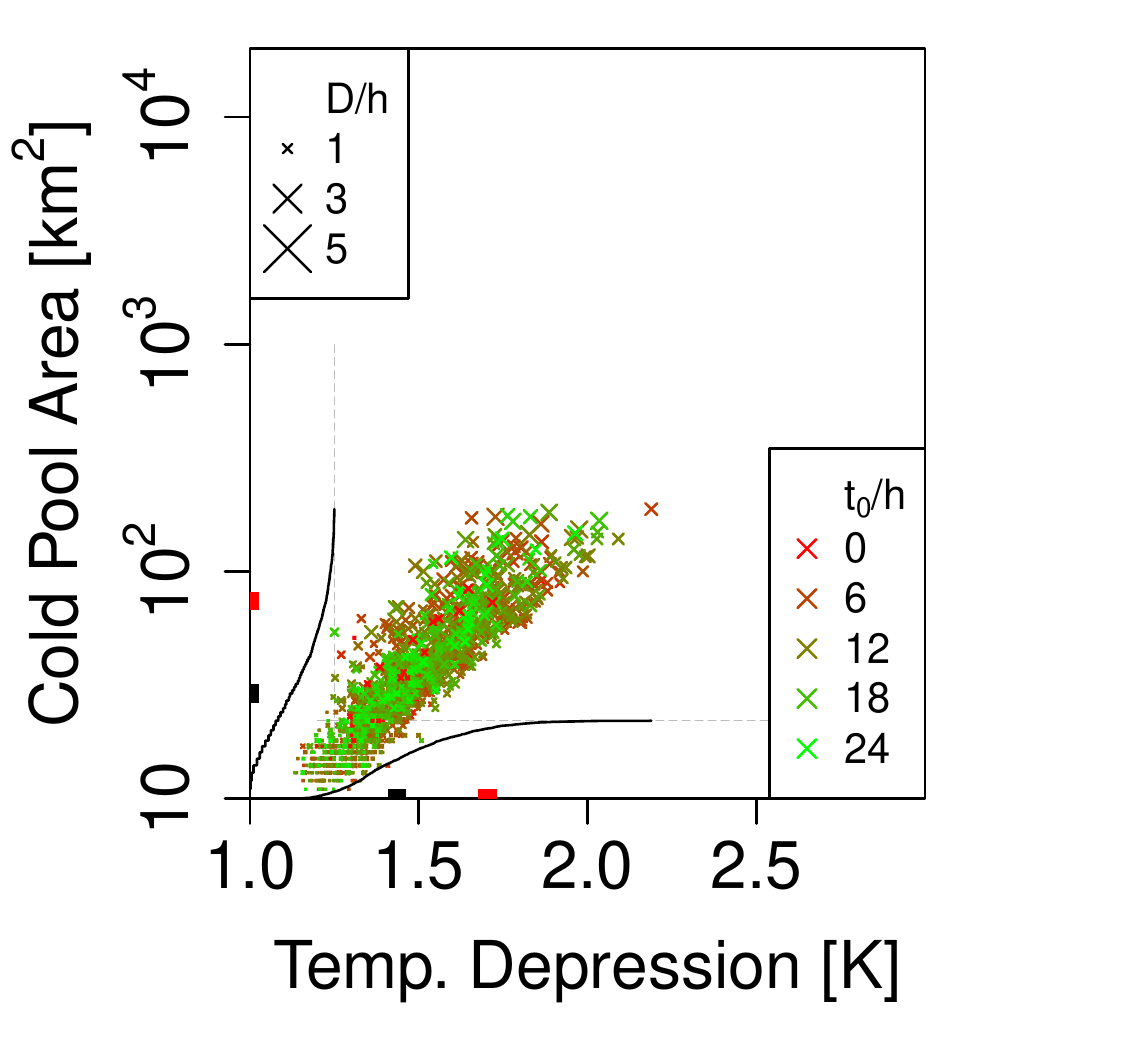}}
\put(75,-30){\includegraphics[trim={.0cm 0cm 0cm 0cm}, clip,width=0.27\linewidth]{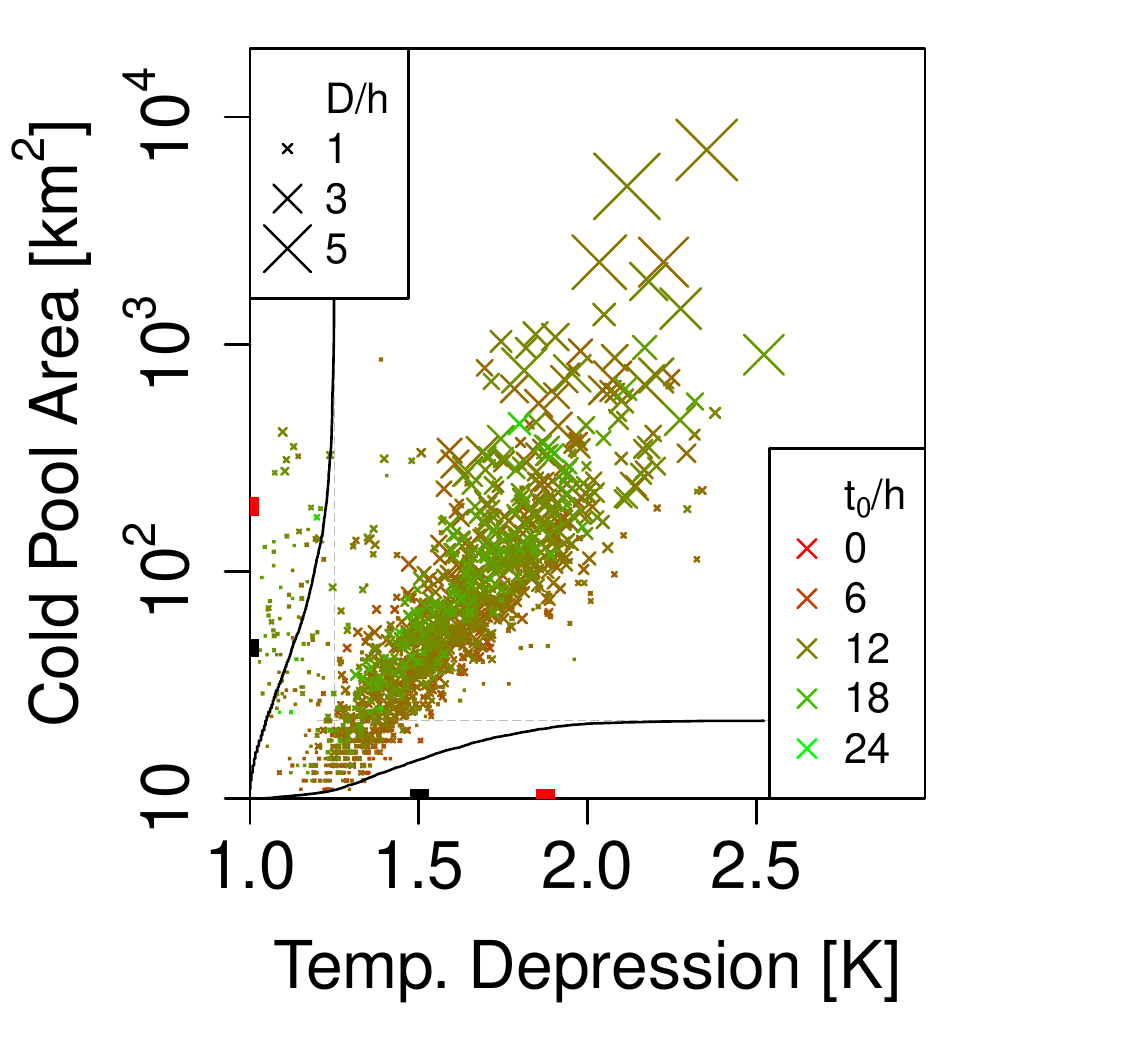}}

\put(-38,50){\bf a}
\put(-21,50){\bf b}
\put(-2,50){\bf c}
\put(31,50){\bf d}
\put(85,50){\bf e}

\put(-38,13){\bf f}
\put(-21,13){\bf g}
\put(-2,13){\bf h}
\put(31,13){\bf i}
\put(85,13){\bf j}

\put(-37,36){day 1}
\put(-19,36){day 2}
\put(-1,36){day 4}

\put(54,50){day 4}
\put(104,50){day 4}
\end{overpic}
\vspace{2.4cm}
\caption{{\bf Cold pool merging and deepening.}
{\bf a}, CP occurrence time, maximum area, and average temperature depression (colours from red to blue, {\it see} legend) on day one of A2b. The black curve indicates the total CP area at each time, whereas the thin coloured curve highlights the timeseries of the largest CP during the respective day. 
{\bf b,c}, Analogous to (a), but for days two and four of A2b.
{\bf d}, Areas covered by CPs during day four of A2b, the colours ({\it see} colourbar) indicate the duration during which CPs were present ({\it compare:} Fig.~\ref{fig:hor_wind_speed}).
{\bf e}, CP area vs.~the corresponding maximum of areal mean temperature depression (day four).
Symbol sizes indicate CP lifetime and colours indicate occurrence time within the model day ({\it see} legends).
{\bf f---j}, Analogous to (a)---(e), but for A5b.
Note the logarithmic vertical axis scale in (a)---(c), (e), and (f)---(h), (j).
}
\label{fig:CP_merging}
\end{figure}
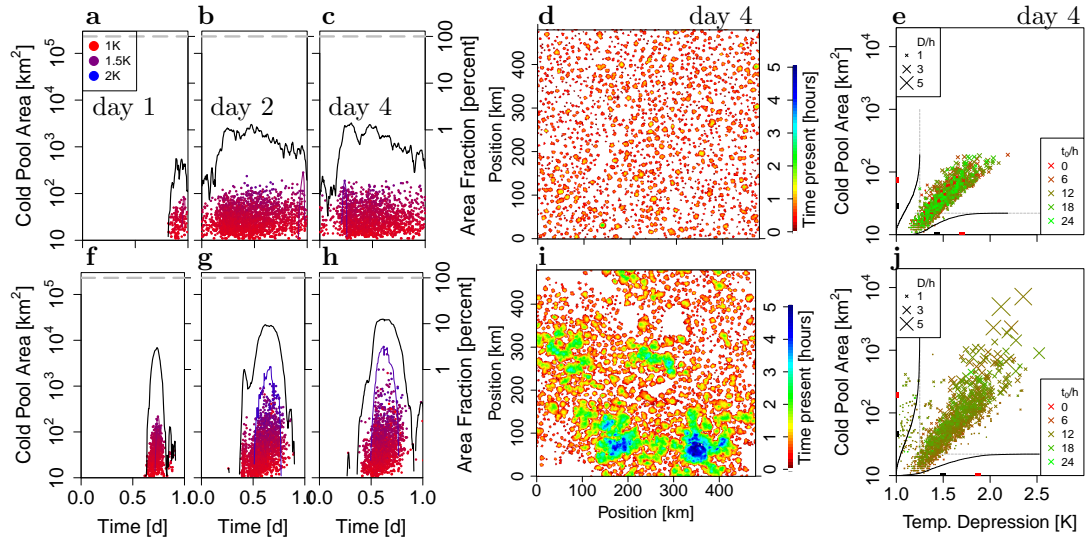

\noindent
{\bf Clustering as a result of cell density difference.}
Areal rain cell density differences between $A2$ and $A5$ are evident from the rainfall diurnal cycle (Fig.~\ref{fig:daily_mean}b), where the initial peak is nearly twice as high for $A5$ compared to $A2$. 
As mentioned, CPs are known to mediate interactions between convective rain cells, and we expect interactions to become more relevant at increased rain cell density.
To quantify density effects on the clustering dynamics in $A2$ vs. $A5$, we first perform a simple cold pool tracking using buoyancy anomalies of threshold $1\;K$ as a measure ({\it Details:} Materials and Methods).
In $A2$, CPs typically do not exceed areas of $500$ $km^2$ (Fig.~\ref{fig:CP_merging}a---d), have modest temperature depressions and lifetimes of generally less than two hours (Fig.~\ref{fig:CP_merging}a---e). 
In $A5$, where CPs only occur during parts of the day, CP areas often exceed $10^3\;km^2$, and such large CPs have much stronger temperature depressions and substantially longer lifetimes (Fig.~\ref{fig:CP_merging}f---j). 
The formation of very large and strongly negatively buoyant CPs suggests that CPs from distinct rain cells often bunch together before each CP has fully expanded. 
We employ the CP tracking to detect merging events, that is, those where two previously separate CP areas combine.
For A5b, CP merging is indeed ubiquitous, whereas it is all but absent in A2b (Fig.~\ref{fig:CP_height_distribution}c).

Additionally, computing the height of CPs, in $A5b$, when CPs first appear, they show heights comparable to those of $A2b$ (Figs~\ref{fig:CP_height_distribution} and \ref{fig:hor_wind_speed}). 
Subsequently, a double peak forms, where groups of CPs develop much larger heights, reaching close to the level of free convection ($\approx 1100\;m$).
Larger CP heights $h$ and deeper temperature depressions $\theta'$ are consistent with higher surface wind speed $v_{cp}\sim (h\theta')^{-1/2}$, \cite{etling2008theoretische} ({\it compare}: panels in Fig.~ \ref{fig:hor_wind_speed}) and surface fluxes (Fig.~\ref{fig:multi-day_timeseries}e---h) during the time of rainfall.

\begin{figure}[ht]
\centering
\begin{overpic}[width=0.4\textwidth ]{dummy.pdf}
\put(-54,57){\includegraphics[trim={.0cm 0cm 0cm 0cm}, clip,height=0.25\linewidth]{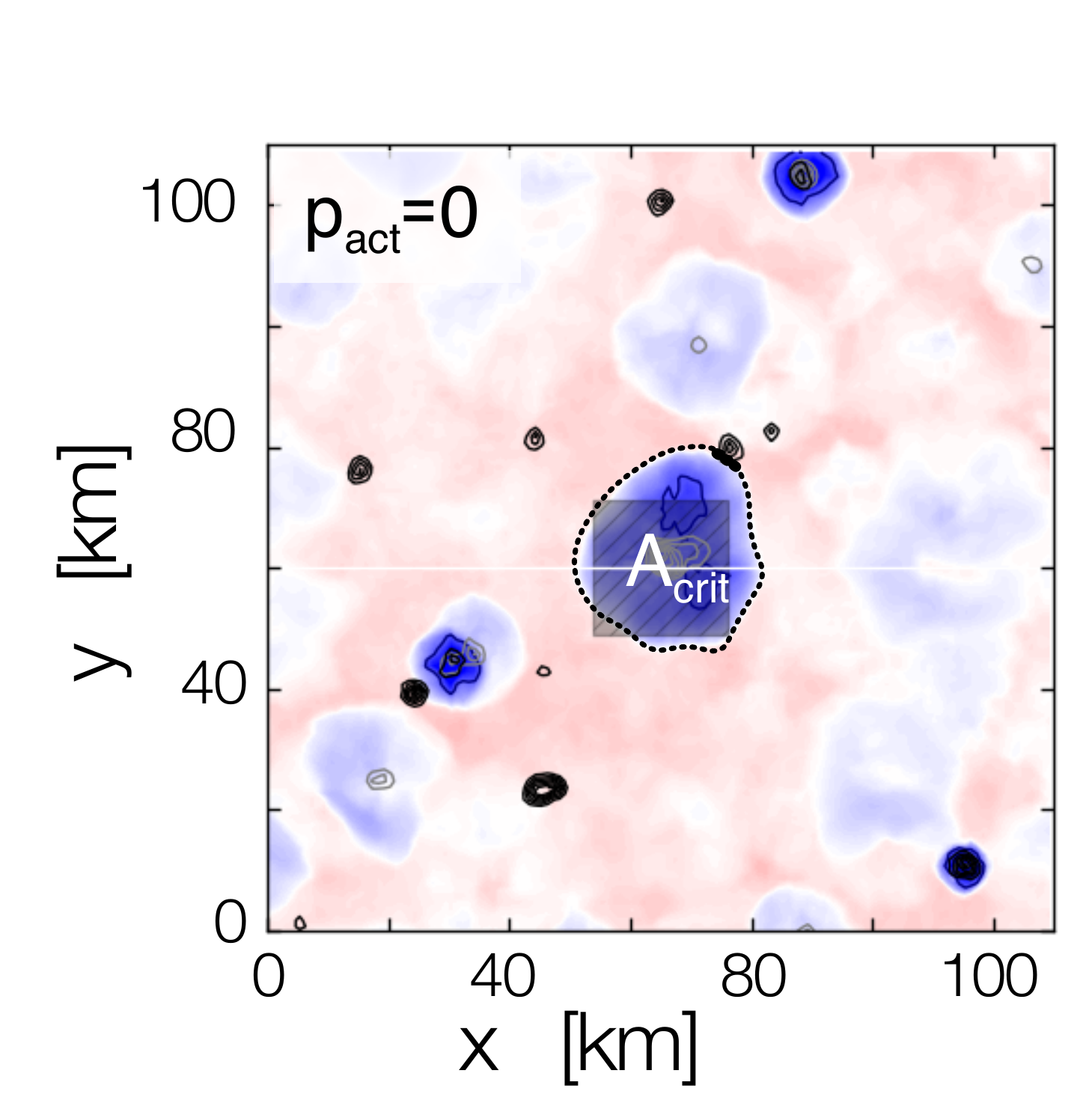}}
\put(-10,57){\includegraphics[trim={.0cm 0cm 0cm 0cm}, clip,height=0.25\linewidth]{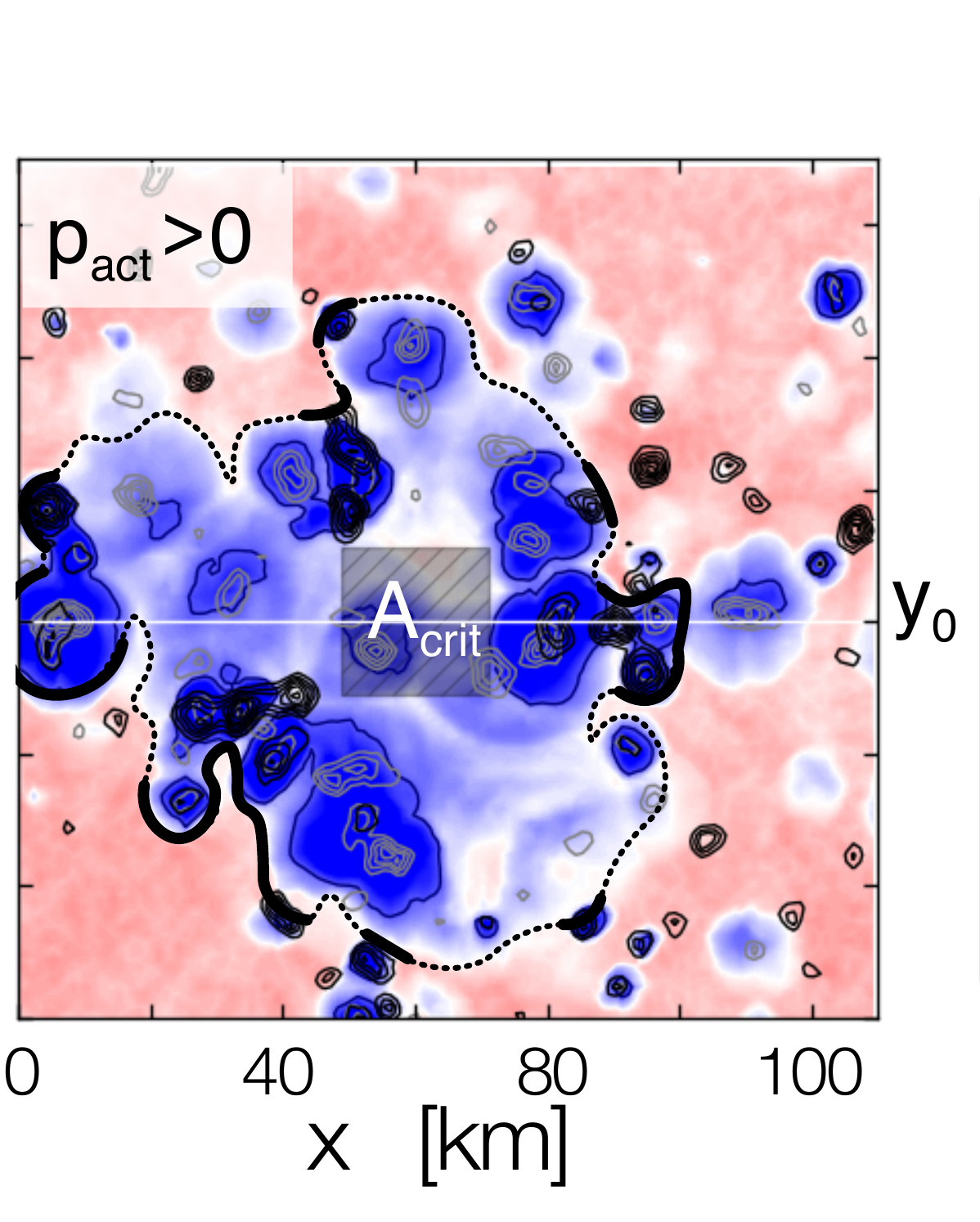}}
\put(31,57){\includegraphics[trim={.0cm 0cm 0cm 0cm}, clip,height=0.25\linewidth]{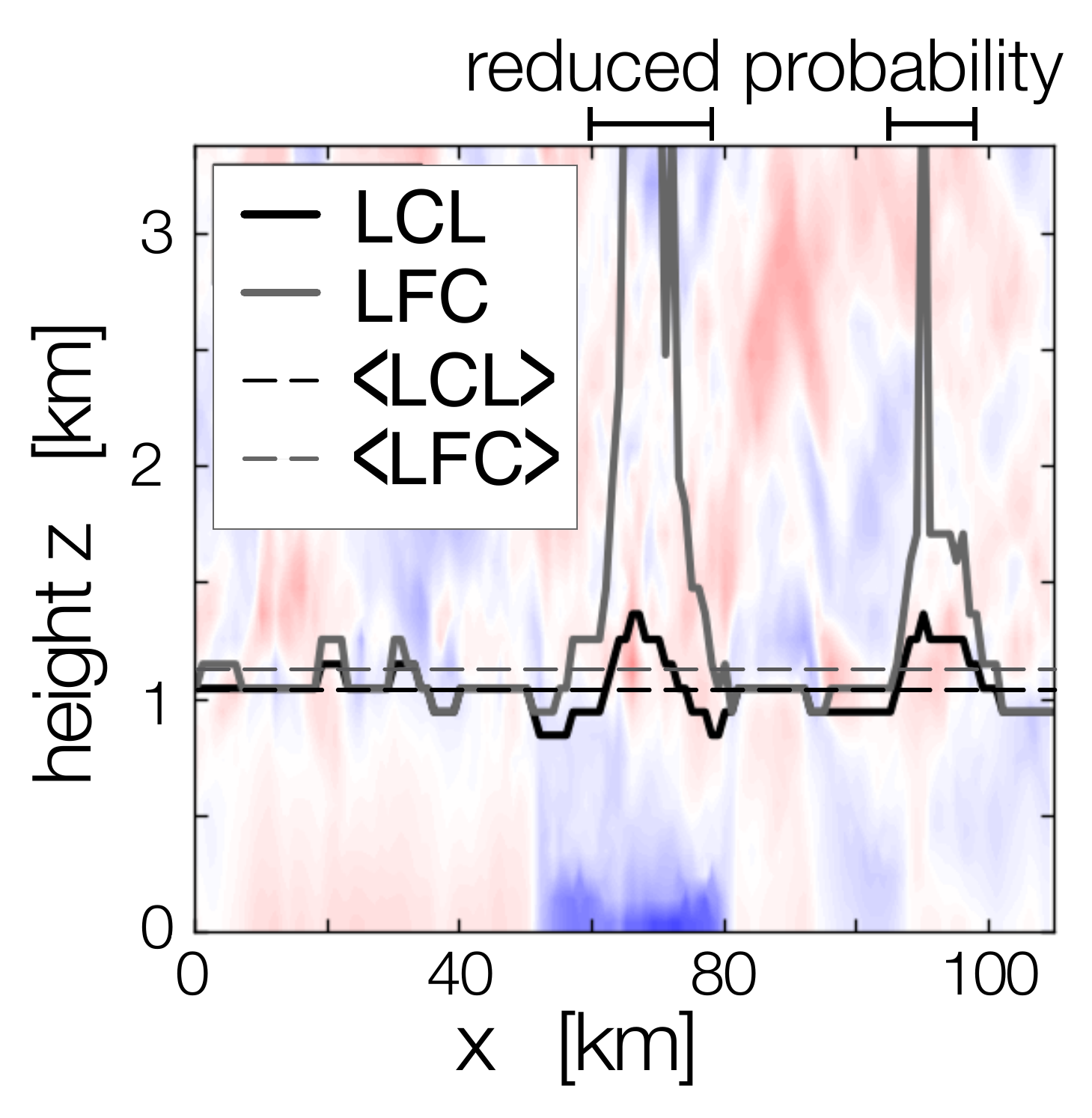}}
\put(75,57){\includegraphics[trim={.0cm 0cm 0cm 0cm}, clip,height=0.25\linewidth]{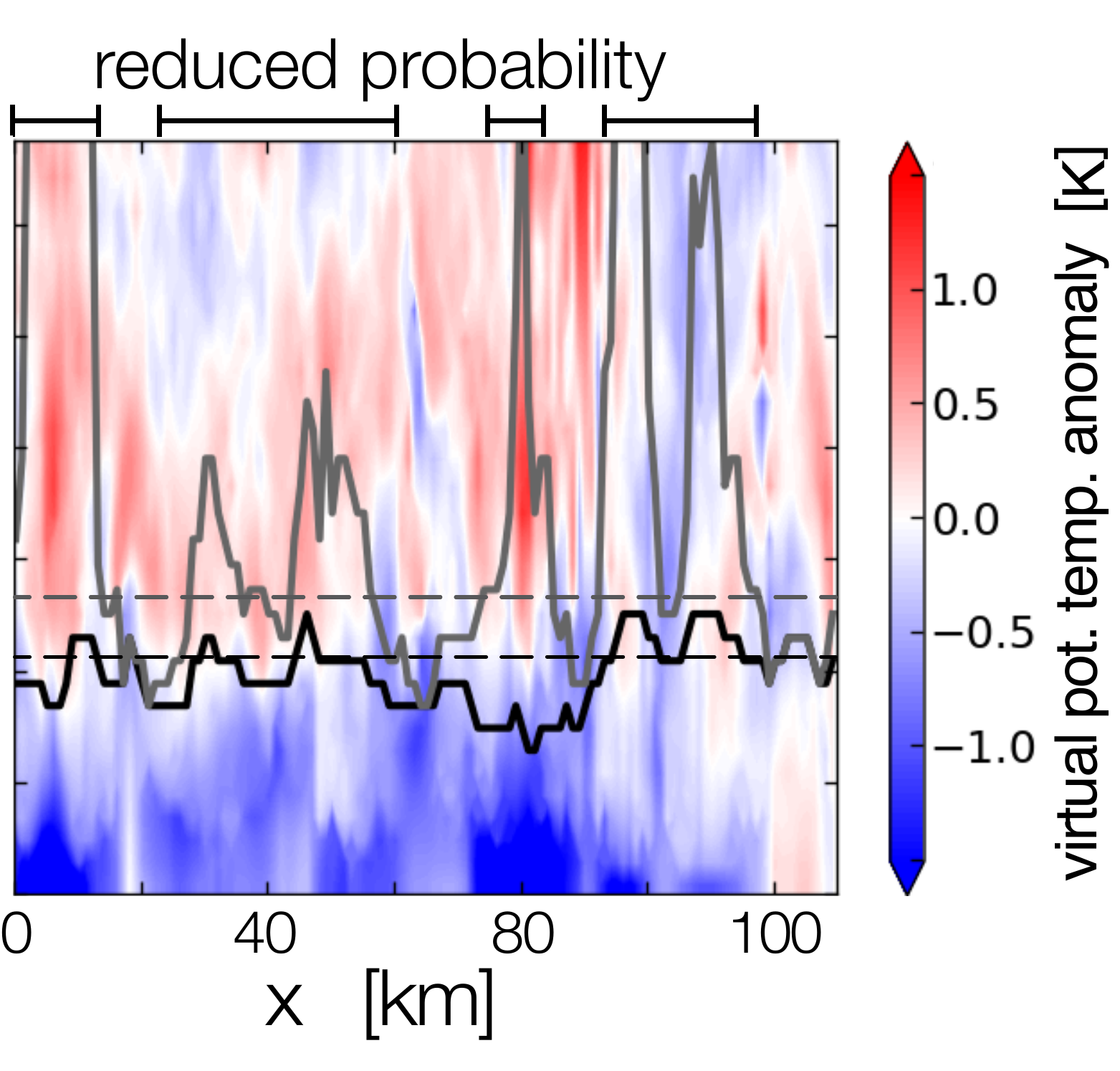}}
\put(-52,100){\large \bf a}
\put(-10,100){\large \bf b}
\put(37,100){\large \bf c}
\put(74,100){\large \bf d}

\put(-53,0){
\includegraphics[height=0.28\linewidth]{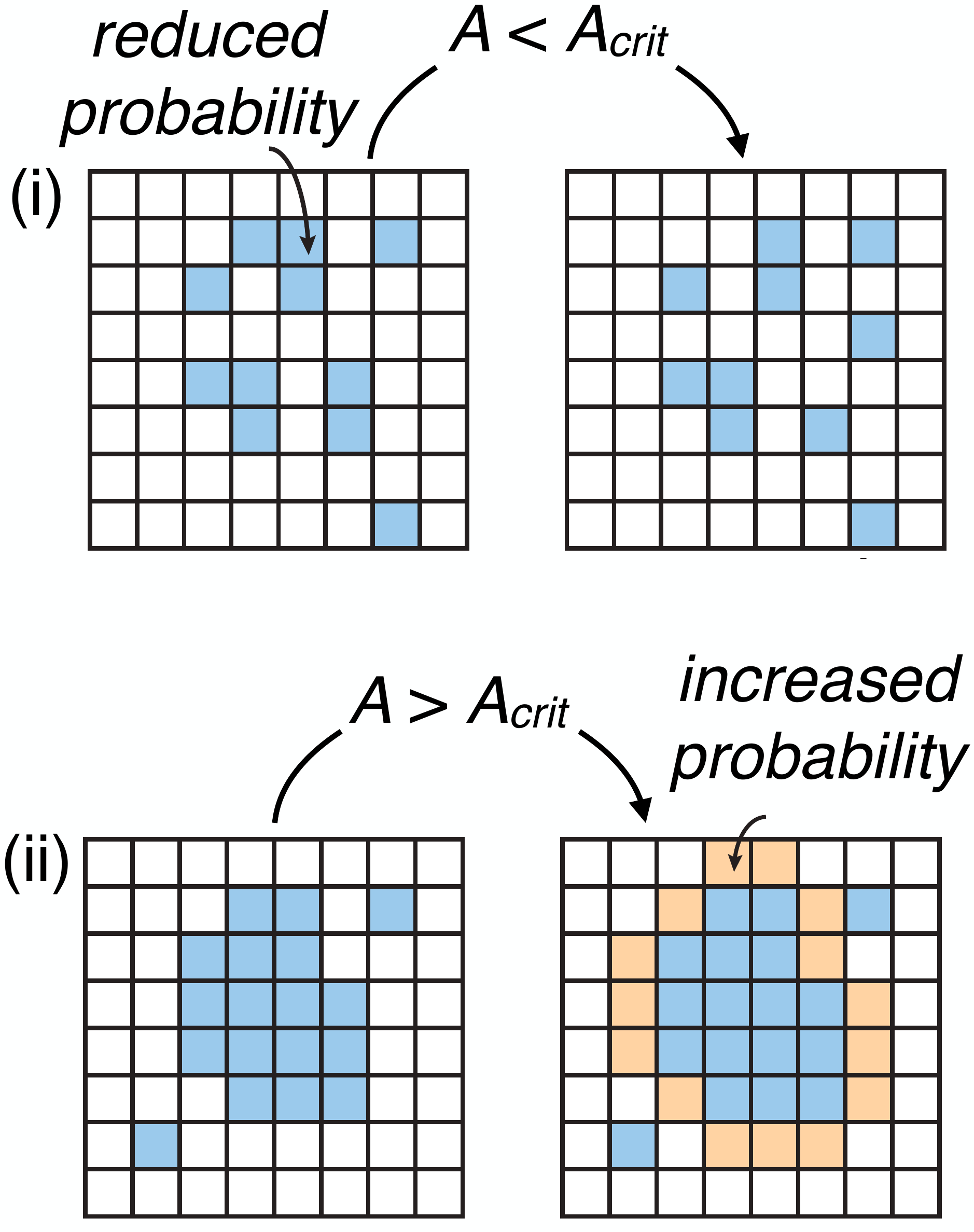}}
\put(76,25){\includegraphics[trim={.0cm 0cm 0cm 0cm}, clip,height=0.165\linewidth]{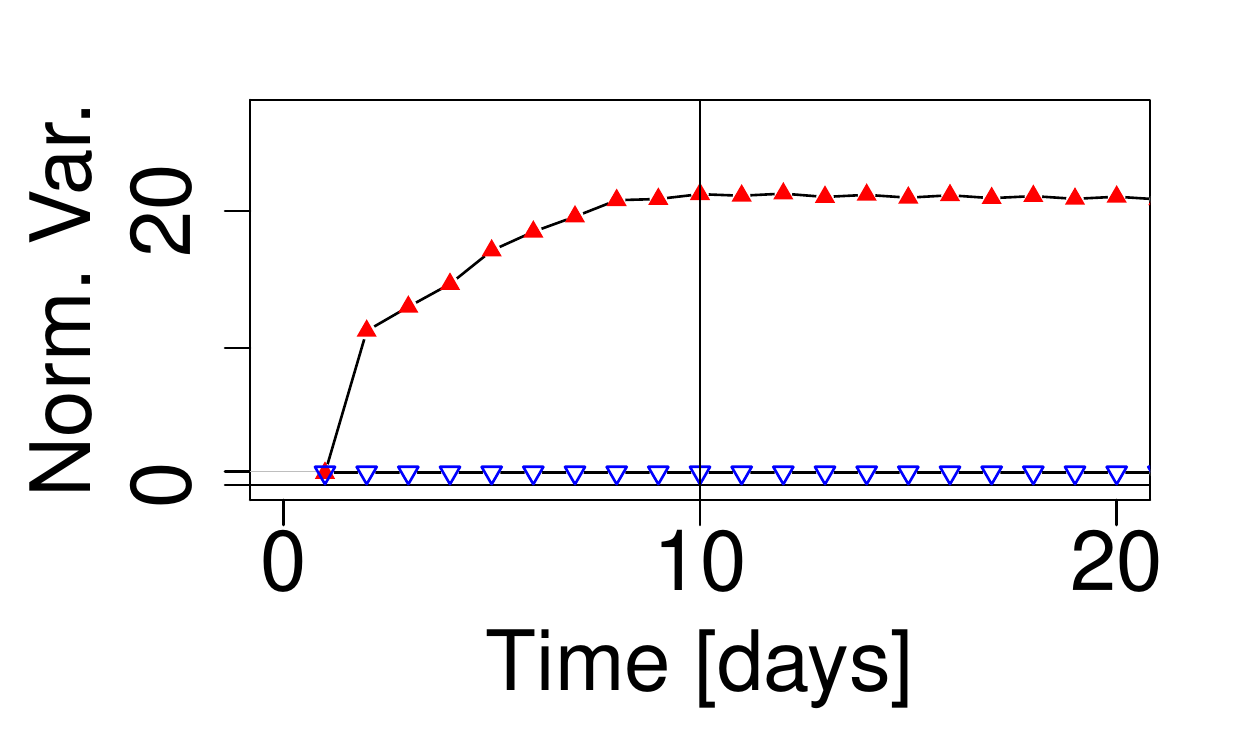}}
\put(78,12){\includegraphics[trim={2cm 2cm 4cm 0cm}, clip,height=0.065\linewidth]{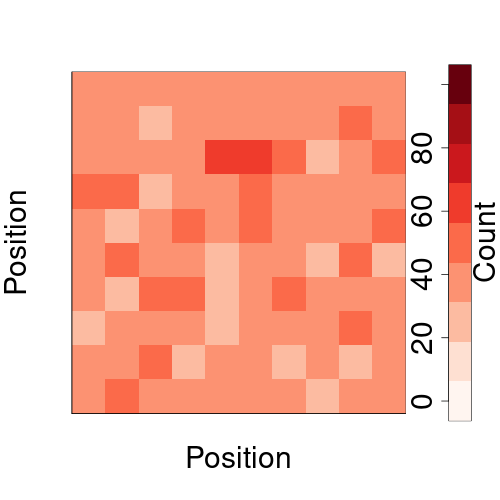}}
\put(87,12){\includegraphics[trim={2cm 2cm 4cm 0cm}, clip,height=0.065\linewidth]{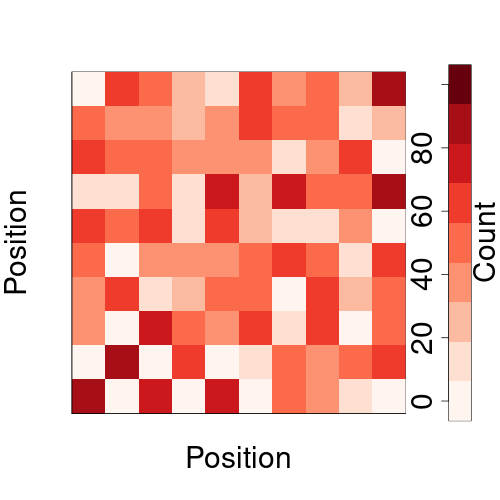}}
\put(96,12){\includegraphics[trim={2cm 2cm 4cm 0cm}, clip,height=0.065\linewidth]{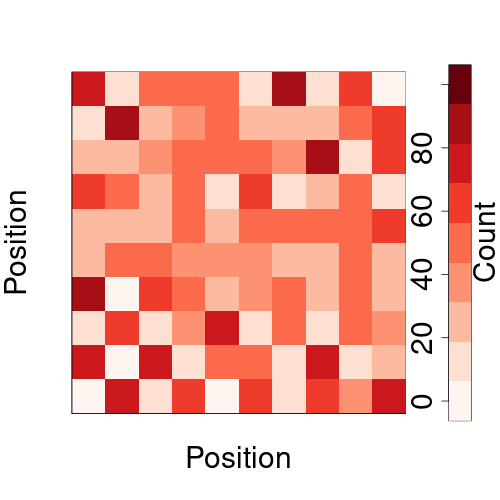}}
\put(105,12){\includegraphics[trim={2cm 2cm 4cm 0cm}, clip,height=0.065\linewidth]{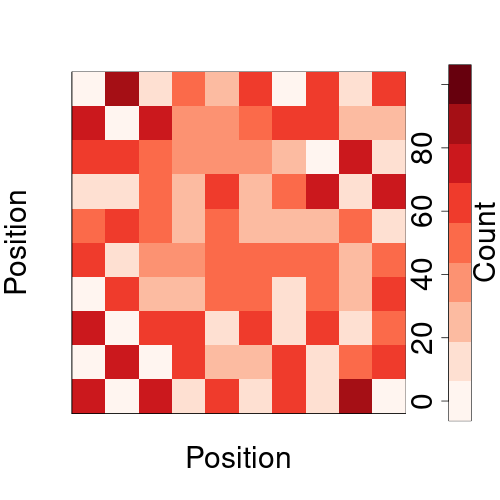}}
\put(114,12){\includegraphics[trim={2cm 2cm 4cm 0cm}, clip,height=0.065\linewidth]{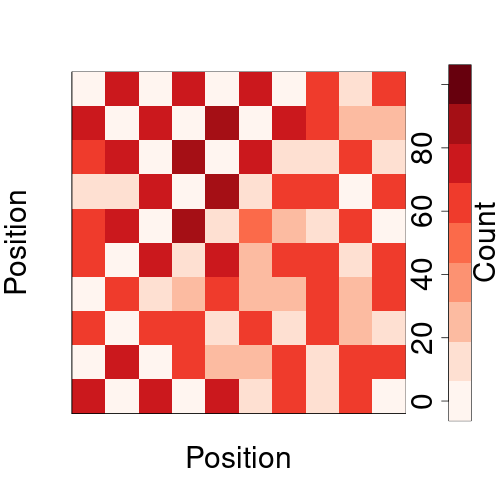}}
\put(78,-2){\includegraphics[trim={2cm 2cm 4cm 0cm}, clip,height=0.065\linewidth]{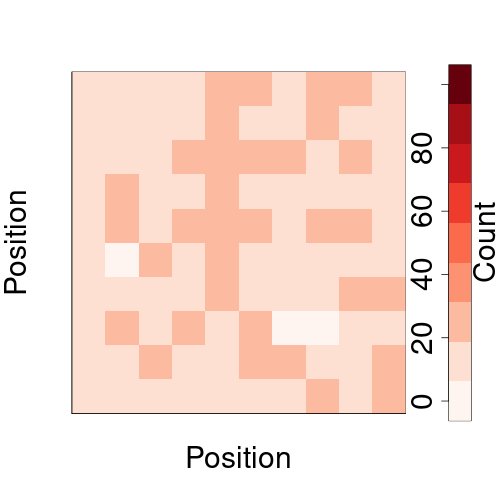}}
\put(87,-2){\includegraphics[trim={2cm 2cm 4cm 0cm}, clip,height=0.065\linewidth]{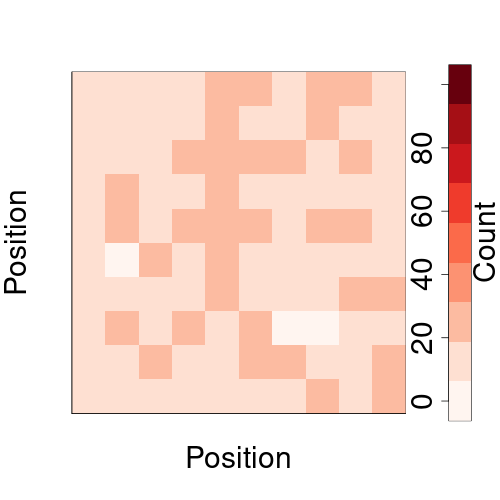}}
\put(96,-2){\includegraphics[trim={2cm 2cm 4cm 0cm}, clip,height=0.065\linewidth]{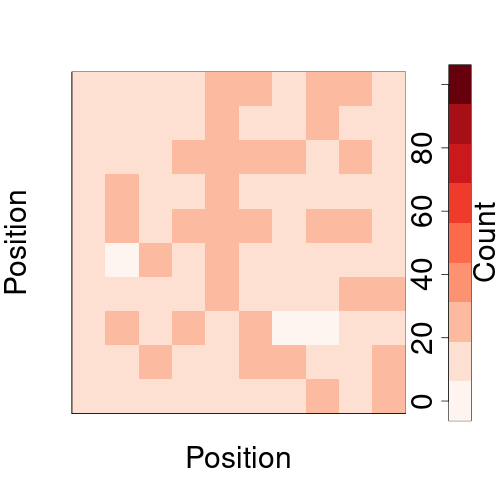}}
\put(105,-2){\includegraphics[trim={2cm 2cm 4cm 0cm}, clip,height=0.065\linewidth]{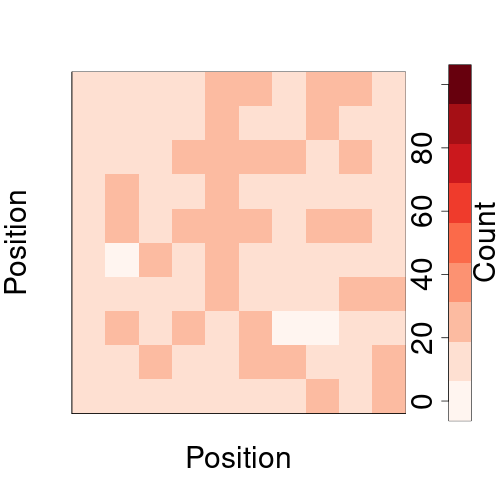}}
\put(114,-2){\includegraphics[trim={2cm 2cm 4cm 0cm}, clip,height=0.065\linewidth]{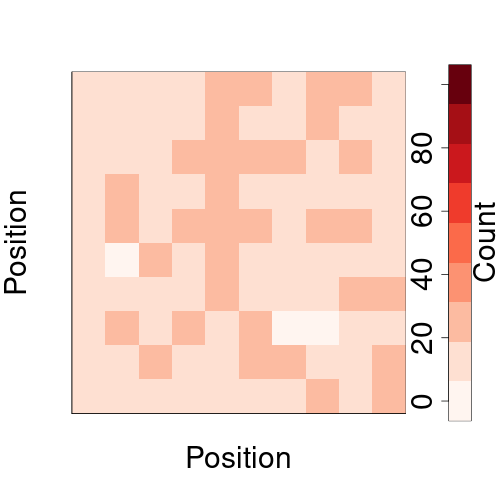}}
\put(-11,-3){\includegraphics[trim={0cm 0cm 0cm 0cm}, clip, height=0.32\linewidth]{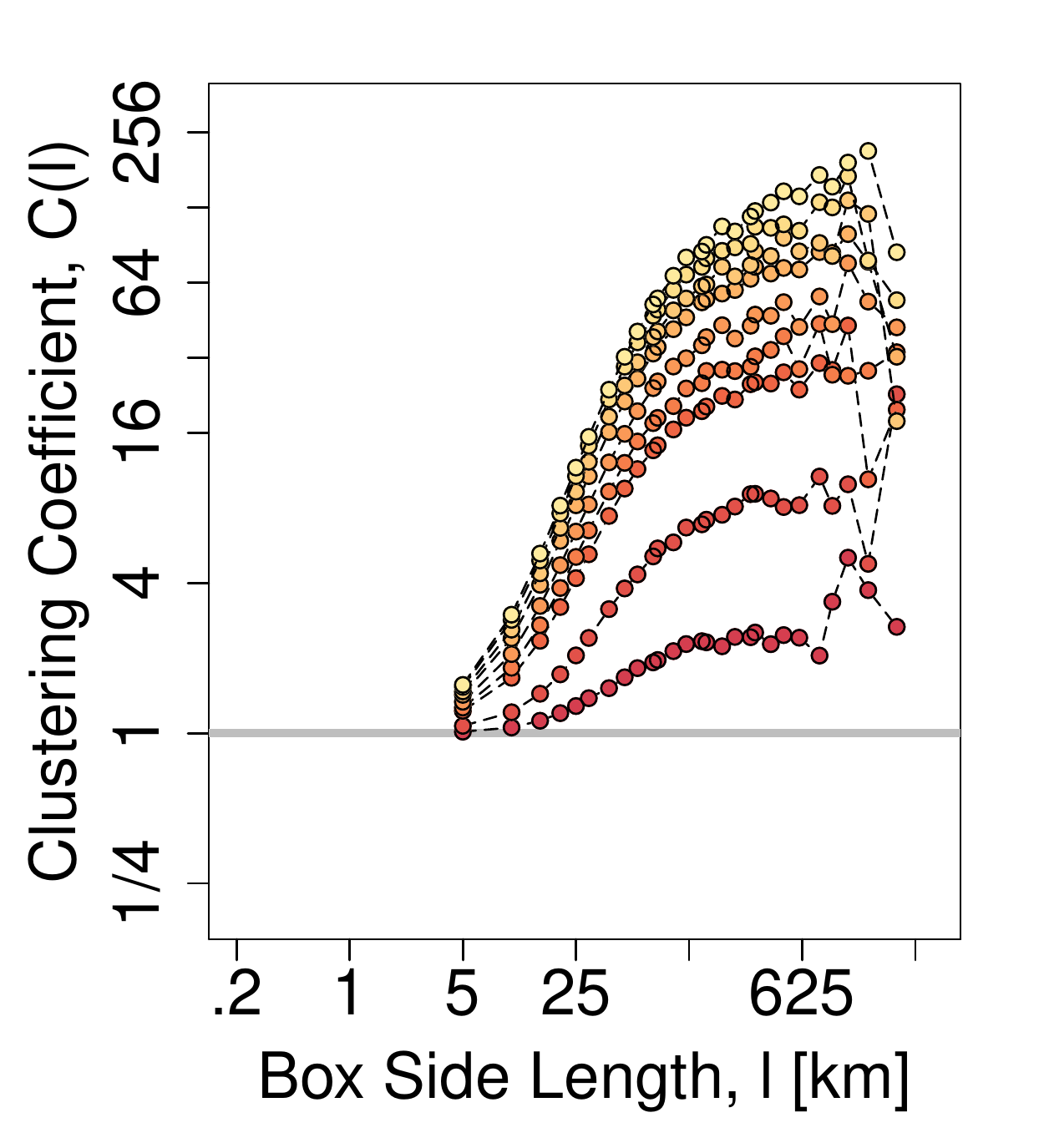}}
\put(37,-3){\includegraphics[trim={2cm 0cm 0cm 0cm}, clip, height=0.32\linewidth]{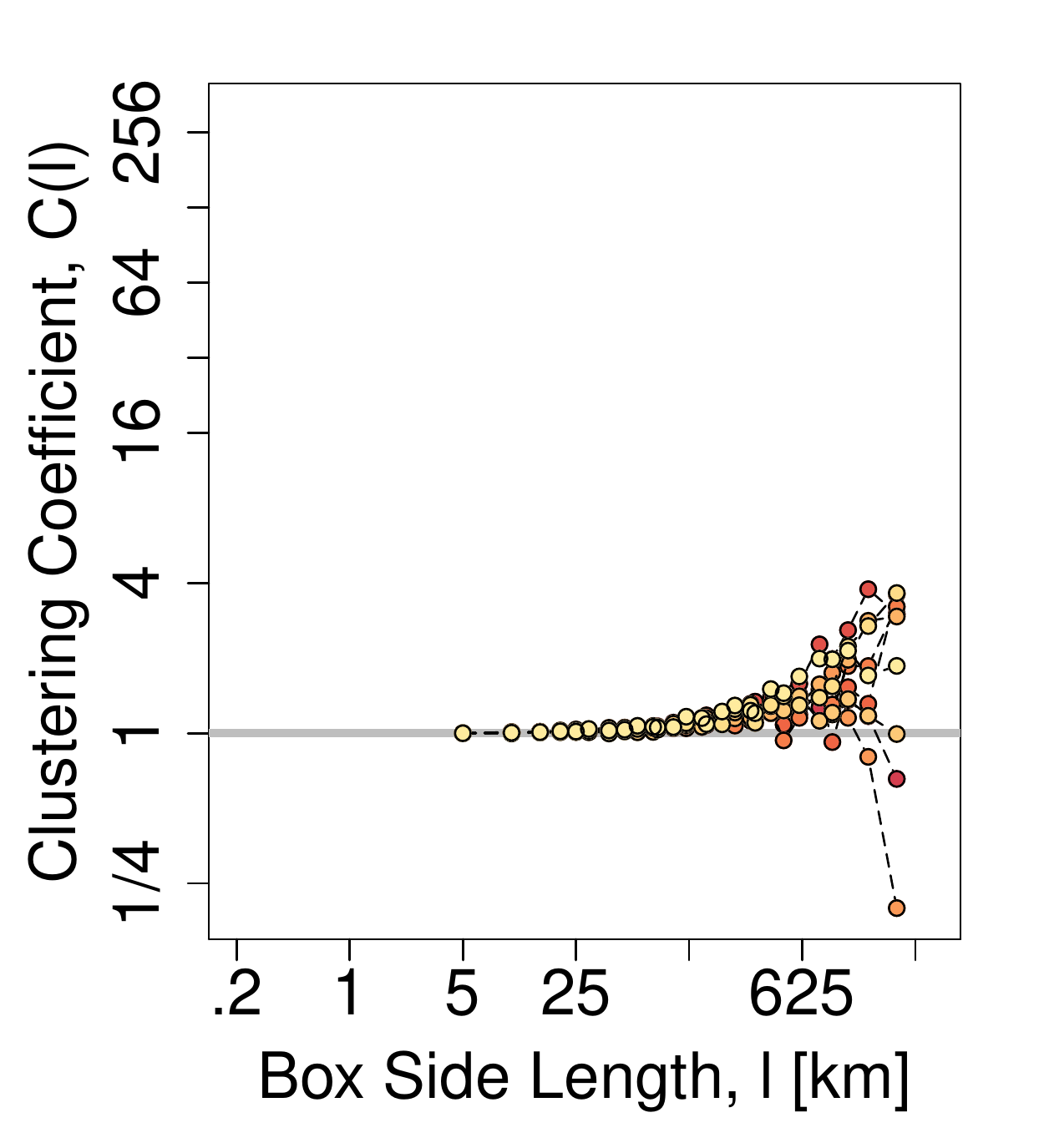}}

\put(-52,51){\large \bf e}
\put(1,51){\large \bf f}
\put(40,51){\large \bf g}
\put(78,51){\large \bf h}

\put(77,22){day \textcolor{red} 1}
\put(90,22){\textcolor{red} 2}
\put(99,22){\textcolor{red} 3}
\put(108,22){\textcolor{red} 4}
\put(117,22){\textcolor{red}{10}}

\put(81,8){\textcolor{blue} 1}
\put(90,8){\textcolor{blue} 2}
\put(99,8){\textcolor{blue} 3}
\put(108,8){\textcolor{blue} 4}
\put(117,8){\textcolor{blue}{10}}

\put(0,12){$t_a=.5\;K$}
\put(40,12){$t_a=.1\;K$}

\put(86,47){\textcolor{red}{large $t_a$}}
\put(93,37){\textcolor{blue}{small $t_a$}}
\put(47.0,43){\textcolor{black}{$t_a [K]$}}
\end{overpic}
\vspace{0.3cm}
\caption{{\bf Simplified model for convective clustering.}
{\bf a}, An example of a typical CP formed during day two for $A2b$ showing virtual potential temperature anomaly at $z=50\;m$.
Thin grey and black contour lines indicate surface precipitation patches formed within the previous and subsequent $60$ $min$, respectively.
Black dotted lines indicate the CP boundary in terms of the zero anomaly contour, marked thicker where it is co-located with subsequent precipitation.
{\bf b}, Analogous to (a), but for $A5b$. An area $A_{crit}$ is marked for later use in panel e.
{\bf c}, $x$-$z$ cross section at $y_0$ corresponding to (a) including the lifting condensation level (LCL) and level of free convection (LFC).
In accordance with high LFC, the regions marked correspond to reduced triggering probability ({\it compare}: panel e).
{\bf d}, Analogous to (c), but for $A5b$.
{\bf e}, Schematic for the simplified model dynamics.
(i) Low-density sub-domain with vacant (white) and active sites (blue, depleted probability);
(ii) Similar to (i) but for high density, showing also boundary sites (orange) with enhanced probability.
{\bf f}, Spatial variance at different box sizes for $t_a=0.5\;K$, 
analogous to Fig.~\ref{fig:quantifying_clustering} but now for the simplified model.
Curves of colours ranging from red to green illustrate increasing days.
For larger scales and later times, events are more strongly clustered.
Note the double-logarithmic axis scaling.
{\bf g}, Analogous to (f), but for smaller $t_a$.
{\it inset}: Diurnal cycle for rain area from simplified model for two values of $t_a$ ({\it compare}: Fig.~\ref{fig:daily_mean}b--c).
{\bf h}, Simple checkerboard model, explaining the increase in variance over time for large $t_a$ (red symbols) and small $t_a$ (blue symbols).
The grids below show the patterns at days 1---4, 10 for large and small $t_a$, respectively.
Box colours white/faint red/red/dark red indicate increasing density.
Note the increasing density variance and anti-correlations for large $t_a$.
}
\label{fig:quantifying_clustering_simplified}
\end{figure}

To formulate a simplified model, consider a concrete comparison of CPs formed in $A2$ and $A5$ (Fig.~\ref{fig:quantifying_clustering_simplified}a,b):
in $A2$, CPs are spatially isolated from one another and the area covered by each CP remains small.
In $A5$, many CPs occur so close to each another, that their temperature anomalies inevitably merge (Fig.~\ref{fig:quantifying_clustering_simplified}b and ~\ref{fig:CP_height_distribution}c), forming a larger patch of dense air.
This combined CP, which we associate with the emergent MCS, reaches to greater height and shows a strong inhibitory effect, as quantified by a divergence of the level of free convection (Fig.~\ref{fig:quantifying_clustering_simplified}d). 
The greater CP height allows environmental air to be forced higher up, setting off new rain cells at the CP gust front.
Indeed, many subsequent rain cells do form near the perimeter of the combined CP (thin black contours in Fig.~\ref{fig:quantifying_clustering_simplified}b), whereas this is not found for $A2$ (Fig.~\ref{fig:quantifying_clustering_simplified}a).

As the emergent MCS spreads outward from the dense region of rain cells, new cells are often triggered at its front --- further feeding the combined CP. 
The MCS leaves behind a relatively cold and dry sub-region, whereas the surroundings of the MCS will benefit from the moisture transported by its front and the additional latent heat provided by enhanced surface fluxes due to the MCS's strong gust front horizontal winds (Fig.~\ref{fig:hor_wind_speed}).
To quantify such moisture re-distribution, we contrast domain sub-regions of $A5$, which receive intense versus weak precipitation during a given model day (Fig.~\ref{fig:moisture_oscillations}).
Regions of intense rainfall are characterised by enhanced moisture near cloud base ($z\sim 1\, km$) before precipitation onset, but marked depletion after rain has occurred.
Conversely, areas of weak rainfall show nearly a "mirror image," with depressed moisture before but enhanced values after precipitation. 
The bi-diurnal dynamics for $A5$ can hence be characterised as an oscillation of cloud-base moisture, driven by the lateral expansion of MCSs, and suggest an inhibitory drying effect on the timescale of approximately one day.
In $A2$, these moisture oscillations are all but lacking (Fig.~\ref{fig:moisture_oscillations}d---f), a finding that falls in line with the absence of organised convection.

In total, the analysis suggests that the growth of a super-CP on one day causes a suppressed region the subsequent day. 
To further check this, we decrease the ventilation coefficients \cite{seifert2006two}, which controls rain evaporation, hence temperature depression $\theta'$ and CP propagation ($\sim \theta'^{1/2}$) ({\it Details}: Methods).
Indeed, a decrease in these coefficients systematically decreases the spatial extent of the rain-free regions (Fig.~\ref{fig:allData}).

\noindent
{\bf Clustering from a two-level atmosphere model.}
A key characteristic of $A5$ is that parts of the day see no rainfall at all, whereas immediately after the onset of rain (near mid-day, Fig.~\ref{fig:daily_mean}b), the area covered by rain is relatively large --- corresponding to a high number density of rain events and cold pools.
Our simplified model mimics the qualitative clustering dynamics, by incorporating that MCSs emerge when a sufficient number of rain events occur nearby. 
We simplify by considering pairs of rain cells and their CPs as one entity, termed "active site," which, for simplicity, occupies the elementary area of the rain cell $a_0\approx 25\;km^2$ (Tab.~\ref{tab:basic_stats}).
Sites not active are considered "vacant."

Assume that, at a given time, the fraction of active sites is $p_0$, and sites are independently populated. 
That is, each site of a square lattice contains a rain event at probability $p_0$.
Now demand that when an area $A>A_{crit}$ is covered by spatially contiguous active sites, vacant sites in their immediate neighbourhood are more likely to become active and the contiguous area may increase in size ({\it compare}: Fig.~\ref{fig:quantifying_clustering_simplified}b). 
This is accomplished by assigning increased probabilities to the neighbourhood sites
(Fig.~\ref{fig:quantifying_clustering_simplified}e).
When $p_0$ is small ($p_0\ll 1$), the system will, however, be unlikely to contain contiguous rain areas exceeding $A_{crit}$ ({\it compare}: shaded box in Fig.~\ref{fig:quantifying_clustering_simplified}a).
To exemplify: the probability of finding two active sites on two neighbouring sites is proportional to $p_0^2$, and this probability will decay exponentially for contiguous areas larger than two. \cite{christensen2005complexity}

But how does $p_0$ emerge from the diurnal cycle dynamics, and how can bi-diurnal temporal correlations be captured (Fig.~\ref{fig:quantifying_clustering})?
To self-consistently incorporates these features, we describe the population of rain cells within a two-layer atmosphere model consisting of a prescribed boundary layer temperature $T_{bl}$, varying sinusoidally with a small amplitude $t_a$ ({\it compare}: Fig.~\ref{fig:daily_mean}a), 
and an interactive free-tropospheric temperature $T_{ft}$.
We compute the probability for a rain cell to become active by coupling it to the temperature difference $\Delta T=T_{bl}-T_{ft}$, as an approximation for atmospheric stability.
When $T_{bl}$ increases in the course of the model day, rain events will eventually be set off.
Two key processes impact on $T_{ft}$: 
thermal radiation to space reduces $T_{ft}$, whereas latent heat transfer by rain events increases it.
$T_{bl}$ is not directly affected by rain cells. However, buoyancy depressions, arising mainly from sustained drying after rain events, are implemented by an inhibitory potential. 
In practice, once an active site transitions back to vacant, it needs to "wait" until it can become active again 
({\it Details:} Methods).

We implement reasonable coefficients for these processes and find the simulations to reach a repetitive diurnal cycle (Fig.~\ref{fig:quantifying_clustering_simplified}d, inset).
Indeed, for small $t_a$, rainfall is present, yet modest, during the entire day, whereas for larger $t_a$, rainfall is either strong or absent.
Time-averaged rain areas for large and small $t_a$ match ({\it compare}: Fig.~\ref{fig:daily_mean}b--c), a result of the radiative constraint and in agreement with the numerical experiments.
Considering the variance of the spatial pattern, the simplified model indeed produces increased clustering over time for large $t_a$, whereas clustering is absent for small $t_a$ (Fig.~\ref{fig:quantifying_clustering_simplified}c,d, and Fig.~\ref{fig:daily_mean_simplified_model}):
when the thermal forcing caused by $T_{bl}$ increases rapidly in the course of the day, many rain events will be set off during a short time period --- leading to large $p_0$ during those times. 
The negative feedback on $T_{ft}$ will then rapidly cause the "budget" of rainfall to be used up. 
MCS will form as long as the increased probability at the edges of the contiguous CP patches counteracts the ongoing increase of $T_{ft}$. 
Hence, MCSs will be able to spread, as long as this is the case, thus setting a time ($\approx 6\;h$) and space scale for MCS ($\approx 100\;km$), which is significantly larger than the scale of a single rain event ($\approx 1\;h$ and $\approx 5\;km$).

\noindent
{\bf Conceptualising further.}
These two scales allow for a simplified conceptual view: 
take the model domain to be broken down into a square lattice consisting of blocks, each the size of an MCS, hence $100\;km\times 100\;km$ or $20\times 20$ locations for single rain cells of area $a_0$.
In each block, an MCS is set off if a sufficient number of rain cells are present. 
We find that a simple set of rules can model the MCS dynamics: 
(1) first, assign an integer, encoding the number of rain cells, drawn from a Binomial distribution, to each block in this square lattice;
(2) to update the system, let all sites above a particular threshold (an MCS forms) hand over their content (the moisture transported by the MCS) to their four neighbouring sites, at equal parts. 
When the mean of the Binomial distribution is low compared to the threshold, which is the case for small $t_a$, no MCS will form and no redistribution will take place (Fig.~\ref{fig:quantifying_clustering_simplified}h, blue points, and lower row of squares). 
In contrast, when $t_a$ is sufficiently large, a sequence of re-allocations will occur, leading to a checkerboard-like clustering, which strengthens in time (red curve and upper row of squares).
This example is sufficient to capture the increase of normalised variance for A5 and the lack of it for A2 (Fig.~\ref{fig:quantifying_clustering_simplified}b).

\section*{Discussion}\label{sec:discussions}
Our study analyses the spontaneous emergence of convective clustering, when the amplitude of the diurnal cycle is varied. 
During mid-latitude summer, extreme convective rainfall events have been associated with higher temperatures \cite{lenderink2008increase,berg2013strong} and flash floods have been found to occur more frequently during long periods of pronounced heating \cite{coumou2012decade}.
The findings here suggest that, when the amplitude of temperature variation is sufficiently large, the clustering, and thereby the potential for flash floods, increases from day-to-day.
We have addressed clustering over a sea surface (Fig.~\ref{fig:domain_mean_timeseries_surface_evap}), finding that
not the mean, but the amplitude of surface temperature gives rise to clustering (A5sea and A2sea in Fig.~\ref{fig:allData}).
Increasing model horizontal resolution to $.5$ or $.2$ $km$ (A5c and A5d in Fig.~\ref{fig:allData})
leads to even stronger clustering effects, in agreement with recent findings with realistic continental topography \cite{rasp2018variability}.
The current findings suggest, that observational studies on extremes should additionally focus on temperature amplitude, not just the mean.

In convective self-aggregation, the standard explanation for clustering invokes arguments based on circulation changes.
Remarkably, the memory enabling clustering in the present work is purely thermodynamically-driven: 
we do not detect any sustained changes of large-scale circulation that harbour a memory from one day to the next.
The findings hence suggest two possible "shortcuts" to self-aggregation over the ocean: 
(i) at times, sea surface temperature oscillations may be sufficient to kick-start clustering;
(ii) clustering may emerge over land surfaces and is then advected over the sea.
Both explanations require that clustering can be sustainable or growing once formed --- a requirement consistent with hysteresis effects suggested earlier. \cite{muller2015favors}
In our conceptual model, hysteresis is, in fact, easy to achieve: 
once an imbalance between neighbouring grid cells is established, sustained re-distributions will be possible, even when the overall event number density is then lowered.
Similarly, it would be interesting to probe, whether a state of self-aggregation can be reached more quickly when sea-surface temperatures transiently oscillate --- thereby exciting initial clustering on the scale of $l_{max}$.

The specific feature of the MCS, formed by higher cell and thus cold pool density at larger temperature amplitude, is to trigger new convective cells at its periphery. 
These new cells form cold pools, feeding the emergent MCS and further forcing updraughts near its boundary.
Notably, new convective cells are also formed within the interiour of the MCS (Fig.~\ref{fig:quantifying_clustering_simplified}b), a finding in line with collision effects of multiple cold pool gust fronts.\cite{haerter2019circling,haerter2019convective} 
These interiour cells could further act to deepen and cool the combined cold pool driving the MCS expansion.
Together, MCSs hence act to excite new convection both within and around the combined cold pool area.

The present work interpolates between the established RCE setup for oceanic convection and typical boundary conditions for continental convection. 
Over ocean, the origin of initial clustering, accelerating self-aggregation, could be found in modest sea surface temperature fluctuations --- in line with those observed. 
Our study suggests that explanations for tropical and summertime mid-latitude flash-floods should also be sought in temperature variations and that the likelihood of extreme rainfall might increase over time, when diurnal temperature variations persist throughout multiple days.

\newpage
\clearpage
\section*{Materials and Methods}
\noindent
{\bf Large-eddy model, boundary, and initial conditions.}
We simulate the convective atmosphere using the University of California, Los Angeles (UCLA) Large Eddy Simulator (LES) with sub-grid scale turbulence parametrised after Smagorinsky \cite{smagorinsky1963general}.
This is combined with a delta four-stream radiation scheme \cite{pincus2009monte} and a two-moment cloud microphysics scheme \cite{stevens2005evaluation}. 
Rain evaporation is implemented after \citeauthor{seifert2006two} (\citeyear{seifert2006two}).
Diurnally oscillating, spatially homogeneous, surface temperature, $T_s(t)$, is prescribed, as
\begin{eqnarray}
    T_s(t)=\overline{T_{s}}-T_{a} \cos{(2\pi\;t/t_0)}\;,
\end{eqnarray}
\noindent
where $\overline{T_{s}}=298\;K$, $t_0=24\;h$ is the duration of the simulated model day, $\overline{T_{s}}$ represents the temporal average and $T_a$ the amplitude of $T_s(t)$.
The insolation diurnal cycle is chosen typical for the equator.
Surface heat fluxes are computed interactively and depend on the vertical temperature and humidity gradients as well as horizontal wind speed, which is approximated using the Monin-Obukhov similarity theory.
Temperature and humidity were initialised using observed profiles that potentially represent convective conditions.
However, due to the repeated diurnal cycle forcing, the system eventually establishes a self-consistent vertical temperature and moisture profile.

\noindent
{\bf Model grid, dynamics, and output.}
The model integrates the anelastic equations of motion on a regular horizontal domain with varying horizontal grid spacing $dx$ and periodic boundary conditions (Tab. \ref{tab:experiments}). 
The vertical model resolution is $100\;m$ below $1\;km$, stretches to $200\;m$ near $6\;km$, and reaches $400\;m$ in the upper layers, with the model top located at $16.5\;km$.
Horizontal resolution $dx$ and domain size vary (Tab.~\ref{tab:experiments}). 
The Coriolis force and the mean wind were set to zero with weak random initial perturbations added as noise to break complete spatial symmetry. 
For all two and three-dimensional model variables, the output time step varies between experiments between $\Delta t_{out}=5\;min$ and $15\;min$. 
At each output time step (Tab.~\ref{tab:experiments}), instantaneous surface precipitation intensity, as well as the three-dimensional moisture and velocity fields, are recorded for the entire model domain. 
Additionally, at 30-second and five-minute intervals, respectively, spatially as well as horizontally averaged time series were extracted from the numerical experiments.

\noindent
{\bf Sensitivity experiments.} 
The principal focus of this study is the response to different values of surface temperature amplitude, $T_a$. 
The main experiments (A5a, A2a, A5b, A2b) were carried out at $T_s=298\;K$, contrasted $T_a=2\;K$ and $T_a=5\;K$ (Tab.~\ref{tab:experiments}) and used $dx=1\;km$ horizontal model resolution.
One intermediate value of $T_a=3.5\;K$ was tested to further constrain the transition to clustering.
These simulations assumed that surface latent heat fluxes occurred at $70$ percent of their potential value. 
This mimics a land surface, which is reasonable given the mean surface latent and sensible heat fluxes of $LHF\approx 57\;W/m^2$ and $SHF\approx 18\;W/m^2$, respectively, yielding a Bowen ratio of $B\approx .30$, realistic for forested land. 
These main experiments already explore domain size effects, as the clustering observed could be influenced by the finite system size. 
We find that results vary little between these domain sizes.
We conducted additional experiments to test the sensitivity to a range of modifications: 
lateral model resolution (A5c, A5d, Fig.~\ref{fig:allData}), 
where $dx=.5\;km$ and $dx=.2\;km$ were used; 
cold pool strength (A5vent05, A5vent001, Fig.~\ref{fig:allData}), 
where the ventilation coefficients $a_v$ and $b_v$ in Eq. 24 of \citeauthor{seifert2006two} (\citeyear{seifert2006two}) were respectively reduced to $.5$ and $.01$ of their default values; 
and surface conditions, where surface evaporation was raised to its potential value, mimicking a sea surface (Fig.~\ref{fig:allData}, 
$B\approx .15$, {\it Details}: Tab.~\ref{tab:experiments}).

\noindent
{\bf Rain cell and cold pool tracking.}
In all experiments, we track rain cells using the Iterative Rain Cell Tracking Method (IRT \cite{moseley2019statistical}).
In the two-dimensional surface precipitation field corresponding to any output timestep, the IRT first detects all spatially contiguous patches of rain intensities exceeding a threshold $I_0=.5\;mm\;h^{-1}$, termed {\it rain objects}.
Tracks and then identified, by determining any rain objects that overlap from one output timestep to the next. 
The threshold value $\theta$ was set to unity, meaning that, in the case of merging of tracks, the track of larger area is continued under the same track ID.
In our variance analysis, we use the set of coordinates formed by the initial positions of each track, which is defined here as the precipitation-weighted centre of mass of the first rain objects belonging to each track.
The locations of all subsequent rain objects of the same track are discarded in our variance analysis, to avoid artefacts from double counting of the positions of the nearly collocated objects.

To account for possible changes in the CP extent between the different simulations, CPs were tracked by a simple temperature depression method.
In this method, the IRT was modified, such that at any timestep, any grid box with a temperature depression, measured relative to the spatial mean at the same time, exceeding a threshold of one Kelvin was recorded.
As for rainfall, spatially contiguous patches of temperature depression were collected into objects and assigned indexes.
To increase the signal, only objects exceeding an area threshold of ten $\;km^2$ were considered.
Going forward in time, overlapping objects were identified and considered to be the same CP track.
The threshold ratio $\theta$ was again set to unity \cite{moseley2019statistical}.

Furthermore, we measure CP height by evaluating the average temperature of the domain and the standard deviation at each output timestep and vertical model level $k$ and comparing the local temperature $T_0$ at each grid cell $(i,j,k)$ with this value.
If $T_0$ is two standard deviations below the domain mean at the same height level $k$, this grid box is considered part of a CP. 
Within each column $(i,j)$ the CP height is determined as the highest level, which fulfils this criterion.

\noindent
{\bf Simplified MCS model.}
A square lattice of $s\times s$ sites is initialised, where the area of each site is taken as the average area $a_0$ occupied by a single convective rain cell, $a_0\approx 5\;km\times 5\;km$ ({\it compare}: Tab.~\ref{tab:basic_stats}). 
The total domain area $A$ hence is $A\equiv a_0 s^2$.
Our model assumes boundary layer processes to be local and fast. 
In contrast, the free troposphere acts as a "bath" of large heat capacity, where heat is quickly redistributed through gravity waves \cite{bretherton1989gravity}, and spatial homogeneity is assumed.
The model incorporates three fundamental processes affecting event formation: 
(1) spontaneous activation due to the moderate drive of the diurnal cycle temperature forcing;
(2) refractory dynamics due to CPs; 
(3) strong activation ahead of MCS fronts.
Below we discuss the estimation of model parameters for these processes.

\noindent
{\it Spontaneous activation.}
To define a vertical temperature gradient, we consider quantities $T_{bl}$ and $T_{ft}$, which represent the deviation of the boundary layer and free-tropospheric temperature from their assumed steady-state values.
$T_{bl}(t)$ is prescribed and oscillates harmonically as $T_{bl}(t)=-t_a \cos (2\pi t/t_0)$, where $t_0=24\;h$.
Free-troposphere temperature $T_{ft}$ is initialised to zero.
We also define the temperature difference $\Delta T\equiv T_{bl}-T_{ft}$ and its normalized version $\Delta T'\equiv \Delta T/\Delta T_0$, where the reference scale $\Delta T_0\approx .3\;K$ is taken as two standard deviations of simulated near-surface temperature.
$\Delta T'$ serves as a proxy for both convective available potential energy (CAPE) and convective inhibition (CIN).
The basic dynamics proceeds in discrete model timesteps of $.5$ $h$, reflecting the typical timescale of convective cloud formation. 
At each timestep, the default probability $p_0$ for a vacant site to become active is taken as
\begin{equation}
    p_0=\begin{cases}
    0 & \text{if $\Delta T'<0$}\\
    \Delta T' & \text{if $0<\Delta T'<1$}\\
    1 & \text{if $\Delta T'>1$}\;.
    \end{cases}
    \label{eq:cases}
\end{equation}
Eq.~\ref{eq:cases} makes the qualitative assumption that there is no activity at all for stratified conditions ($\Delta T'<0$), and events occur with certainty when the temperature gradient is much larger than typical fluctuations. 
Both of these limits could be softened, as could the assumption of linearity at intermediate $\Delta T'$. 
One could equally argue for a smooth function of $\Delta T'$, such as an error function. 
Nonetheless, Eq.~\ref{eq:cases} captures the observed fact that more initial activity occurs when surface temperature changes quickly.
Without any further perturbations, the initiation probability $p_{ij}$ at each site $(i,j)$ will be equal, $p_{ij}=p_0$.

\noindent
{\it Spatial structure.} 
Notably, Eq.~\ref{eq:cases} does not depend on the position within the lattice. 
Local modifications cause spatial structure ({\it compare}: Fig.~\ref{fig:quantifying_clustering_simplified}a).
Cold pools have two effects: reduction of local temperature and reduction of local humidity. 
However, the recovery timescale for the former is fast ($\tau_{cp}\approx 3\;h$), whereas that of the latter can be slower ($\tau_{inh}\approx 24\;h$, {\it compare}: Fig.~\ref{fig:moisture_oscillations}).
Temperature reduction dominates the density change and thereby the mechanical cold pool properties, whereas moisture crucially impacts on the local initiation probability.
To consider both, we take active sites to persist for $\tau_{cp}$, during which no further raincell initiation is possible at the same site.
Simultaneously, the local probability $p_{ij}$ is reduced as $p_{ij}=p_0-p_{inh}\exp(-\delta t/\tau_{inh})$, where $\delta t$ is the time after the occurrence of the rain event and $p_{inh}>0$. 
$p_{inh}$ essentially controls the fidelity of the anticorrelation from day-to-day, and results are not qualitatively affected by it.

An MCS can occur when an active cluster exceeds the threshold area $A_{crit}\equiv n_0a_0$, with $n_0=20$ (Fig.~\ref{fig:quantifying_clustering_simplified}). 
While this is the case, and $\Delta T'>0$, the probability $p$ at each of the surrounding sites $(i,j)$ becomes $p_{ij}\rightarrow p_{ij}+p_{act}$, which acts to lower the barrier presented by CIN.
When $p_{ij}$ is no longer a surrounding site, the term $p_{act}$ is no longer applied for this site.
The initiation probability $p_{ij}$ is then defined analogously to the one in Eq.~\ref{eq:cases}.

\noindent
{\it Radiative constraint.}
Generally, as $T_{bl}$ rises during the day as the surface is heated by insolation. 
$p_{ij}$ will then eventually become positive at some sites, and rain cells can be produced there.
When this occurs, latent heat is transferred to the free troposphere, increasing $T_{ft}$, thus subsequently lowering $\Delta T$ and thereby $p_{ij}$.
Simultaneously, the site $(i,j)$ is shut down by the adverse buoyancy effects through CP formation.
$T_{ft}$ will relax by heat loss ($P_{out}\approx 200\;W\;m^{-2}$) through outgoing thermal radiation. 
According to the Stefan-Boltzmann law, $P_{out}=\sigma T_{eff}^4$, with $\sigma\approx 5.7\times 10^{-8}W\;m^{-2}\;K^{-4}$, the Stefan-Boltzmann constant --- translating to an effective emission temperature $T_{eff}\approx 250\;K$. 
Through $P_{out}$, the free troposphere would hence cool by approximately $2\;K\;d^{-1}$.
As a crude estimate of the change in $T_{ft}$ through latent heat transfer, we estimate the free-tropospheric heat capacity $C_{ft}=M_{ft}c_{pd}$, assuming dry air.
$c_{pd}\approx 1\;kJ/kg$ and the mass of the free troposphere $M_{ft}=\int_{z=z_{LCL}}^{z_{top}}dz\;\rho(z)\approx 8\times 10^3\;kg$, using $z_{LCL}\approx 1\;km$ and $z_{top}\approx 16\;km$ for the lifting condensation level (LCL) and top of the troposphere, respectively.
Hence, $C_{ft}\approx 8\times 10^6\;Jm^{-2}K^{-1}$.
From rain cell tracking of our LES simulations, we obtain the average rain cell lifetime to be $\approx 1\;h$ (Tab.~\ref{tab:basic_stats}), the mean cell area $a_0\approx 20\;km^2$ and the average cell rain rate of $\approx 4\;mm\;h^{-1}$ (Tab.~\ref{tab:basic_stats}); hence, each rain cell can be assumed to yield $M_{event}=4\;kg\;m^{-2}$ of liquid water.
Assuming that the corresponding latent heat was previously deposited in the free troposphere upon ascent, each rain event heats the free troposphere by $Q_{event}=L_v\;M_{event}$.
We take this heat to increase $T_{eff}$ accordingly by $\delta\;T_{eff}=Q_{event}C_{ft}^{-1}s^{-2}$, where $s^{-2}$ results from the ratio $a_0/A$, yielding $\delta T_{eff}s^2\approx 1\;K$: if one rain event occurred instantaneously at each site, the free tropospheric temperature would rise by $1\;K$.


\begin{table}[ht]
\begin{tabular}{lll}
    Model Parameter & Numerical Value & Description \\
    \hline
    $A_{crit}$ & $500\;km^2$ & critical area for MCS formation \\
    $p_{act}$ & $.8$ & activation potential along MCS fronts \\ 
    $\tau_{cp}$ & $3\;h$ & duration when CPs can activate new raincells\\ 
    $p_{inh}$ & $-.8$ & refractory potential underneath cold pools\\
    $\tau_{inh}$ & $24\;h$ & humidity inhibitory time \\
    $t_a$ & $0.1$ K, $0.5$ K & near-surface temperature amplitudes \\
    \hline
\end{tabular}
\caption{{\bf Parameters in the simplified model.}}
\label{tab:parameters}
\end{table}

\begin{table}[ht]
\begin{tabular}{lllll}
    Experiment & Forcing Amplitude & Horizontal Resolution & Domain Size & Days with\\
    Name & $T_a$ [$K$] & $dx$ [$km$] & $L$ [$km$] & 3D output\\
    \hline
    A5a & $5$ & $1$ & $960$ & $1$---$6$ \\
    A2a & $2$ & $1$ & $960$ & $1,4$---$7$ \\
    A5b & $5$ & $1$ & $480$ & $1$---$8$\\
    A2b & $2$ & $1$ & $480$ & $1$---$4$,$8^*$,$9^*$\\
    \hline
    A5c & $5$ & $.5$ & $240$ & $1$---$3$ \\
    A5d & $5$ & $.2$ & $240$ & $1$---$3$ \\
    A3.5 & $3.5$ & $1$ & $480$ & $3$---$5$\\
    A5sea & $5$ & $1$ & $480$ & $1$---$4$\\
    A2sea & $2$ & $1$ & $480$ & $1$, $8$\\
    A5vent05 & $5$ & $1$ & $480$ & $4$\\
    A5vent001 & $5$ & $1$ & $480$ & $4$\\
    A5p2K & $5$ & $1$ & $480$ & $4$, $8$\\
    \hline
\end{tabular}
\caption{{\bf Summary of numerical experiments.}
The main four experiments are listed above the horizontal line (A5a, A2a, A5b, A2b). 
The experiments labelled by a star (*) are equivalent to A2b but constitute an additional, longer-duration run (A2long).
All the above experiments were carried out at $T_s=298\;K$, except A5p2K, which was carried out at $T_s=300\;K$. 
A5sea and A2sea were carried out, by increasing surface evaporation to $100$ percent potential evaporation.
In all other experiments, surface evaporation was set to 70 percent of the potential value.
A5vent05 and A5vent001 denote experiments, where the ventilation coefficients \cite{seifert2006two} were reduced to $.5$ and $.01$ of their default values. }
\label{tab:experiments}
\end{table}

\section*{Acknowledgements}
We thank Kim Sneppen for fruitful discussions on the simplified modeling.
JOH and BM gratefully acknowledge funding by a grant from the VILLUM Foundation (grant number: 13168) and the European Research Council (ERC) under the European Union's Horizon 2020 research and innovation program (grant number: 771859). 
SBN acknowledges funding through the Danish National Research Foundation (grant number: DNRF116).
The authors acknowledge the Danish Climate Computing Center (DC3) and the German Climate Computing Center (DKRZ).

\section*{Competing Interests}
The authors declare that they have no competing financial interests.

\section*{Correspondence}
Correspondence and requests for materials should be addressed to J.O.H. (email: haerter@nbi.ku.dk).

\section*{Author Contributions}
\noindent
J.O.H. ran and processed the large-eddy simulations (LES) and wrote the manuscript. B.M. produced the case study and commented on the manuscript. S.B.N. contributed in model development and implementation and revised the manuscript.

\clearpage
\newpage
\pagebreak
\bibliography{aaa_text}

\newcommand{\noop}[1]{}
\begin{thebibliography}{56}
\providecommand{\natexlab}[1]{#1}
\providecommand{\url}[1]{\texttt{#1}}
\expandafter\ifx\csname urlstyle\endcsname\relax
  \providecommand{\doi}[1]{doi: #1}\else
  \providecommand{\doi}{doi: \begingroup \urlstyle{rm}\Url}\fi

\bibitem[Tan et~al.(2015)Tan, Jakob, Rossow, and Tselioudis]{tan2015increases}
Jackson Tan, Christian Jakob, William~B Rossow, and George Tselioudis.
\newblock Increases in tropical rainfall driven by changes in frequency of
  organized deep convection.
\newblock \emph{Nature}, 519\penalty0 (7544):\penalty0 451--454, 2015.
\newblock \doi{https://doi.org/10.1038/nature14339}.

\bibitem[Stevenson and Schumacher(2014)]{stevenson201410}
Stephanie~N Stevenson and Russ~S Schumacher.
\newblock A 10-year survey of extreme rainfall events in the central and
  eastern united states using gridded multisensor precipitation analyses.
\newblock \emph{Monthly Weather Review}, 142\penalty0 (9):\penalty0 3147--3162,
  2014.
\newblock \doi{https://doi.org/10.1175/MWR-D-13-00345.1}.

\bibitem[Feng et~al.(2018)Feng, Leung, Houze~Jr, Hagos, Hardin, Yang, Han, and
  Fan]{feng2018structure}
Zhe Feng, L~Ruby Leung, Robert~A Houze~Jr, Samson Hagos, Joseph Hardin, Qing
  Yang, Bin Han, and Jiwen Fan.
\newblock {Structure and evolution of mesoscale convective systems: Sensitivity
  to cloud microphysics in convection-permitting simulations over the United
  States}.
\newblock \emph{Journal of Advances in Modeling Earth Systems}, 10\penalty0
  (7):\penalty0 1470--1494, 2018.
\newblock \doi{https://doi.org/10.1029/2018MS001305}.

\bibitem[Moeng and LeMone(1995)]{moeng1995atmospheric}
Chin-Hoh Moeng and Margaret~A LeMone.
\newblock Atmospheric planetary boundary-layer research in the us: 1991--1994.
\newblock \emph{Reviews of Geophysics}, 33\penalty0 (S2):\penalty0 923--931,
  1995.
\newblock \doi{https://doi.org/10.1029/95RG00185}.

\bibitem[Bretherton et~al.(2005)Bretherton, Blossey, and
  Khairoutdinov]{bretherton2005energy}
Christopher~S Bretherton, Peter~N Blossey, and Marat Khairoutdinov.
\newblock An energy-balance analysis of deep convective self-aggregation above
  uniform {SST}.
\newblock \emph{Journal of the {A}tmospheric {S}ciences}, 62\penalty0
  (12):\penalty0 4273--4292, 2005.

\bibitem[Khairoutdinov and Emanuel(2010)]{khairoutdinov2010aggregation}
Marat~F Khairoutdinov and Kerry~A Emanuel.
\newblock Aggregated convection and the regulation of tropical climate.
\newblock In \emph{29th Conference on Hurricanes and Tropical Meteorology,
  Amer. Meteorol. Soc., Tucson, AZ}, 2010.

\bibitem[Muller and Held(2012)]{muller2012detailed}
Caroline~J Muller and Isaac~M Held.
\newblock Detailed investigation of the self-aggregation of convection in
  cloud-resolving simulations.
\newblock \emph{Journal of the Atmospheric Sciences}, 69\penalty0 (8):\penalty0
  2551--2565, 2012.
\newblock \doi{https://doi.org/10.1175/JAS-D-11-0257.1}.

\bibitem[Held et~al.(1993)Held, Hemler, and Ramaswamy]{held1993radiative}
Isaac~M Held, Richard~S Hemler, and V~Ramaswamy.
\newblock Radiative-convective equilibrium with explicit two-dimensional moist
  convection.
\newblock \emph{Journal of the Atmospheric Sciences}, 50\penalty0
  (23):\penalty0 3909--3927, 1993.

\bibitem[Tompkins and Craig(1998)]{tompkins1998radiative}
Adrian~M Tompkins and George~C Craig.
\newblock Radiative--convective equilibrium in a three-dimensional
  cloud-ensemble model.
\newblock \emph{Quarterly Journal of the Royal Meteorological Society},
  124\penalty0 (550):\penalty0 2073--2097, 1998.
\newblock \doi{https://doi.org/10.1002/qj.49712455013}.

\bibitem[Hohenegger and Stevens(2016)]{hohenegger2016coupled}
Cathy Hohenegger and Bjorn Stevens.
\newblock Coupled radiative convective equilibrium simulations with explicit
  and parameterized convection.
\newblock \emph{Journal of Advances in Modeling Earth Systems}, 8\penalty0
  (3):\penalty0 1468--1482, 2016.
\newblock \doi{https://doi.org/10.1002/2016MS000666}.

\bibitem[Muller and Bony(2015)]{muller2015favors}
Caroline Muller and Sandrine Bony.
\newblock What favors convective aggregation and why?
\newblock \emph{Geophysical Research Letters}, 42\penalty0 (13):\penalty0
  5626--5634, 2015.
\newblock \doi{https://doi.org/10.1002/2015GL064260}.

\bibitem[Jeevanjee and Romps(2013)]{jeevanjee2013convective}
Nadir Jeevanjee and David~M Romps.
\newblock Convective self-aggregation, cold pools, and domain size.
\newblock \emph{Geophysical Research Letters}, 40\penalty0 (5):\penalty0
  994--998, 2013.
\newblock \doi{https://doi.org/10.1002/grl.50204}.

\bibitem[Haerter(2019)]{haerter2019convective}
Jan~O Haerter.
\newblock Convective self-aggregation as a cold pool-driven critical
  phenomenon.
\newblock \emph{Geophysical Research Letters}, 46\penalty0 (7):\penalty0
  4017--4028, 2019.
\newblock \doi{https://doi.org/10.1029/2018GL081817}.

\bibitem[Weller and Anderson(1996)]{weller1996surface}
RA~Weller and SP~Anderson.
\newblock Surface meteorology and air-sea fluxes in the western equatorial
  pacific warm pool during the toga coupled ocean-atmosphere response
  experiment.
\newblock \emph{Journal of Climate}, 9\penalty0 (8):\penalty0 1959--1990, 1996.

\bibitem[Johnson et~al.(1999)Johnson, Rickenbach, Rutledge, Ciesielski, and
  Schubert]{johnson1999trimodal}
Richard~H Johnson, Thomas~M Rickenbach, Steven~A Rutledge, Paul~E Ciesielski,
  and Wayne~H Schubert.
\newblock Trimodal characteristics of tropical convection.
\newblock \emph{Journal of Climate}, 12\penalty0 (8):\penalty0 2397--2418,
  1999.

\bibitem[Kawai and Wada(2007)]{kawai2007diurnal}
Yoshimi Kawai and Akiyoshi Wada.
\newblock Diurnal sea surface temperature variation and its impact on the
  atmosphere and ocean: A review.
\newblock \emph{Journal of Oceanography}, 63\penalty0 (5):\penalty0 721--744,
  2007.
\newblock \doi{https://doi.org/10.1007/s10872-007-0063-0}.

\bibitem[Liu and Moncrieff(1998)]{liu1998numerical}
Changhai Liu and Mitchell~W Moncrieff.
\newblock A numerical study of the diurnal cycle of tropical oceanic
  convection.
\newblock \emph{Journal of the Atmospheric Sciences}, 55\penalty0
  (13):\penalty0 2329--2344, 1998.

\bibitem[Tian et~al.(2006)Tian, Waliser, and Fetzer]{tian2006modulation}
Baijun Tian, Duane~E Waliser, and Eric~J Fetzer.
\newblock {Modulation of the diurnal cycle of tropical deep convective clouds
  by the MJO}.
\newblock \emph{Geophysical Research Letters}, 33\penalty0 (20), 2006.
\newblock \doi{https://doi.org/10.1029/2006GL027752}.

\bibitem[Suzuki(2009)]{suzuki2009diurnal}
Tsuneaki Suzuki.
\newblock {Diurnal cycle of deep convection in super clusters embedded in the
  Madden-Julian Oscillation}.
\newblock \emph{Journal of Geophysical Research: Atmospheres}, 114\penalty0
  (D22), 2009.
\newblock \doi{https://doi.org/10.1029/2008JD011303}.

\bibitem[Chen et~al.(1996)Chen, Houze~Jr, and Mapes]{chen1996multiscale}
Shuyi~S Chen, Robert~A Houze~Jr, and Brian~E Mapes.
\newblock {Multiscale variability of deep convection in realation to
  large-scale circulation in TOGA COARE}.
\newblock \emph{Journal of the Atmospheric Sciences}, 53\penalty0
  (10):\penalty0 1380--1409, 1996.

\bibitem[Droegemeier and Wilhelmson(1985)]{droegemeier1985three}
Kelvin~K Droegemeier and Robert~B Wilhelmson.
\newblock {Three-dimensional numerical modeling of convection produced by
  interacting thunderstorm outflows. Part I: Control simulation and low-level
  moisture variations}.
\newblock \emph{Journal of the Atmospheric Sciences}, 42\penalty0
  (22):\penalty0 2381--2403, 1985.

\bibitem[Rotunno et~al.(1988)Rotunno, Klemp, and Weisman]{rotunno1988theory}
Richard Rotunno, Joseph~B Klemp, and Morris~L Weisman.
\newblock {A Theory for Strong, Long-Lived Squall Lines}.
\newblock \emph{Journal of the Atmospheric Sciences}, 45\penalty0 (3):\penalty0
  463--485, 1988.

\bibitem[Tompkins(2001)]{tompkins2001organizationCold}
Adrian~M Tompkins.
\newblock Organization of tropical convection in low vertical wind shears: The
  role of cold pools.
\newblock \emph{Journal of the Atmospheric Sciences}, 58\penalty0
  (13):\penalty0 1650--1672, 2001.

\bibitem[B{\"o}ing et~al.(2012)B{\"o}ing, Jonker, Siebesma, and
  Grabowski]{boing2012influence}
Steven~J B{\"o}ing, Harm~JJ Jonker, A~Pier Siebesma, and Wojciech~W Grabowski.
\newblock Influence of the subcloud layer on the development of a deep
  convective ensemble.
\newblock \emph{Journal of the Atmospheric Sciences}, 69\penalty0 (9):\penalty0
  2682--2698, 2012.
\newblock \doi{https://doi.org/10.1175/JAS-D-11-0317.1}.

\bibitem[Schlemmer and Hohenegger(2015)]{schlemmer2015modifications}
Linda Schlemmer and Cathy Hohenegger.
\newblock Modifications of the atmospheric moisture field as a result of
  cold-pool dynamics.
\newblock \emph{Quarterly Journal of the Royal Meteorological Society},
  142:\penalty0 30--42, 2015.
\newblock \doi{https://doi.org/10.1002/qj.2625}.

\bibitem[Feng et~al.(2015)Feng, Hagos, Rowe, Burleyson, Martini, and
  Szoeke]{feng2015mechanisms}
Zhe Feng, Samson Hagos, Angela~K Rowe, Casey~D Burleyson, Matus~N Martini, and
  Simon~P Szoeke.
\newblock Mechanisms of convective cloud organization by cold pools over
  tropical warm ocean during the amie/dynamo field campaign.
\newblock \emph{Journal of Advances in Modeling Earth Systems}, 7\penalty0
  (2):\penalty0 357--381, 2015.
\newblock \doi{https://doi.org/10.1002/2014MS000384}.

\bibitem[Moseley et~al.(2016)Moseley, Hohenegger, Berg, and
  Haerter]{moseley2016intensification}
Christopher Moseley, Cathy Hohenegger, Peter Berg, and Jan~O Haerter.
\newblock Intensification of convective extremes driven by cloud--cloud
  interaction.
\newblock \emph{Nature Geoscience}, 9\penalty0 (10):\penalty0 748, 2016.
\newblock \doi{https://doi.org/10.1038/ngeo2789}.

\bibitem[Haerter et~al.(2019)Haerter, B{\"o}ing, Henneberg, and
  Nissen]{haerter2019circling}
Jan~O Haerter, Steven~J B{\"o}ing, Olga Henneberg, and Silas~Boye Nissen.
\newblock Circling in on convective organization.
\newblock \emph{Geophysical Research Letters}, 46\penalty0 (12):\penalty0
  7024--7034, 2019.
\newblock \doi{https://doi.org/10.1029/2019GL082092}.

\bibitem[Lochbihler et~al.(2019)Lochbihler, Lenderink, and
  Siebesma]{lochbihler2019response}
Kai Lochbihler, Geert Lenderink, and A~Pier Siebesma.
\newblock Response of extreme precipitating cell structures to atmospheric
  warming.
\newblock \emph{Journal of Geophysical Research: Atmospheres}, 124\penalty0
  (13):\penalty0 6904--6918, 2019.
\newblock \doi{https://doi.org/10.1029/2018JD029954}.

\bibitem[Chen and Houze(1997)]{chen1997diurnal}
Shuyi~S Chen and Robert~A Houze.
\newblock Diurnal variation and life-cycle of deep convective systems over the
  tropical pacific warm pool.
\newblock \emph{Quarterly Journal of the Royal Meteorological Society},
  123\penalty0 (538):\penalty0 357--388, 1997.
\newblock \doi{https://doi.org/10.1002/qj.49712353806}.

\bibitem[Smith et~al.(2001)Smith, Baeck, Zhang, and
  Doswell~III]{smith2001extreme}
James~A Smith, Mary~Lynn Baeck, Yu~Zhang, and Charles~A Doswell~III.
\newblock Extreme rainfall and flooding from supercell thunderstorms.
\newblock \emph{Journal of Hydrometeorology}, 2\penalty0 (5):\penalty0
  469--489, 2001.

\bibitem[Cintineo and Stensrud(2013)]{cintineo2013predictability}
Rebecca~M Cintineo and David~J Stensrud.
\newblock On the predictability of supercell thunderstorm evolution.
\newblock \emph{Journal of the Atmospheric Sciences}, 70\penalty0 (7):\penalty0
  1993--2011, 2013.
\newblock \doi{https://doi.org/10.1175/JAS-D-12-0166.1}.

\bibitem[Petch et~al.(2002)Petch, Brown, and Gray]{petch2002impact}
JC~Petch, AR~Brown, and MEB Gray.
\newblock The impact of horizontal resolution on the simulations of convective
  development over land.
\newblock \emph{Quarterly Journal of the Royal Meteorological Society},
  128\penalty0 (584):\penalty0 2031--2044, 2002.
\newblock \doi{https://doi.org/10.1256/003590002320603511}.

\bibitem[Guichard et~al.(2004)Guichard, Petch, Redelsperger, Bechtold,
  Chaboureau, Cheinet, Grabowski, Grenier, Jones, K{\"o}hler, Piriou, Tailleux,
  and Tomasini]{guichard2004modelling}
F~Guichard, JC~Petch, J-L Redelsperger, P~Bechtold, J-P Chaboureau, S~Cheinet,
  W~Grabowski, H~Grenier, CG~Jones, M~K{\"o}hler, J‐M Piriou, R~Tailleux, and
  M~Tomasini.
\newblock Modelling the diurnal cycle of deep precipitating convection over
  land with cloud-resolving models and single-column models.
\newblock \emph{Quarterly Journal of the Royal Meteorological Society},
  130\penalty0 (604):\penalty0 3139--3172, 2004.
\newblock \doi{https://doi.org/10.1256/qj.03.145}.

\bibitem[Schlemmer(2011)]{schlemmer2011diurnal}
Linda Schlemmer.
\newblock The diurnal cycle of midlatitude, summertime moist convection over
  land in an idealized cloud-resolving model.
\newblock \emph{ETH Zurich}, 2011.

\bibitem[Haerter and Schlemmer(2018)]{haerter2018intensified}
Jan~O Haerter and Linda Schlemmer.
\newblock Intensified cold pool dynamics under stronger surface heating.
\newblock \emph{Geophysical Research Letters}, 45\penalty0 (12):\penalty0
  6299--6310, 2018.
\newblock \doi{https://doi.org/10.1029/2017GL076874}.

\bibitem[Ban et~al.(2015)Ban, Schmidli, and Sch{\"a}r]{ban2015heavy}
Nikolina Ban, Juerg Schmidli, and Christoph Sch{\"a}r.
\newblock Heavy precipitation in a changing climate: Does short-term summer
  precipitation increase faster?
\newblock \emph{Geophysical Research Letters}, 42\penalty0 (4):\penalty0
  1165--1172, 2015.
\newblock \doi{https://doi.org/10.1002/2014GL062588}.

\bibitem[Prein et~al.(2017)Prein, Liu, Ikeda, Bullock, Rasmussen, Holland, and
  Clark]{prein2017simulating}
Andreas~F Prein, Changhai Liu, Kyoko Ikeda, Randy Bullock, Roy~M Rasmussen,
  Greg~J Holland, and Martyn Clark.
\newblock {Simulating North American mesoscale convective systems with a
  convection-permitting climate model}.
\newblock \emph{Climate Dynamics}, pages 1--16, 2017.
\newblock \doi{https://doi.org/10.1007/s00382-017-3993-2}.

\bibitem[Rasp et~al.(2018)Rasp, Selz, and Craig]{rasp2018variability}
Stephan Rasp, Tobias Selz, and George~C Craig.
\newblock Variability and clustering of midlatitude summertime convection:
  Testing the craig and cohen theory in a convection-permitting ensemble with
  stochastic boundary layer perturbations.
\newblock \emph{Journal of the Atmospheric Sciences}, 75\penalty0 (2):\penalty0
  691--706, 2018.
\newblock \doi{https://doi.org/10.1175/JAS-D-17-0258.1}.

\bibitem[Satoh et~al.(2019)Satoh, Stevens, Judt, Khairoutdinov, Lin, Putman,
  and D{\"u}ben]{satoh2019global}
Masaki Satoh, Bjorn Stevens, Falko Judt, Marat Khairoutdinov, Shian-Jiann Lin,
  William~M Putman, and Peter D{\"u}ben.
\newblock Global cloud-resolving models.
\newblock \emph{Current Climate Change Reports}, pages 1--13, 2019.
\newblock \doi{https://doi.org/10.1007/s40641-019-00131-0}.

\bibitem[Houze~Jr(2004)]{houze2004mesoscale}
Robert~A Houze~Jr.
\newblock Mesoscale convective systems.
\newblock \emph{Reviews of Geophysics}, 42\penalty0 (4), 2004.
\newblock \doi{https://doi.org/10.1029/2004RG000150}.

\bibitem[Yang and Slingo(2001)]{yang2001diurnal}
Gui-Ying Yang and Julia Slingo.
\newblock The diurnal cycle in the tropics.
\newblock \emph{Monthly Weather Review}, 129\penalty0 (4):\penalty0 784--801,
  2001.

\bibitem[Held and Soden(2006)]{held2006robust}
Isaac~M Held and Brian~J Soden.
\newblock Robust responses of the hydrological cycle to global warming.
\newblock \emph{Journal of Climate}, 19\penalty0 (21):\penalty0 5686--5699,
  2006.
\newblock \doi{https://doi.org/10.1175/JCLI3990.1}.

\bibitem[Feller(1957)]{feller1957introduction}
William Feller.
\newblock \emph{An Introduction to Probability Theory and Its Applications}.
\newblock John Wiley \& Sons, 1957.

\bibitem[Tompkins and Semie(2017)]{tompkins2017organization}
Adrian~M Tompkins and Addisu~G Semie.
\newblock Organization of tropical convection in low vertical wind shears: Role
  of updraft entrainment.
\newblock \emph{Journal of Advances in Modeling Earth Systems}, 9\penalty0
  (2):\penalty0 1046--1068, 2017.
\newblock \doi{https://doi.org/10.1002/2016MS000802}.

\bibitem[Etling(2008)]{etling2008theoretische}
Dieter Etling.
\newblock \emph{Theoretische Meteorologie: Eine Einf{\"u}hrung}.
\newblock Springer-Verlag, 2008.

\bibitem[Seifert and Beheng(2006)]{seifert2006two}
A~Seifert and KD~Beheng.
\newblock A two-moment cloud microphysics parameterization for mixed-phase
  clouds. {P}art 1: Model description.
\newblock \emph{Meteorology and {A}tmospheric {P}hysics}, 92\penalty0
  (1-2):\penalty0 45--66, 2006.
\newblock \doi{https://doi.org/10.1007/s00703-005-0113-3}.

\bibitem[Christensen and Moloney(2005)]{christensen2005complexity}
Kim Christensen and Nicholas~R Moloney.
\newblock \emph{Complexity and Criticality}, volume~1.
\newblock World Scientific Publishing Company, 2005.

\bibitem[Lenderink and Van~Meijgaard(2008)]{lenderink2008increase}
Geert Lenderink and Erik Van~Meijgaard.
\newblock Increase in hourly precipitation extremes beyond expectations from
  temperature changes.
\newblock \emph{Nature Geoscience}, 1\penalty0 (8):\penalty0 511--514, 2008.
\newblock \doi{https://doi.org/10.1038/ngeo262}.

\bibitem[Berg et~al.(2013)Berg, Moseley, and Haerter]{berg2013strong}
Peter Berg, Christopher Moseley, and Jan~O Haerter.
\newblock Strong increase in convective precipitation in response to higher
  temperatures.
\newblock \emph{Nature Geoscience}, 6\penalty0 (3):\penalty0 181--185, 2013.
\newblock \doi{https://doi.org/10.1038/ngeo1731}.

\bibitem[Coumou and Rahmstorf(2012)]{coumou2012decade}
Dim Coumou and Stefan Rahmstorf.
\newblock A decade of weather extremes.
\newblock \emph{Nature Climate Change}, 2\penalty0 (7):\penalty0 491, 2012.
\newblock \doi{https://doi.org/10.1038/nclimate1452}.

\bibitem[Smagorinsky(1963)]{smagorinsky1963general}
Joseph Smagorinsky.
\newblock General circulation experiments with the primitive equations: I. the
  basic experiment.
\newblock \emph{Monthly Weather Review}, 91\penalty0 (3):\penalty0 99--164,
  1963.

\bibitem[Pincus and Stevens(2009)]{pincus2009monte}
Robert Pincus and Bjorn Stevens.
\newblock {M}onte {C}arlo spectral integration: {A} consistent approximation
  for radiative transfer in large eddy simulations.
\newblock \emph{Journal of Advances in Modeling Earth Systems}, 1\penalty0 (2),
  2009.
\newblock \doi{https://doi.org/10.3894/JAMES.2009.1.1}.

\bibitem[Stevens et~al.(2005)Stevens, Moeng, Ackerman, Bretherton, Chlond,
  de~Roode, Edwards, Golaz, Jiang, Khairoutdinov, Kirkpatrick, Lewellen, Lock,
  Müller, Stevens, Whelan, and Zhu]{stevens2005evaluation}
Bjorn Stevens, Chin-Hoh Moeng, Andrew~S Ackerman, Christopher~S Bretherton,
  Andreas Chlond, Stephan de~Roode, James Edwards, Jean-Christophe Golaz,
  Hongli Jiang, Marat Khairoutdinov, Michael~P Kirkpatrick, David~C. Lewellen,
  Adrian Lock, Frank Müller, David~E. Stevens, Eoin Whelan, and Ping Zhu.
\newblock Evaluation of large-eddy simulations via observations of nocturnal
  marine stratocumulus.
\newblock \emph{Monthly Weather Review}, 133\penalty0 (6):\penalty0 1443--1462,
  2005.
\newblock \doi{https://doi.org/10.1175/MWR2930.1}.

\bibitem[Moseley et~al.(2019)Moseley, Henneberg, and
  Haerter]{moseley2019statistical}
Christopher Moseley, Olga Henneberg, and Jan~O Haerter.
\newblock A statistical model for isolated convective precipitation events.
\newblock \emph{Journal of Advances in Modeling Earth Systems}, 11\penalty0
  (1):\penalty0 360--375, 2019.
\newblock \doi{https://doi.org/10.1029/2018MS001383}.

\bibitem[Bretherton and Smolarkiewicz(1989)]{bretherton1989gravity}
Christopher~S Bretherton and Piotr~K Smolarkiewicz.
\newblock {Gravity Waves, Compensating Subsidence and Detrainment around
  Cumulus Clouds}.
\newblock \emph{Journal of the Atmospheric Sciences}, 46\penalty0 (6):\penalty0
  740--759, 1989.

\end{thebibliography}

\pagebreak
\clearpage
\newpage

\renewcommand{\theequation}{S\arabic{equation}}
\renewcommand{\thesection}{S\arabic{section}}
\renewcommand{\thefigure}{S\arabic{figure}}
\renewcommand{\thetable}{S\arabic{table}}

\setcounter{equation}{0}
\setcounter{figure}{0}
\setcounter{section}{0}
\setcounter{table}{0}

\section*{\Large Diurnal Self-Aggregation }\label{sec:supp}
\noindent
{\bf \large --- Supplementary Information ---}

\noindent
In this supplementary information, we provide further analysis of interest to some expert readers, but not directly relevant for the understanding of the main text.

\begin{table}[h]
\centering
\begin{tabular}{lllll}
    Experiment & Mean Event & Mean Event & Mean Event & Mean Track\\
    Name & Density & Intensity & Area & Duration\\
     & $N/A$ [$km^{-2}$] & $\langle I_e\rangle$ [$mm\;h^{-1}$] & $a_0$ [$km^2$] & $D$ [$min$]\\
    \hline
    A5a & $.016$ & $4.6$ & $22.4$ & 61\\
    A2a & $.013$ & $4.7$ & $19.8$ & 76\\
    A5b & $.019$ & $4.5$ & $22.0$ & $55$\\
    A2b & $.016$ & $4.75$ & $19.3$ & $65$\\
    \hline
    A5c & $.027$ & $4.73$ & $16.1$ & $45$\\
    A5d & $.12$ & $4.98$ & $19.2$ & $48$\\
    A3.5 & $.017$ & $4.41$ & $19.2$ & $62$\\
    A5sea & $.016$ & $5.32$ & $30.0$ & $65$\\
    A2sea & $.021$ & $4.96$ & $20.1$ & $75$\\
    A5vent05 & $.021$ & $4.27$ & $16.6$ & $62$\\
    A5vent001 & $.026$ & $4.05$ & $13.8$ & $63$\\
    A5p2K & $.016$ & $4.94$ & $28.4$ & $55$\\
    \hline
\end{tabular}
\caption{{\bf Basic statistics for all numerical experiments.}
The table lists basic statistics as a general overview of the events and tracks in each experiment. 
Mean values were computed for all available days except days one and two, which were considered transient. 
The main four experiments are listed above the horizontal line (A5a, A2a, A5b, A2b). 
Sensitivity experiments are listed below the horizontal line. 
Note that some caution should be exercised in interpreting means for experiments with only a single day of data.
}
\label{tab:basic_stats}
\end{table}

\begin{figure*}[b]
\centering
\vspace{.5cm}
\begin{overpic}[width=0.4\textwidth ]{dummy.pdf}
\put(-55,70){\includegraphics[trim={0 0cm 0cm 0}, clip, width=0.45\linewidth]{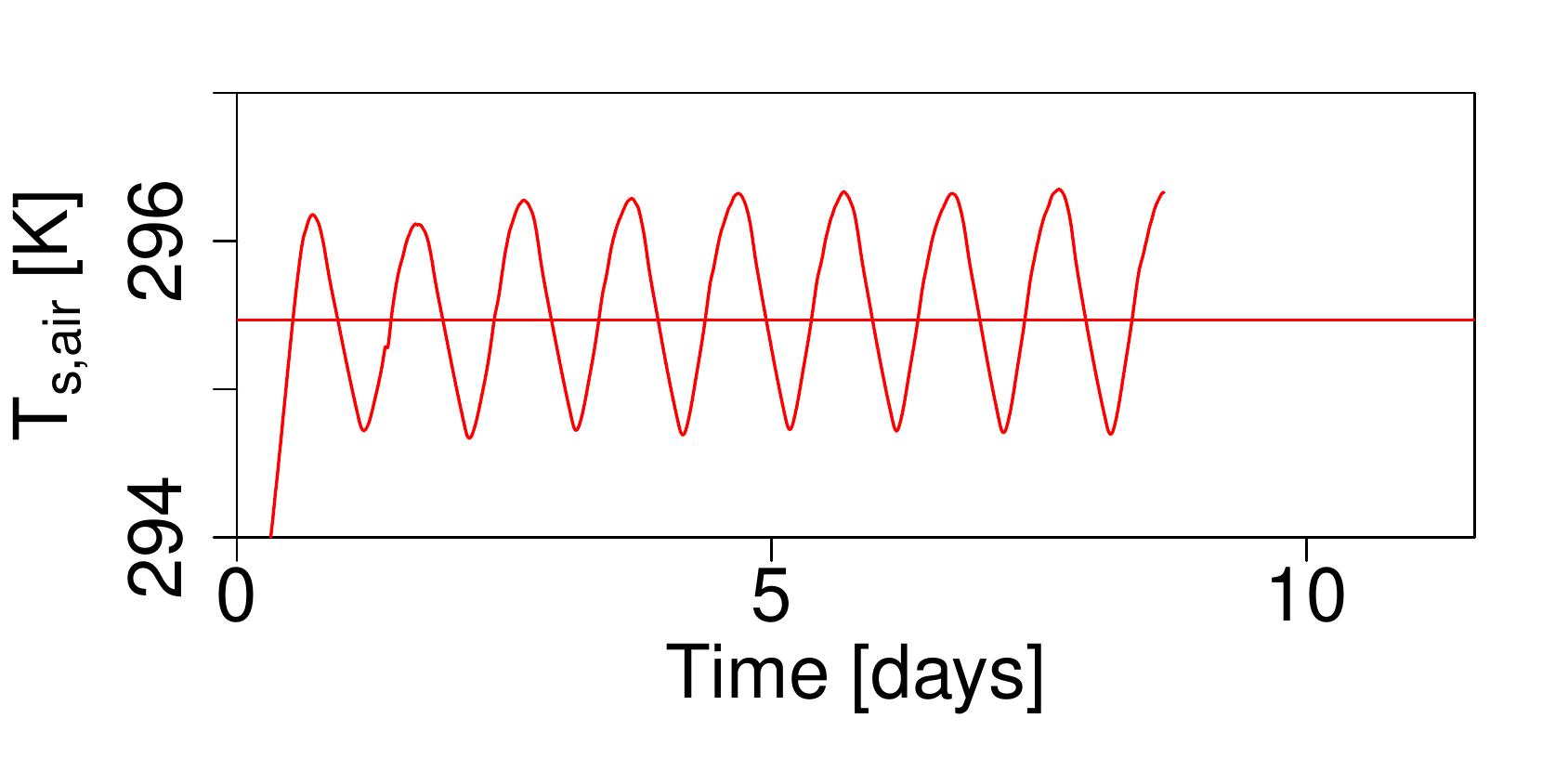}}
\put(30,70){\includegraphics[trim={0 0cm 0cm 0}, clip, width=0.45\linewidth]{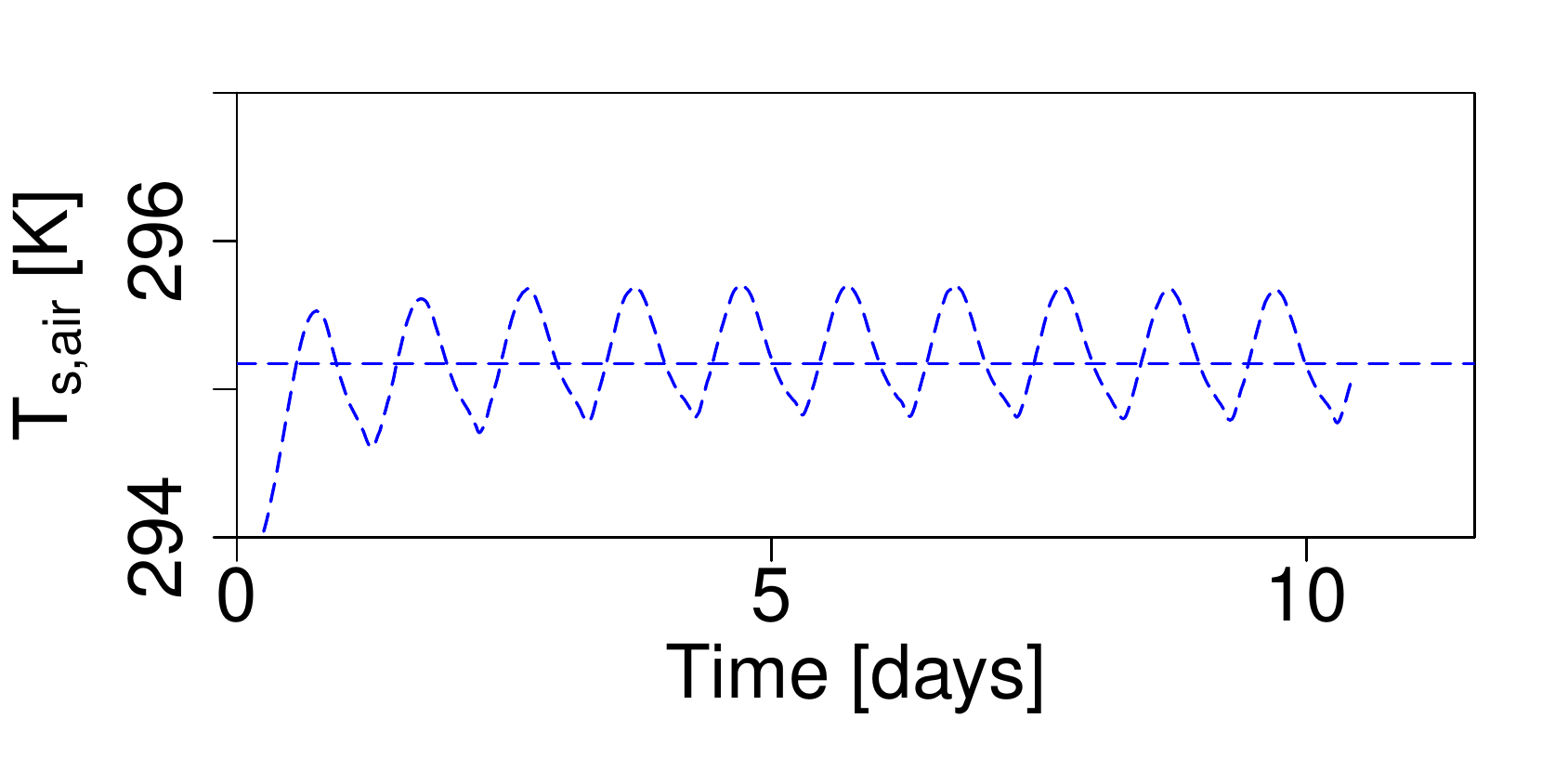}}

\put( -55,35){
\includegraphics[trim={0 0cm 0cm 0}, clip, width=0.45\linewidth]{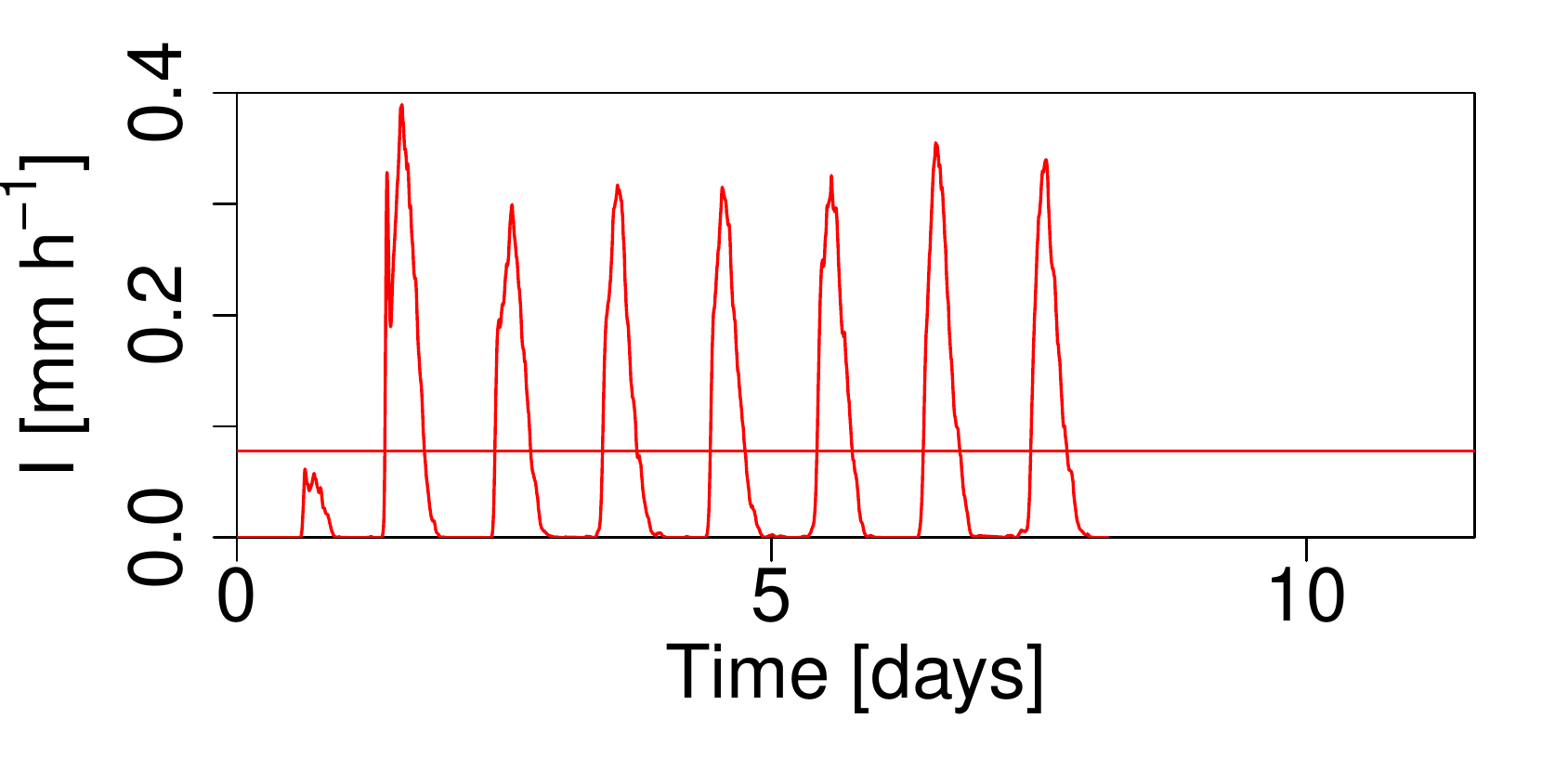}}
\put( 30,35){
\includegraphics[trim={0 0 0cm 0}, clip, width=0.45\linewidth]{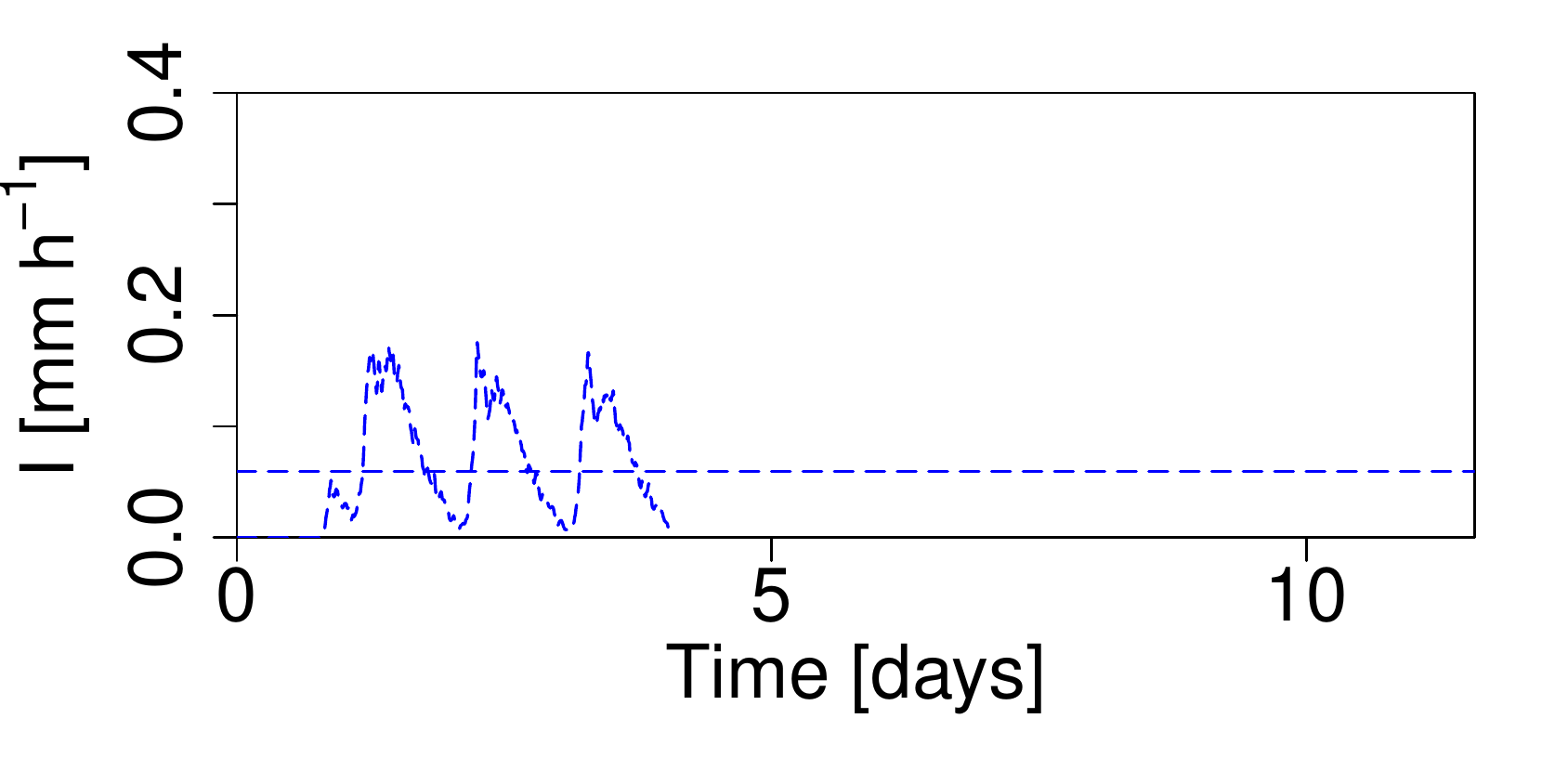}}

\put( -55,0){
\includegraphics[trim={0 0cm 0cm 0}, clip, width=0.45\linewidth]{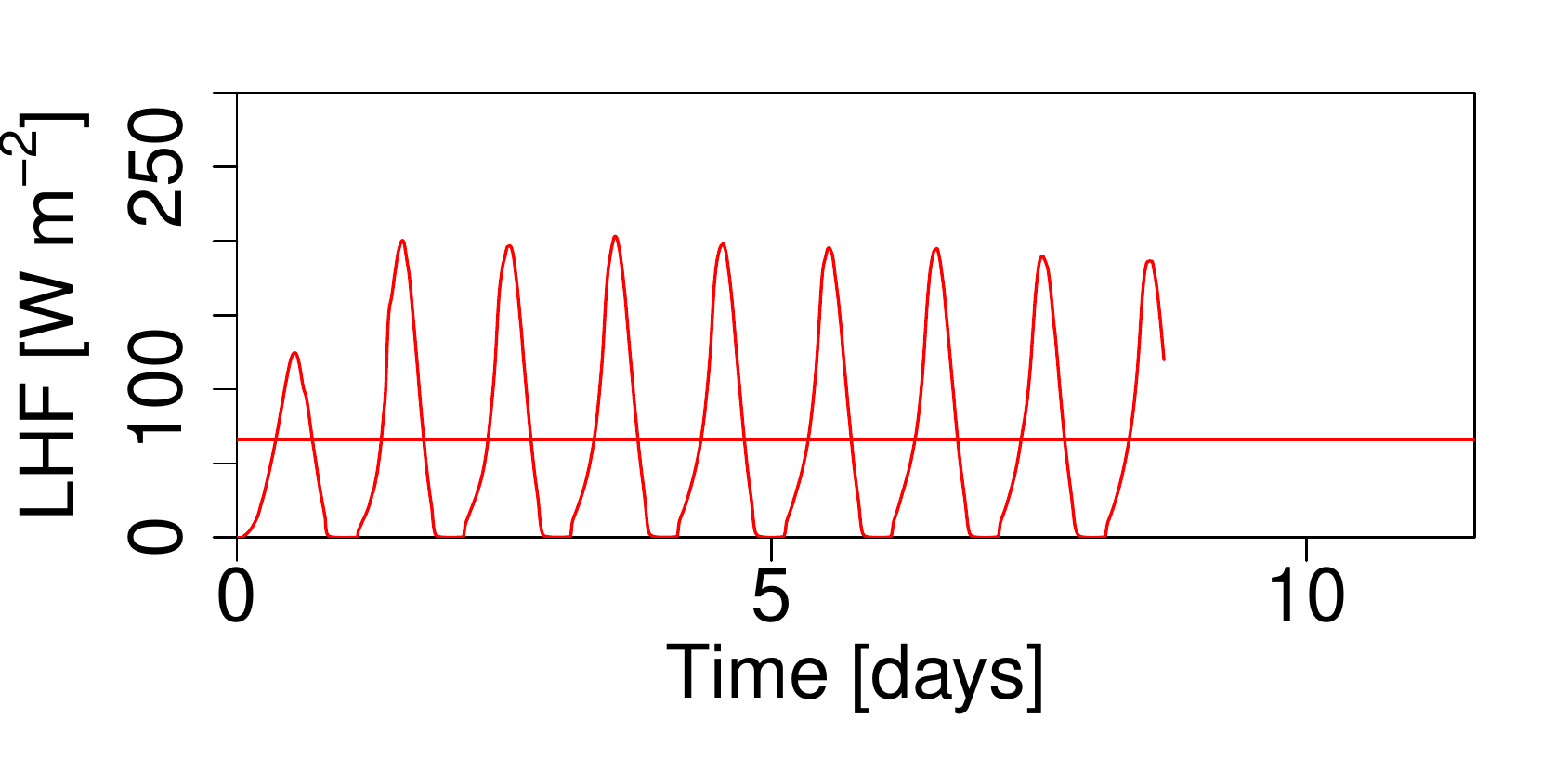}}
\put( 30,0){
\includegraphics[trim={0 0 0cm 0}, clip, width=0.45\linewidth]{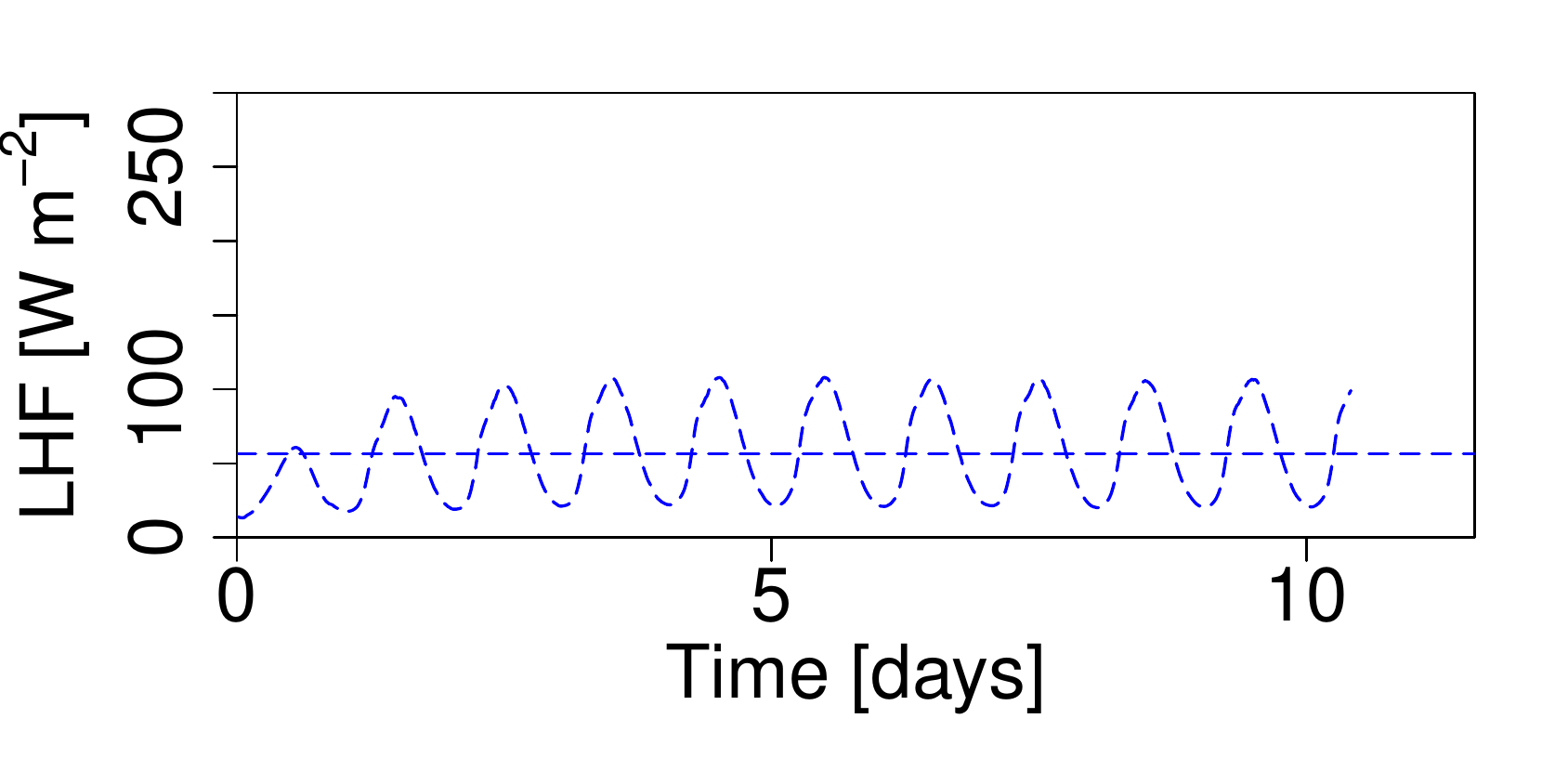}}

\put( -55,-35){
\includegraphics[trim={0 0cm 0cm 0}, clip, width=0.45\linewidth]{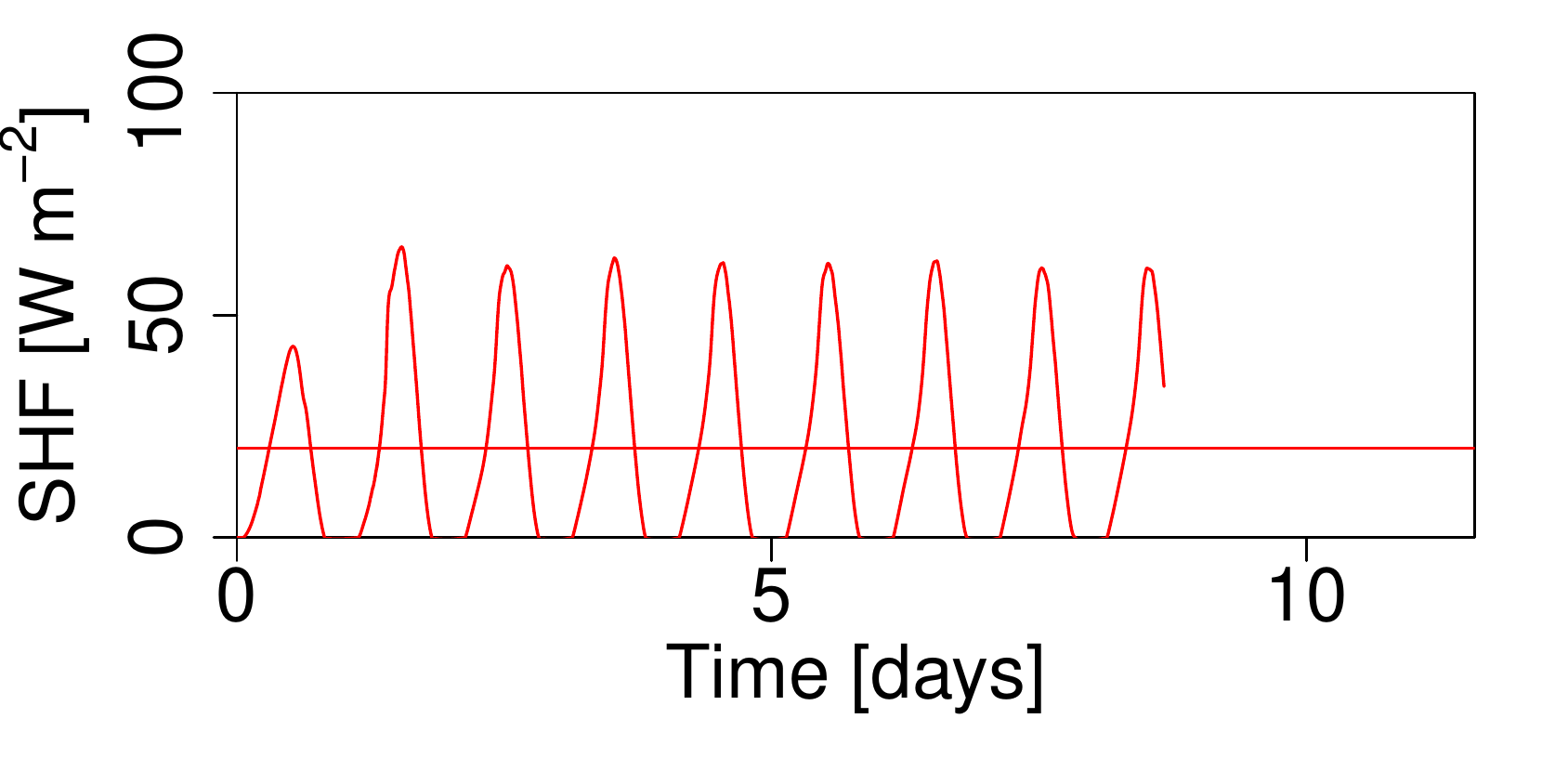}}
\put( 30,-35){
\includegraphics[trim={0 0 0cm 0}, clip, width=0.45\linewidth]{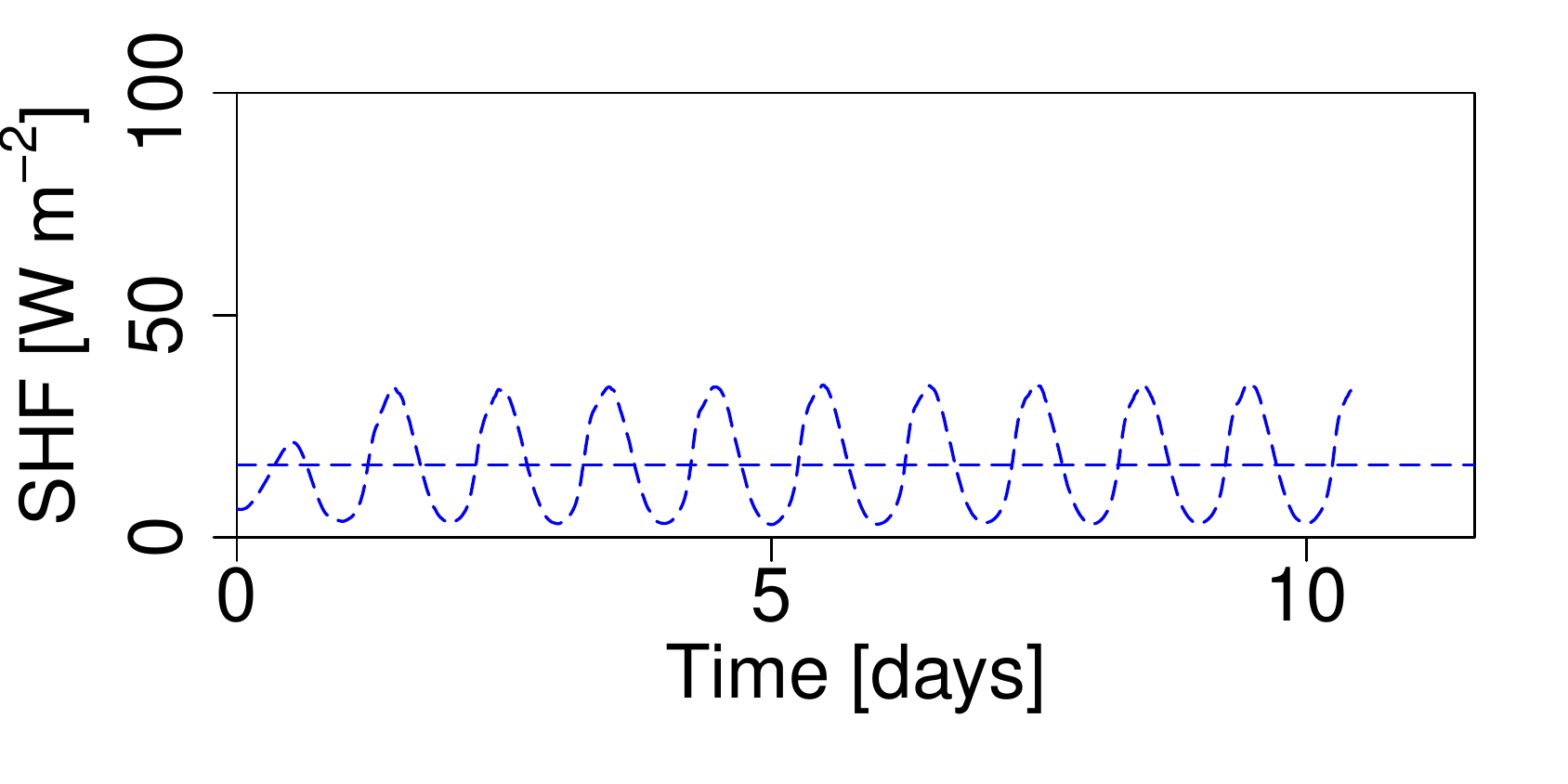}}
\put(-41,106){\bf a}
\put( 44,106){\bf b}
\put(-41,71){\bf c}
\put( 44,71){\bf d}
\put(-41,36){\bf e}
\put( 44,36){\bf f}
\put(-41,1){\bf g}
\put( 44,1){\bf h}
\end{overpic}
\vspace{2.3cm}
\caption{{\bf Multi-day time-series of domain averaged quantities}. 
{\bf a}, Domain-mean near-surface temperature ($T(z=50\;m)$) for the simulation A5b. 
The horizontal line indicates the time average over the entire time-series;
{\bf b}, Analogous to (a), but for $A2b$ ({\it compare}: Tab.~\ref{tab:experiments});
{\bf c,d}, Analogous to (a),(b), but for domain mean precipitation intensity;
{\bf e,f}, Analogous, but for the surface latent heat flux;
{\bf g,h}, Analogous, but for the surface sensible heat flux.}
\label{fig:multi-day_timeseries}
\end{figure*}


\begin{figure*}[t]
\centering
\begin{overpic}[width=0.4\textwidth ]{dummy.pdf}
\put(-60,0){\includegraphics[trim={0 0 0cm 0}, clip, height=0.32\linewidth]{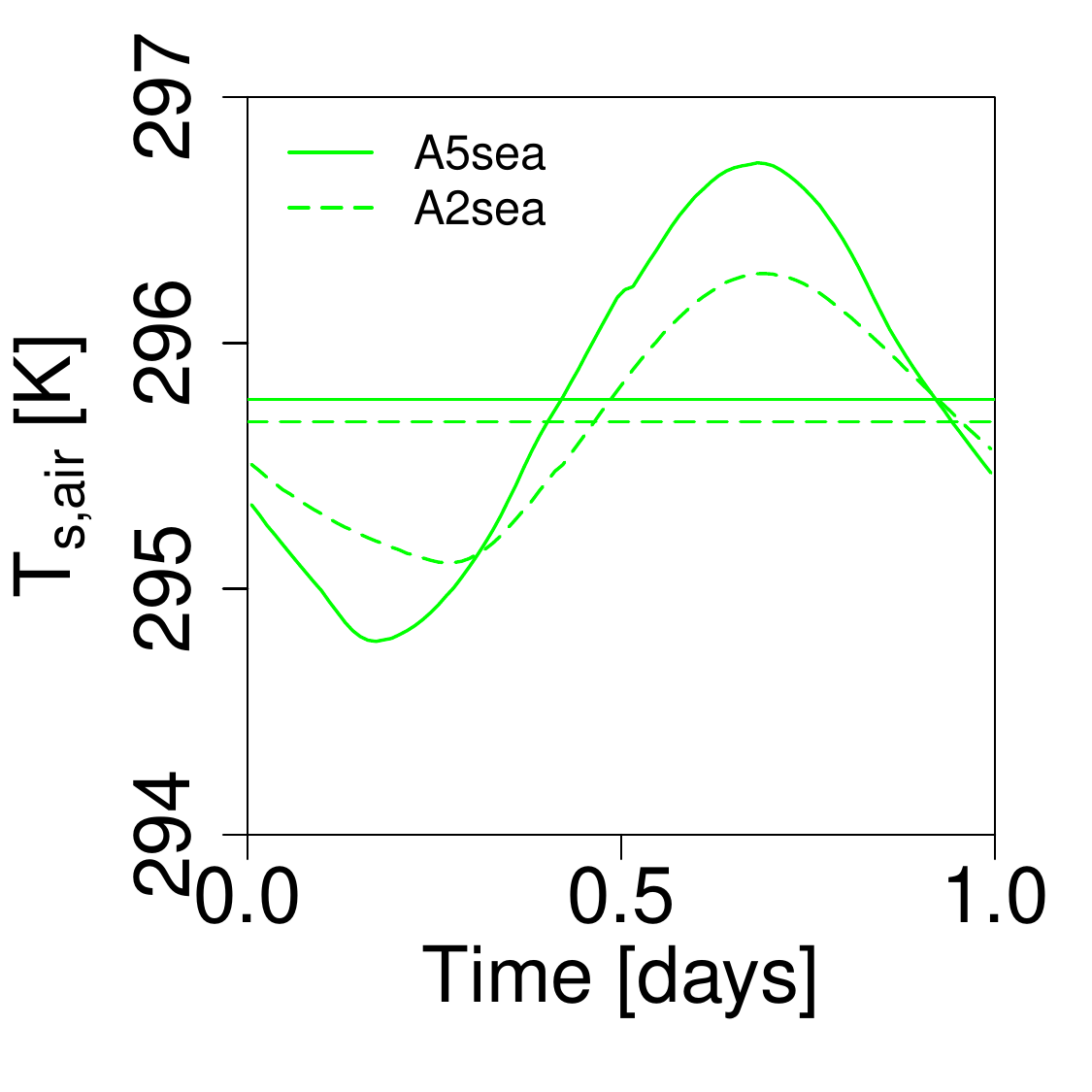}}
\put(-5,0){
\includegraphics[trim={0 0 0cm 0}, clip, height=0.32\linewidth]{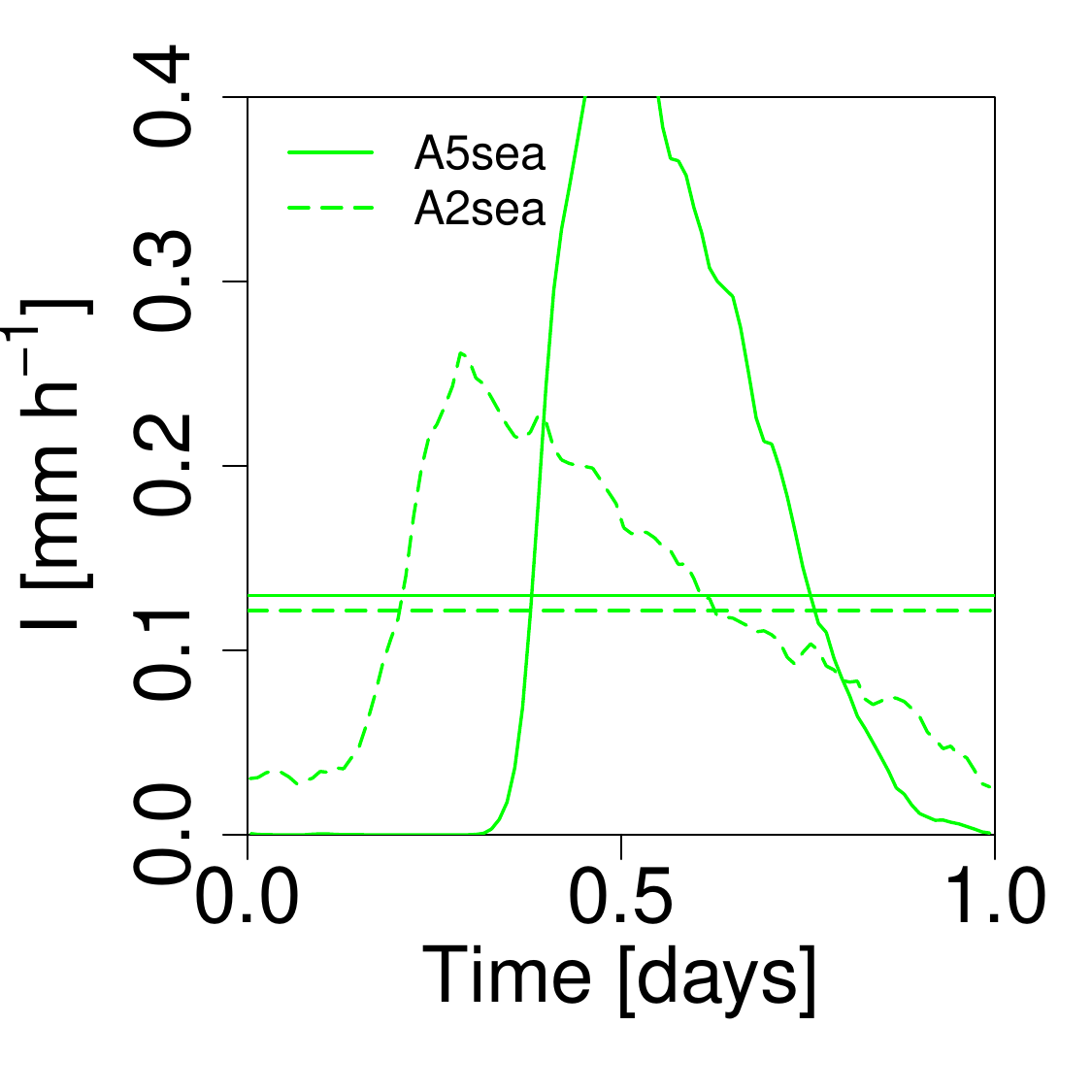}}
\put(50,0){\includegraphics[trim={0 0 0cm 0}, clip, height=0.32\linewidth]{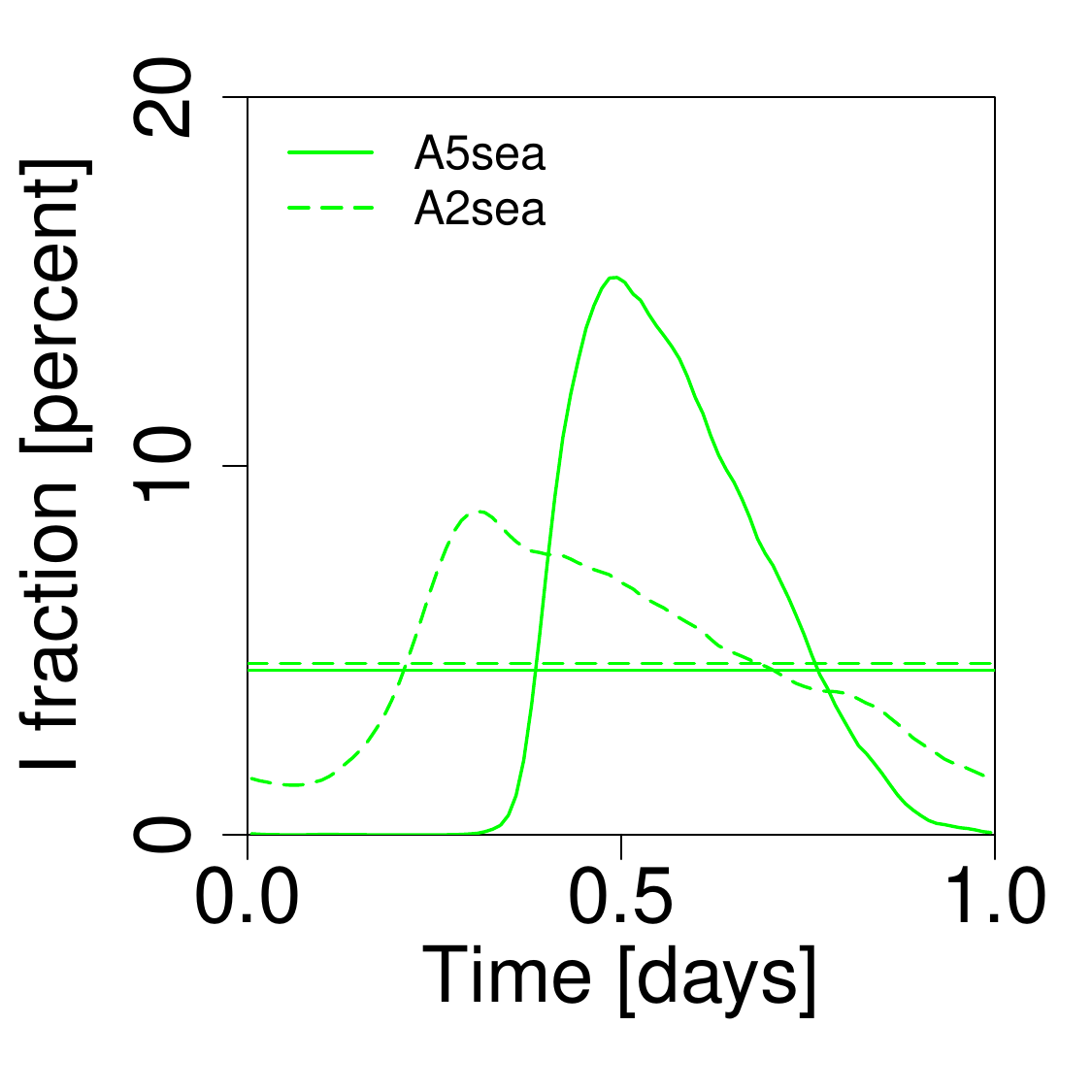}}
\put(-46,53){\bf a}
\put( 10,53){\bf b}
\put( 63,53){\bf c}
\end{overpic}
\caption{{\bf Diurnal cycles of domain averaged quantities for a sea surface}. 
Similar to Fig.~\ref{fig:daily_mean} but using potential evaporation at the surface, mimicking a sea surface ($A2sea$ and $A5sea$, respectively, {\it compare}: Tab.~\ref{tab:experiments}).
}
\label{fig:domain_mean_timeseries_surface_evap}
\end{figure*}

\begin{figure}[ht]
\centering
\includegraphics[width=0.83\linewidth]{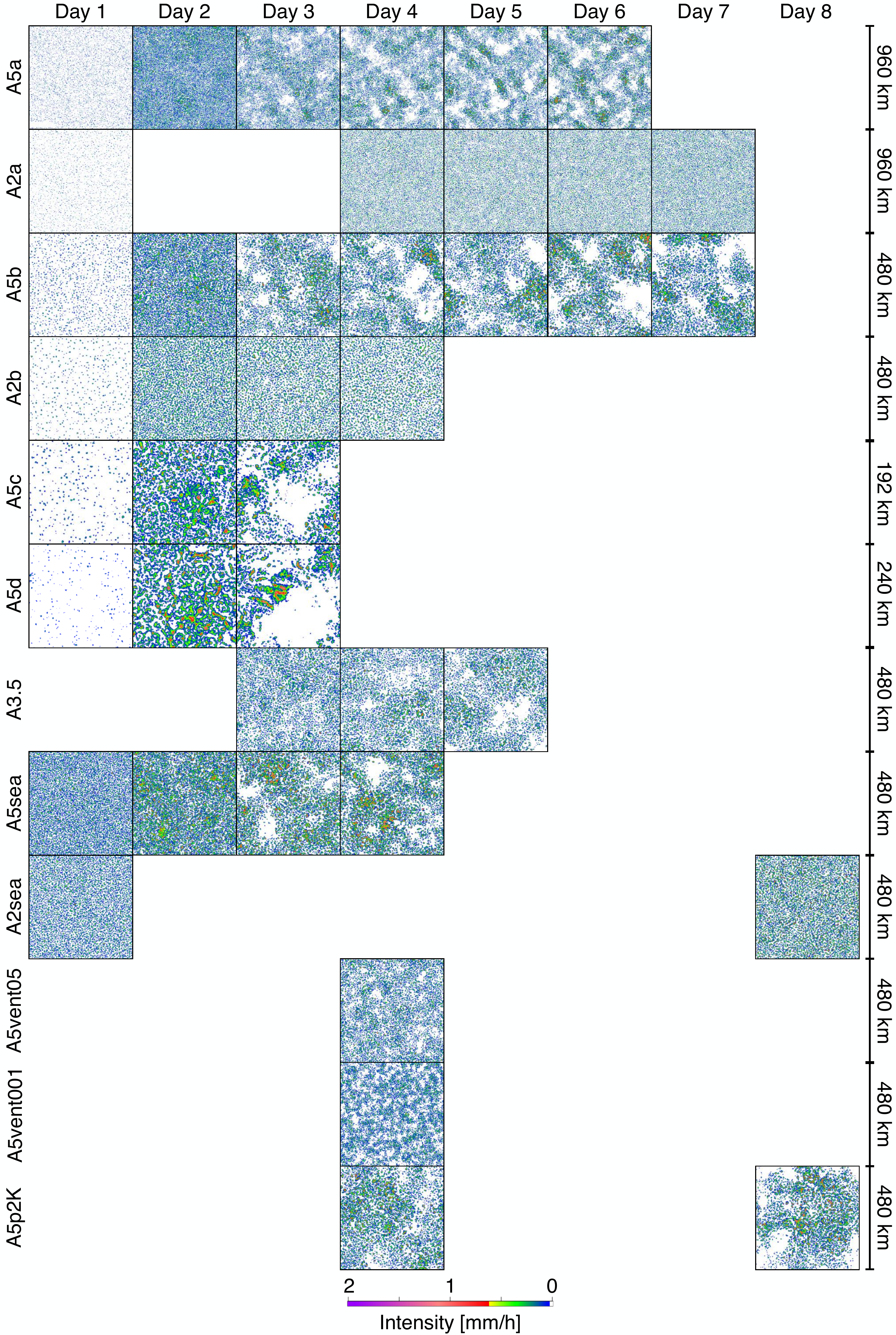}
\caption{{\bf Daily average precipitation intensity (all available days).} Similar to the heatmaps shown in Fig.~\ref{fig:daily_mean} but for all available data ({\it compare}: Tab.~\ref{tab:experiments}).}
\label{fig:allData}
\end{figure}

\begin{figure}[ht]
\centering
\begin{overpic}[width=0.4\textwidth]{dummy.pdf}
\put(-55,0){\includegraphics[trim={0cm 0cm 0cm 0cm}, clip, height=0.37\linewidth]{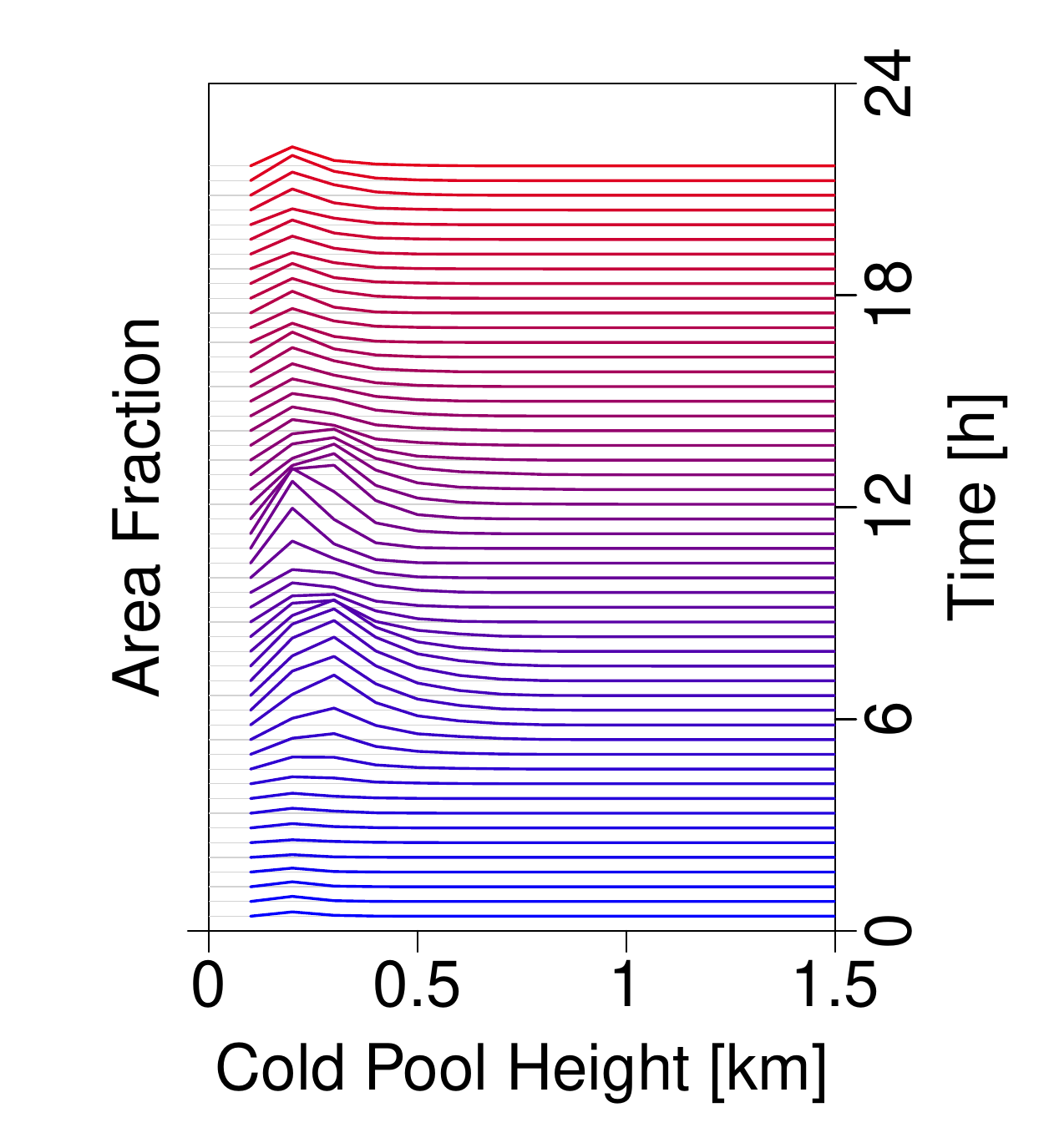}}
\put(0,0){\includegraphics[trim={0cm 0cm 0cm 0cm}, clip, height=0.37\linewidth]{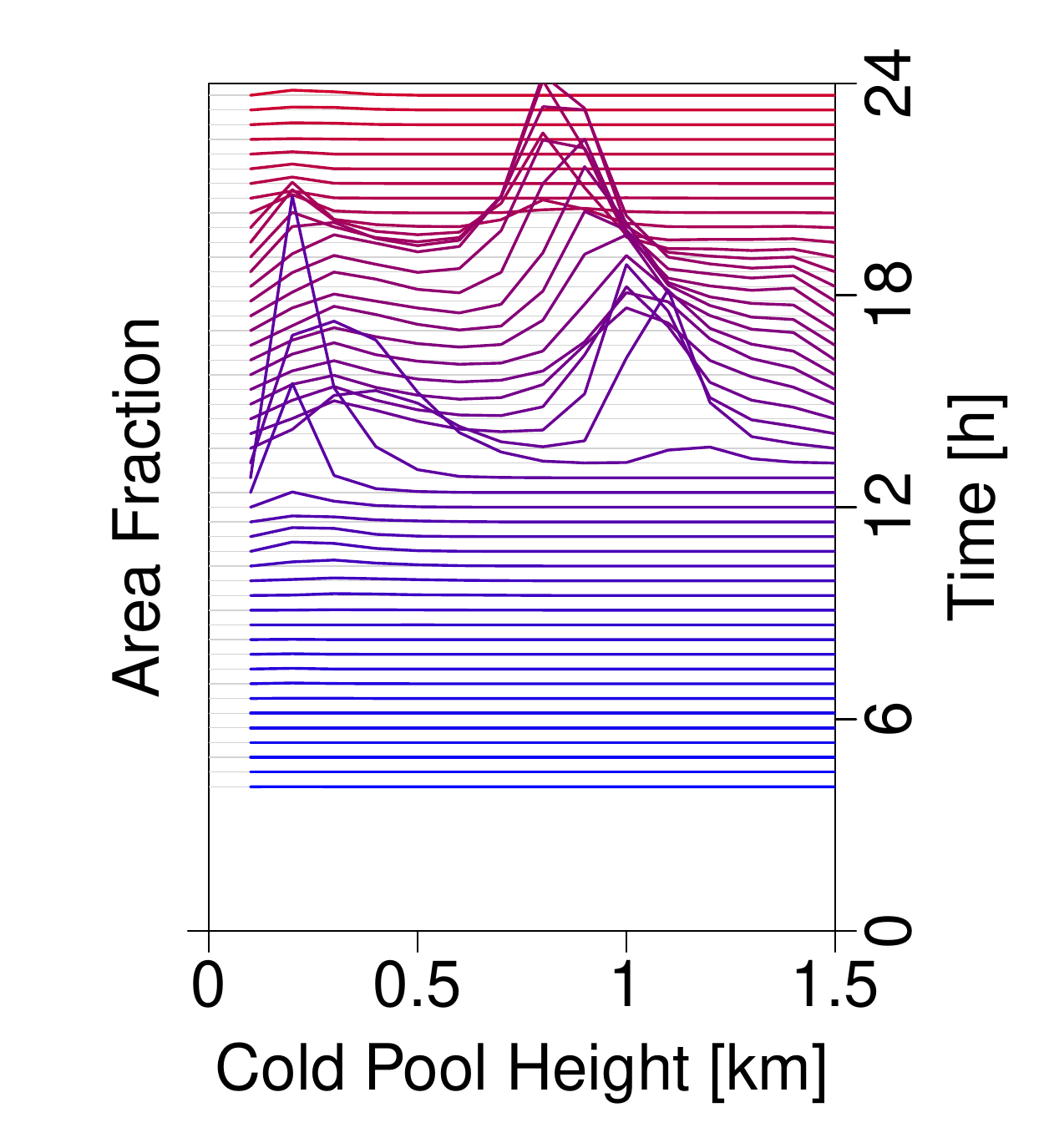}}
\put(60,0){\includegraphics[trim={.0cm 0cm 0cm 0cm}, clip,height=0.3\linewidth]{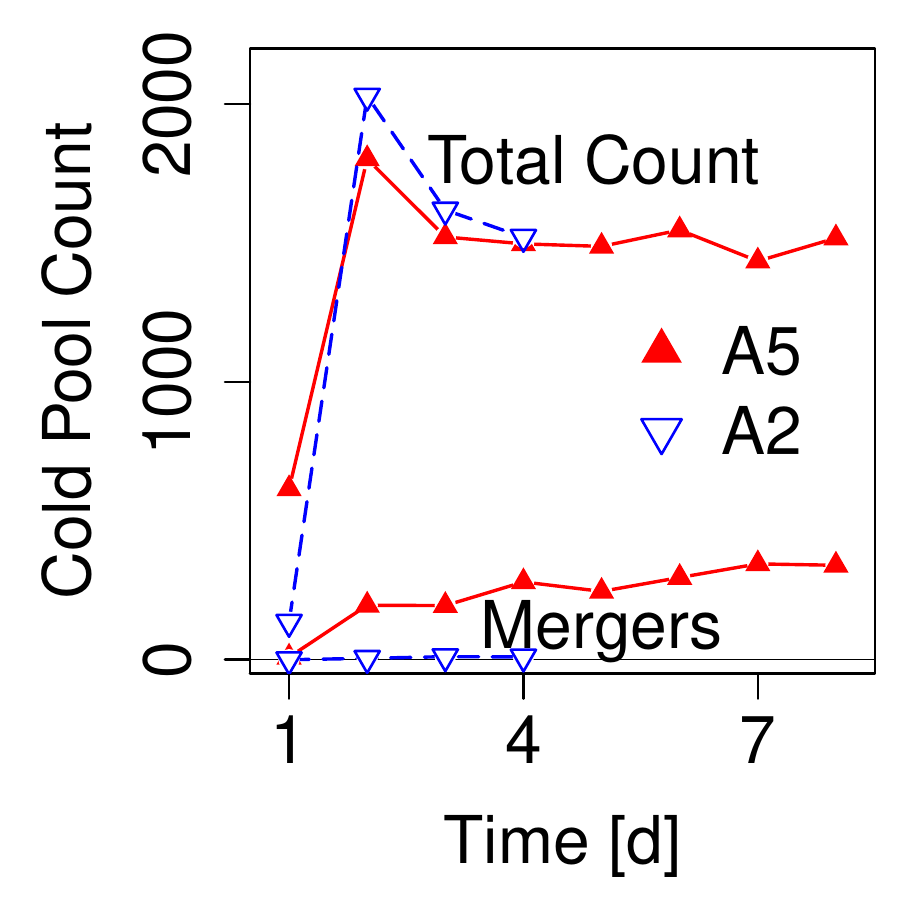}}
\put(-50,59){\bf a}
\put(5,59){\bf b}
\put(65,52){\bf c}
\end{overpic}
\vspace{0cm}
\caption{{\bf Cold pool height distribution functions.}
{\bf a}, Cold pool height distribution function for day four of A2b ({\it compare}: Fig.~\ref{fig:CP_merging}d). 
Each curve shows the histogram for all cold pool heights within five model timesteps (right vertical axis), that is, $480\times 480\times 5$ individual height values. 
A dominant peak occurs at the surface, which is not shown, as it corresponds to regions without cold pools.
The probability origin of each curve is shifted and marked by a grey horizontal line. 
{\bf b}, Analogous to (a), but for A5b during day four ({\it compare}: Fig.~\ref{fig:CP_merging}g).
Note the pronounced double-peak structure that emerges for A5b near mid-day.
{\bf c}, Statistics of the number of CPs and mergers as function of time, for $A5b$ and $A2b$.
Note that mergers are much more frequent for $A5b$, whereas the total count of CPs is similar for the two simulations.
}
\label{fig:CP_height_distribution}
\end{figure}

\begin{figure}[ht]
\centering
\includegraphics[trim={0 0 2.6cm 0},clip,height=.28\linewidth]{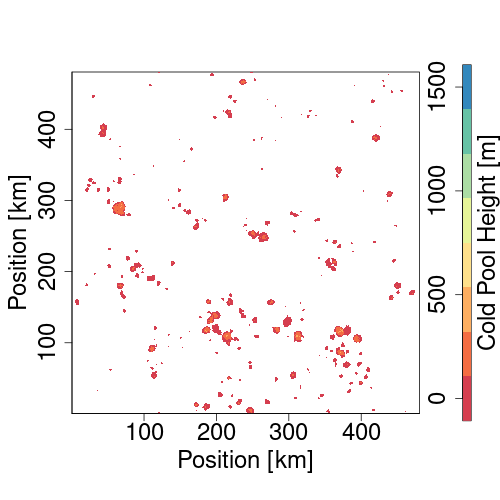}
\includegraphics[trim={2.3cm 0 2.6cm 0},clip,height=.28\linewidth]{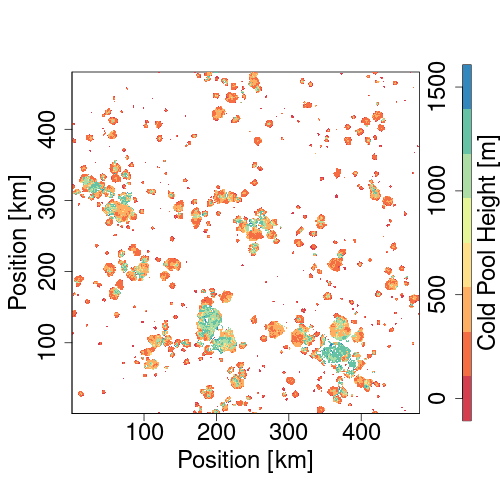}
\includegraphics[trim={2.3cm 0 2.6cm 0},clip,height=.28\linewidth]{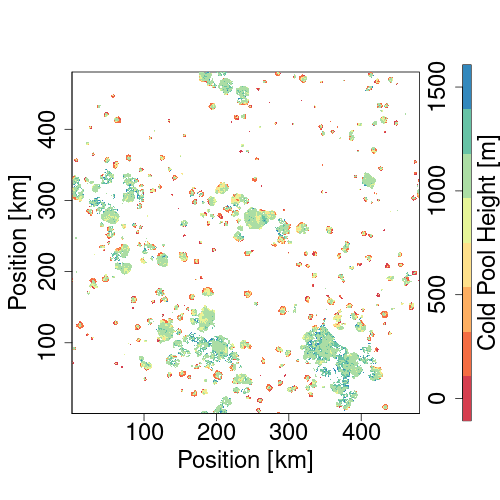}
\includegraphics[trim={2.3cm 0 0 0},clip,height=.28\linewidth]{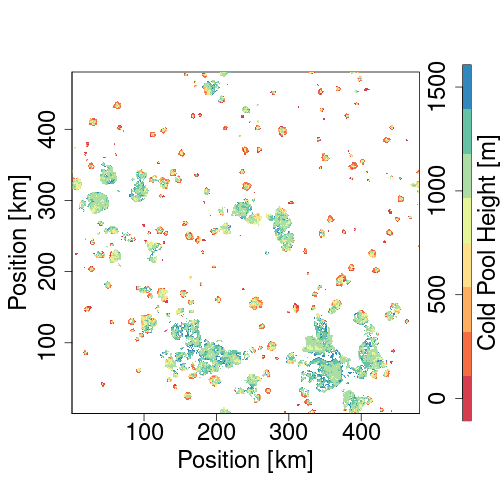}
\\
\includegraphics[trim={0 0 2.6cm 0},clip,height=.28\linewidth]{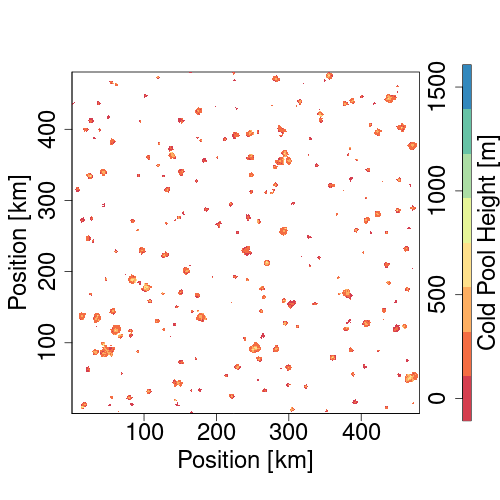}
\includegraphics[trim={2.5cm 0 2.6cm 0},clip,height=.28\linewidth]{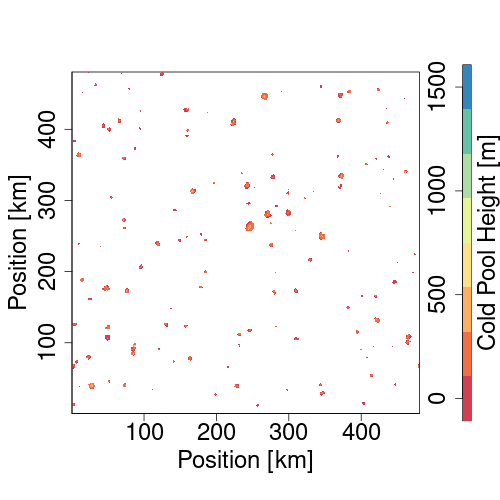}
\includegraphics[trim={2.5cm 0 2.6cm 0},clip,height=.28\linewidth]{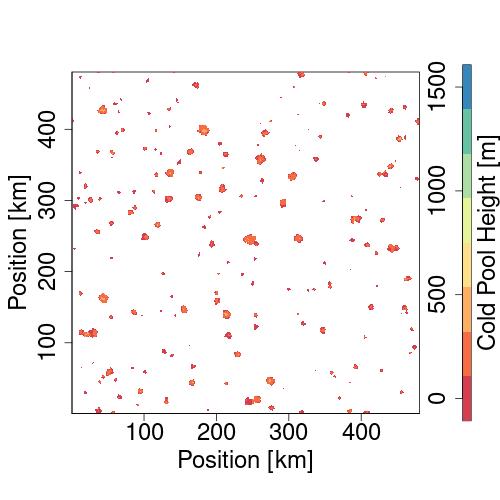}
\includegraphics[trim={2.5cm 0 0 0},clip,height=.28\linewidth]{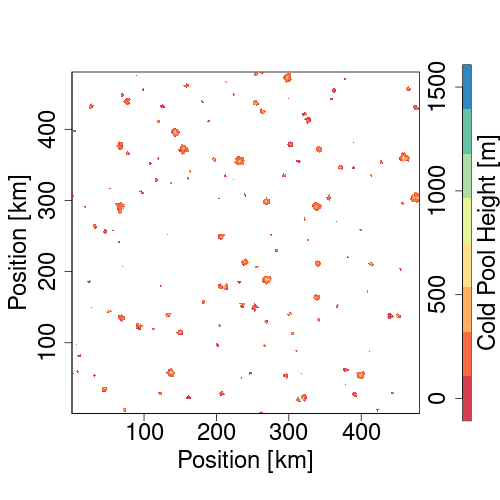}
\vspace{0cm}
\includegraphics[trim={0 0 2.6cm 0},clip,height=.28\linewidth]{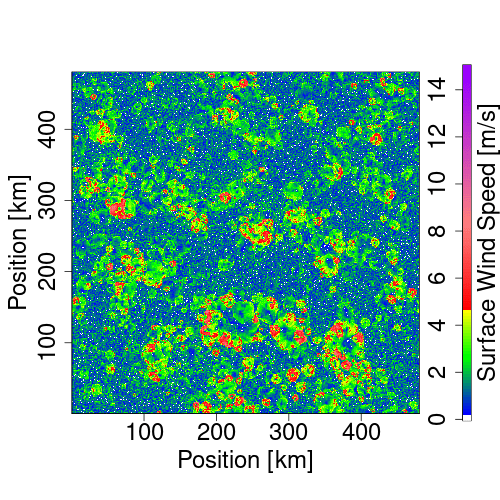}
\includegraphics[trim={2.3cm 0 2.6cm 0},clip,height=.28\linewidth]{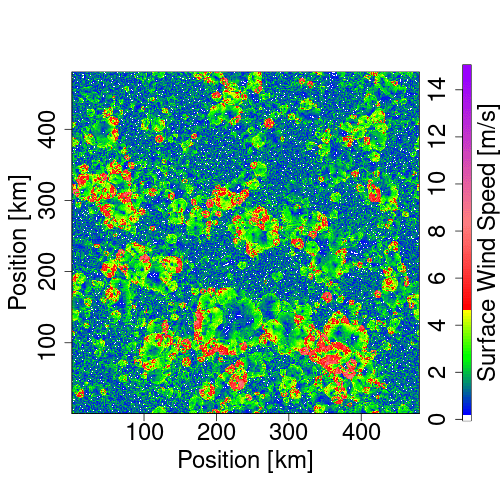}
\includegraphics[trim={2.3cm 0 2.6cm 0},clip,height=.28\linewidth]{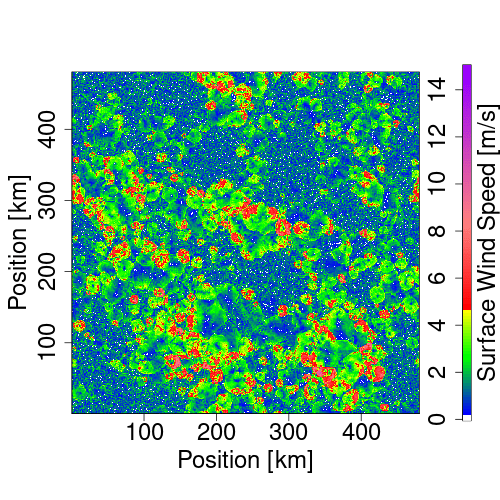}
\includegraphics[trim={2.3cm 0 0 0},clip,height=.28\linewidth]{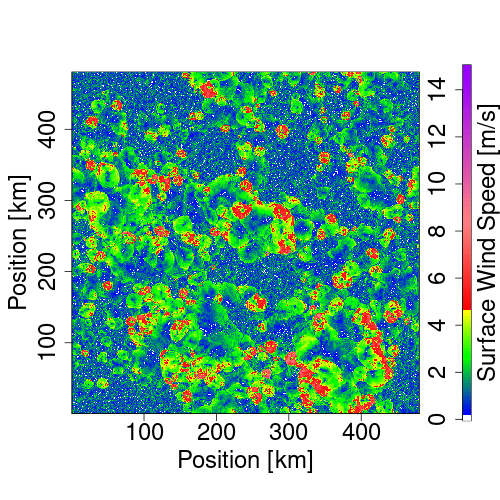}
\\
\includegraphics[trim={0 0 2.6cm 0},clip,height=.28\linewidth]{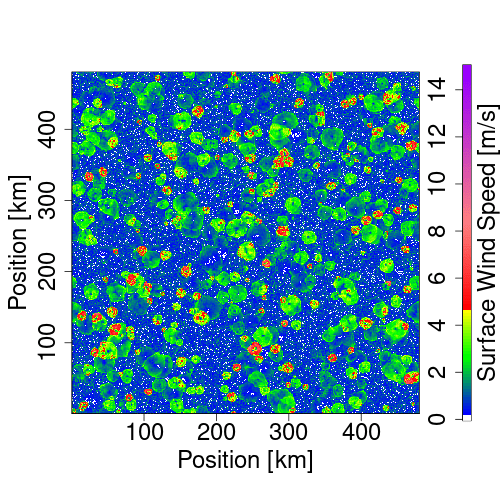}
\includegraphics[trim={2.5cm 0 2.6cm 0},clip,height=.28\linewidth]{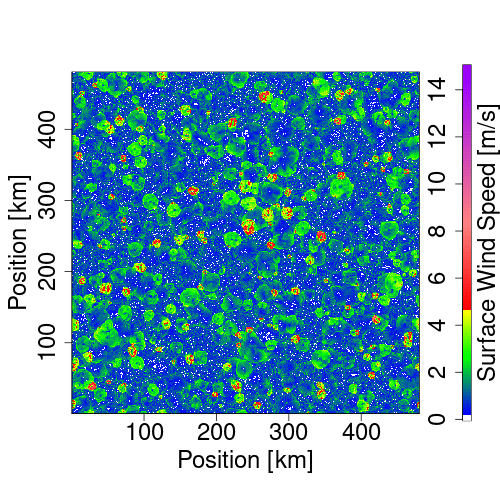}
\includegraphics[trim={2.5cm 0 2.6cm 0},clip,height=.28\linewidth]{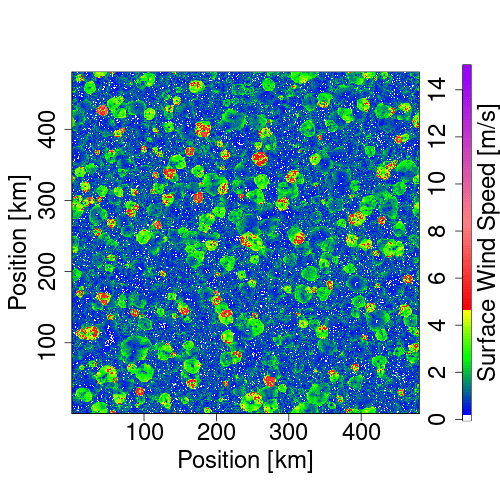}
\includegraphics[trim={2.5cm 0 0 0},clip,height=.28\linewidth]{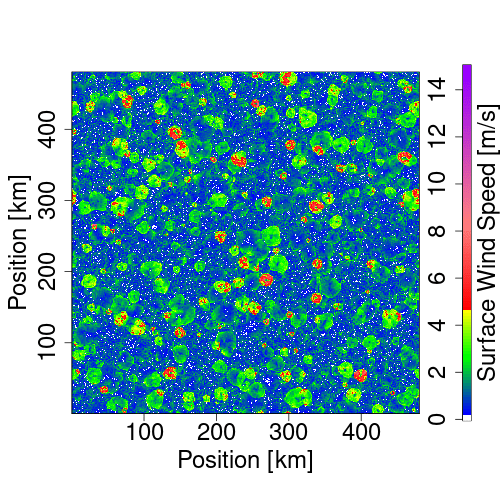}
\vspace{0cm}
\caption{{\bf Cold pool height and near surface horizontal wind speed.}
{\it top row}, CP heights for A5b, day 4, at times $t/h$ in $\{12.5, 13.5, 14.5, 15.5\}$. 
{\it second row}, CP heights for A2b, day 4, at times $t/h$ $\{7, 9,, 11, 13\}$. ({\it compare}: Fig.~\ref{fig:CP_height_distribution}).
{\it third row}, wind speed for A5b, day 4, at times $t/h$ in $\{12.5, 13.5, 14.5, 15.5\}$. 
{\it bottom row}, wind speed for A2b, day 4, at times $t/h$ $\{7, 9,, 11, 13\}$.
}
\label{fig:hor_wind_speed}
\end{figure}

\begin{figure}[ht]
\centering
\begin{overpic}[width=0.4\textwidth]{dummy.pdf}
\put(-55,16){\includegraphics[trim={0cm 2.1cm 0cm 0cm}, clip, height=0.19\linewidth]{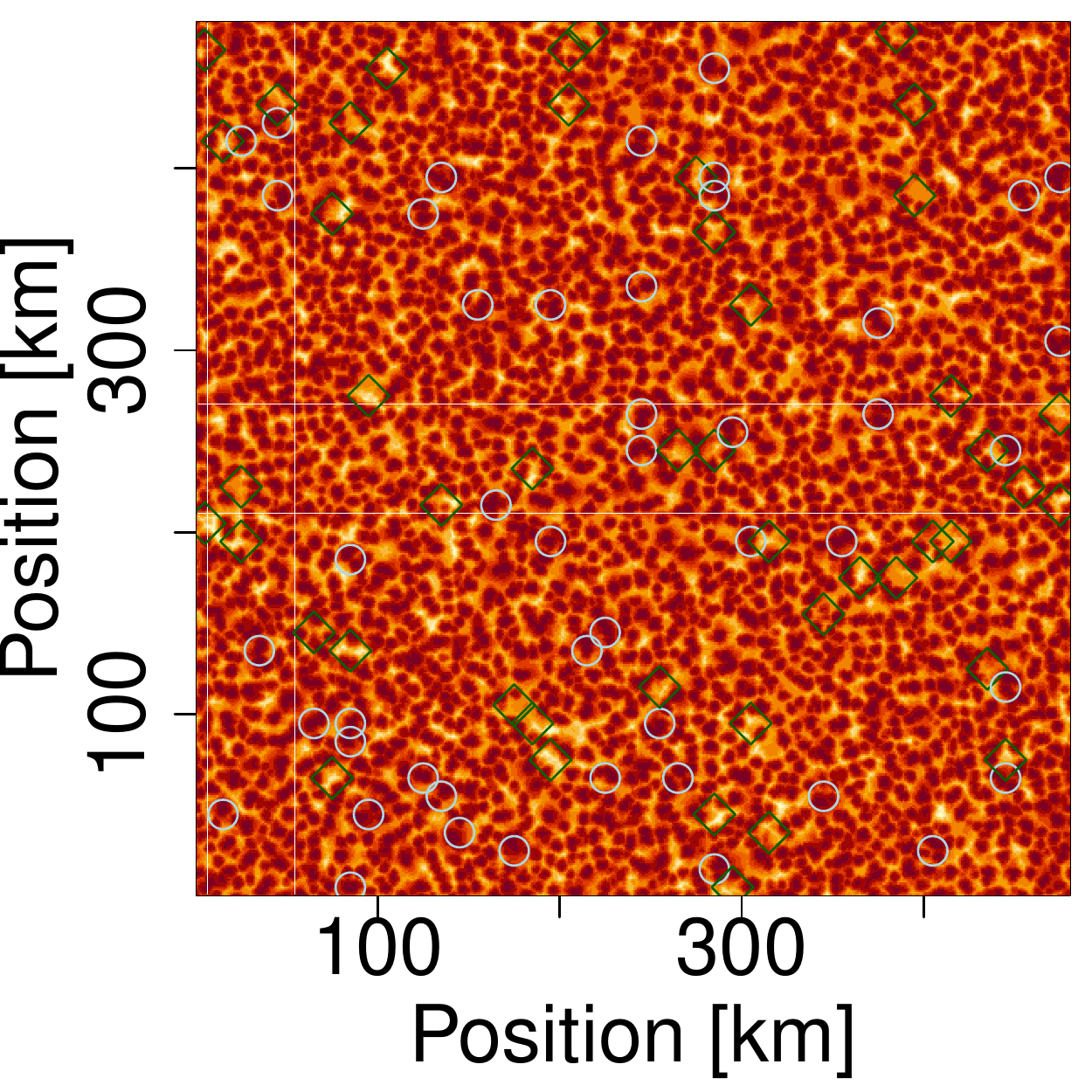}}
\put(-14,16){\includegraphics[trim={0cm 2.1cm 0cm 0cm}, clip, height=0.19\linewidth]{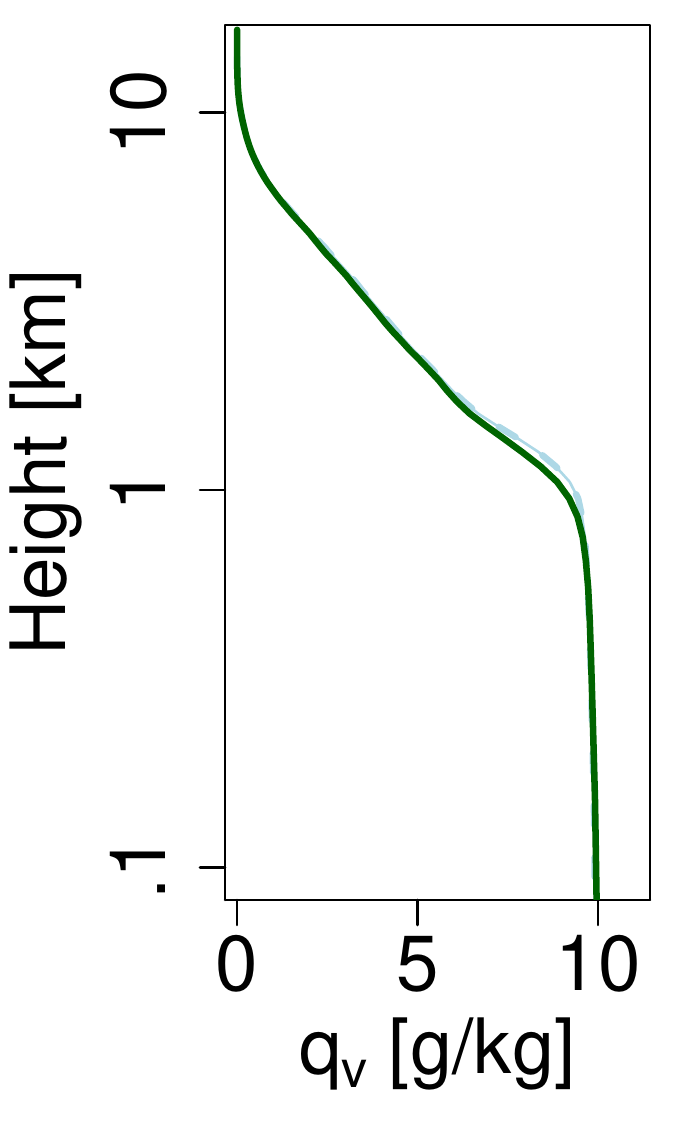}}
\put(10,16){\includegraphics[trim={2cm 2.1cm 0cm 0cm}, clip, height=0.19\linewidth]{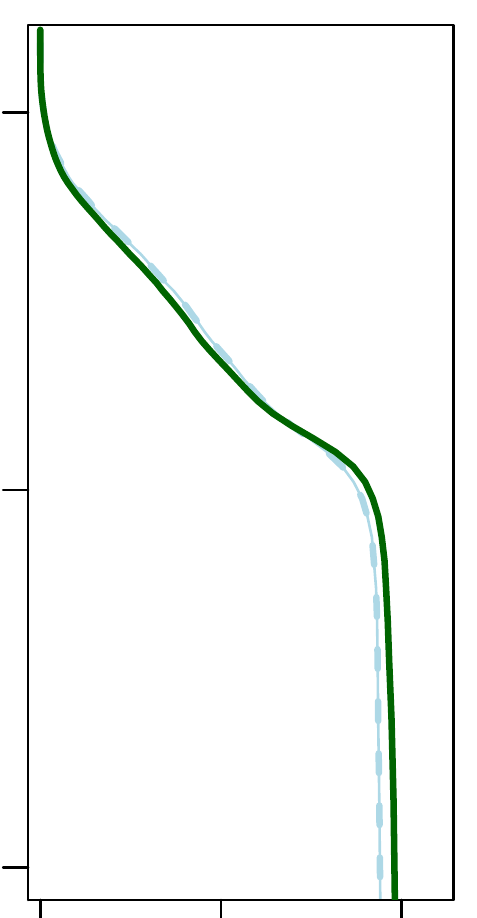}}
\put(-55,-27.0){\includegraphics[trim={0cm 0cm 0cm 0cm}, clip, height=0.225\linewidth]{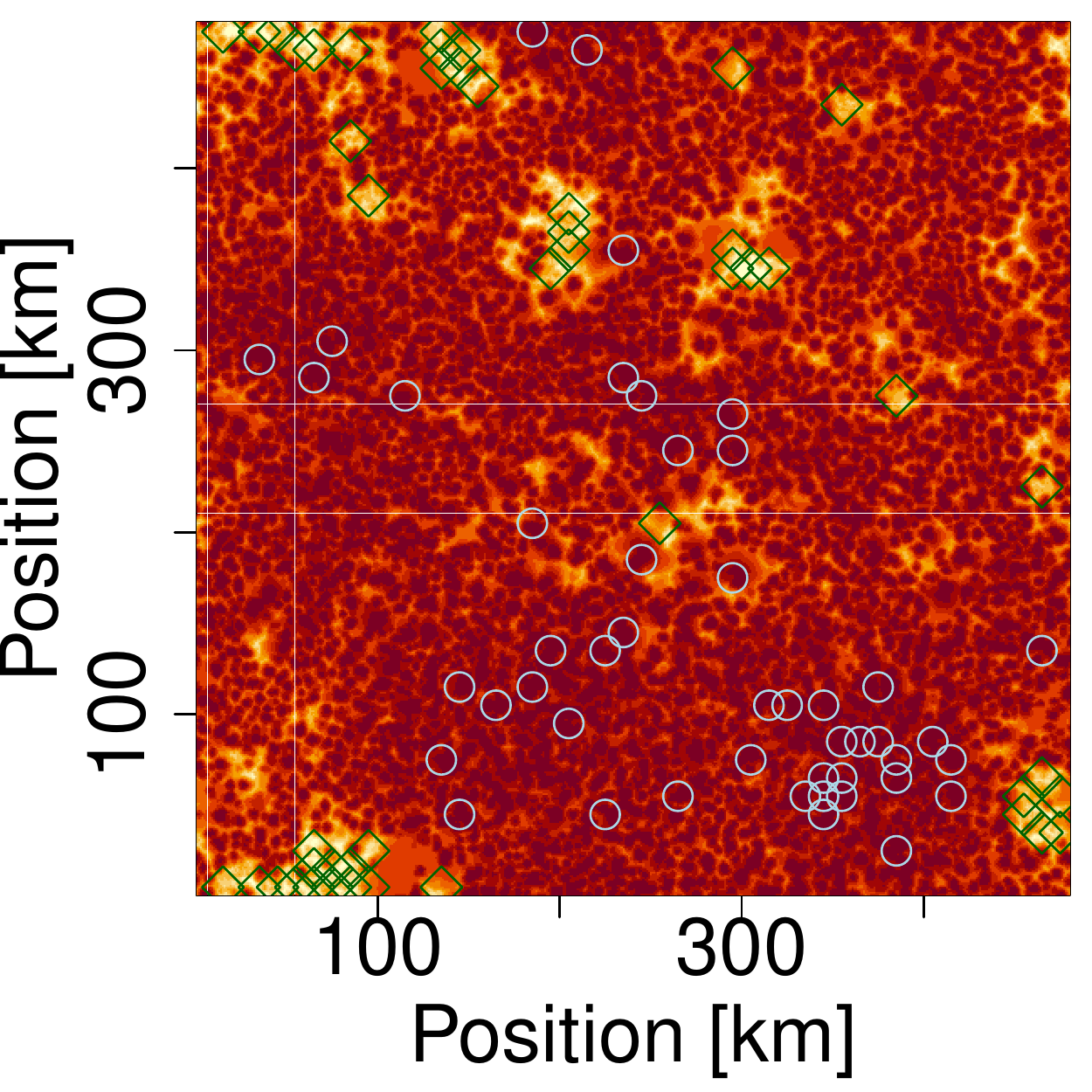}}
\put(-14,-28){\includegraphics[trim={0cm 0cm 0cm 0cm}, clip, height=0.23\linewidth]{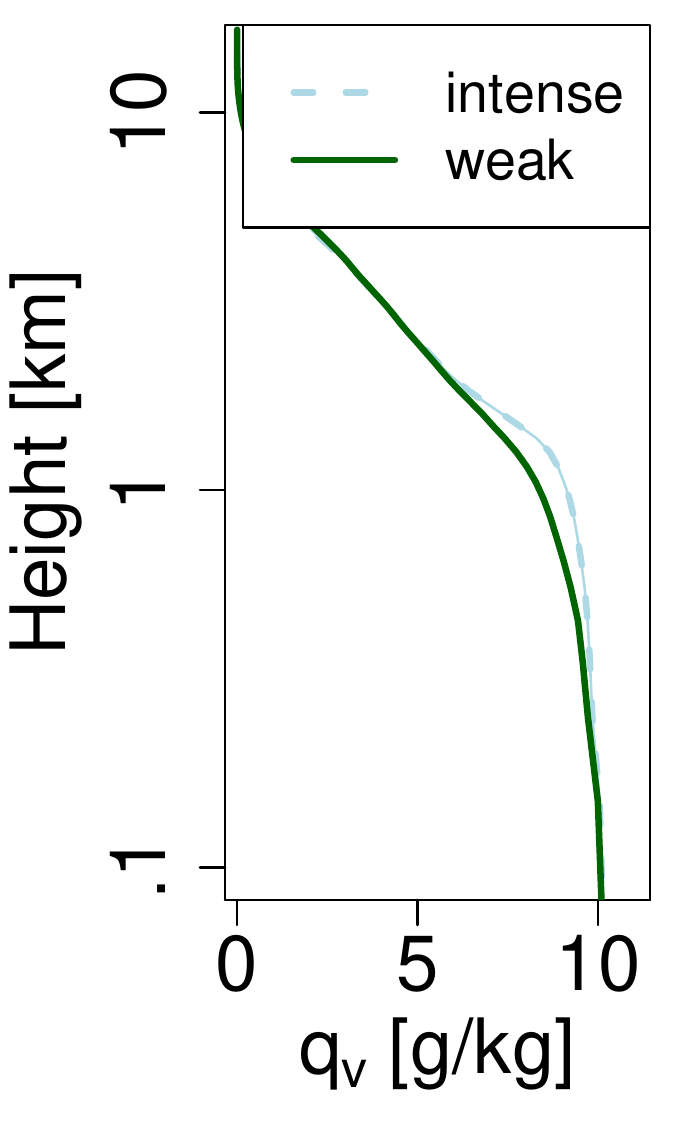}}
\put(10,-28){\includegraphics[trim={2cm 0cm 0cm 0cm}, clip, height=0.23\linewidth]{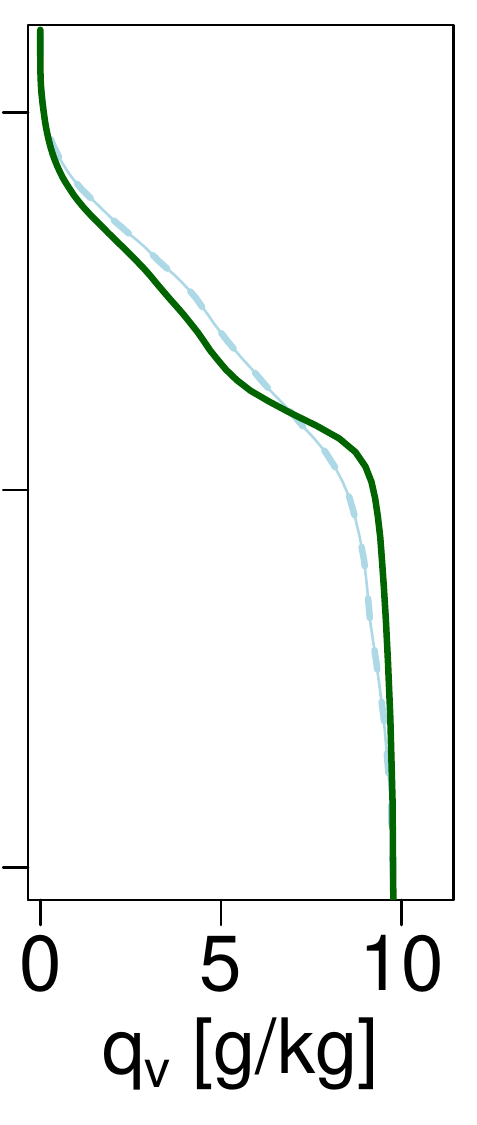}}

\put(-48,49){\bf a}
\put(-6,49){\bf b}
\put(11,49){\bf c}
\put(-48,12){\bf d}
\put(-6,12){\bf e}
\put(11,12){\bf f}
\end{overpic}
\vspace{2cm}
\caption{{\bf Moisture oscillations.}
For each sub-region, we compute the vertical specific humidity profile, $q_v(z)$, during the early morning and in the evening, that is, before and after the onset of precipitation in $A5$.
{\bf a}, Simulation $A2b$ day four. Spatial rain distribution (red) with high and low values marked in light blue circles and green diamonds, respectively. 
{\bf b}, Specific humidity vs. height for intense (top 10 percent) and weak (lowest 10 percent) precipitation regions before the onset of precipitation (early morning).
{\bf c}, Analogous to (b), but at the end of the model day.
{\bf d---f}, Analogous to (a)---(c), but for $A5b$.
Note the moisture oscillations and the logarithmic vertical axis scaling in panels b,c,e,f.
}
\label{fig:moisture_oscillations}
\end{figure}


\begin{figure*}[ht]
\centering
\begin{overpic}[width=0.4\textwidth ]{dummy.pdf}
\put(-65,40){\includegraphics[height=0.37\linewidth,trim=0cm 2cm 0cm 0cm, clip]{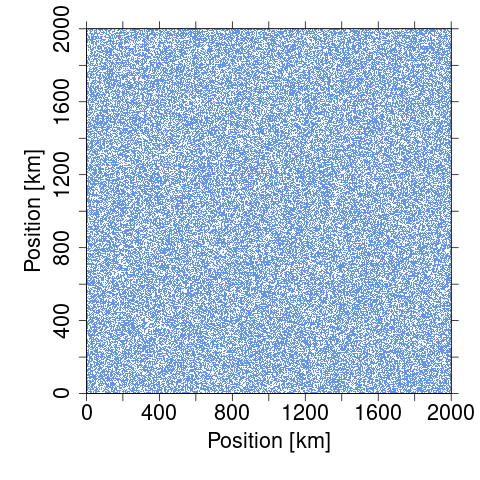}}
\put(5,40){\includegraphics[trim={2.5cm 2cm 0cm 0cm}, clip, height=0.37\linewidth½]{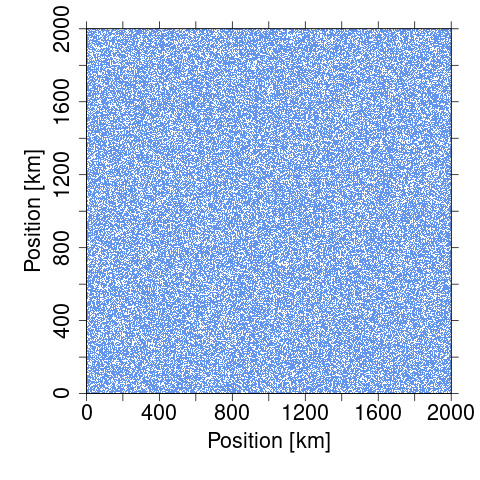}}
\put(65,40){\includegraphics[trim={2.5cm 2cm 0cm 0cm}, clip, height=0.37\linewidth]{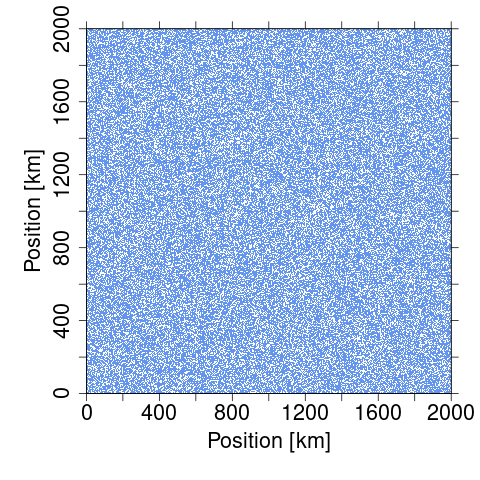}}
\put(-65,-35){\includegraphics[trim={0cm 0cm 0cm 0cm}, clip, height=0.42\linewidth]{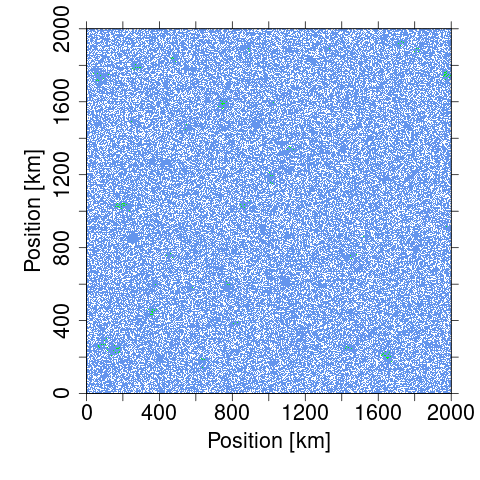}}
\put(5,-35){\includegraphics[trim={2.5cm 0cm 0cm 0cm}, clip, height=0.42\linewidth]{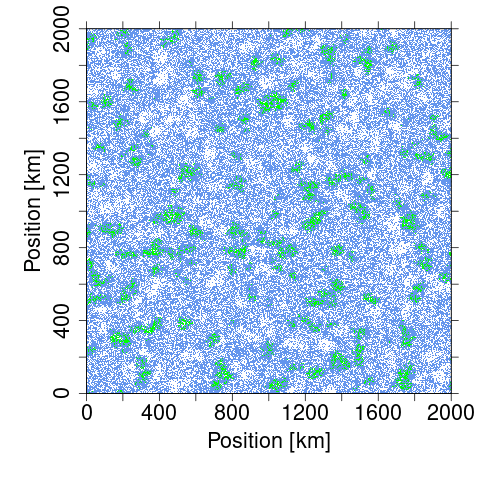}}
\put(65,-35){\includegraphics[trim={2.5cm 0cm 0cm 0cm}, clip, height=0.42\linewidth]{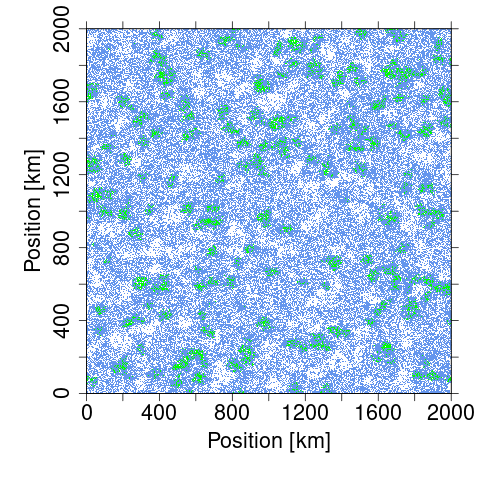}}
\put(-52,102){\large \bf a}
\put(6,102){\large \bf b}
\put(68,102){\large \bf c}
\put(-52,36){\large \bf d}
\put(6,36){\large \bf e}
\put(68,36){\large \bf f}

\put(-50,90){\large small $t_a$}
\put(-50,27){\large large $t_a$}

\put(10,77){\color{black}\line(0,-0.5){5}}
\put(10,72){\color{black}\line(.5,0){5}}
\put(10,77){\color{black}\line(.5,0){5}}
\put(15,77){\color{black}\line(0,-0.5){5}}

\put(10,10){\color{black}\line(0,-1){5}}
\put(10,5){\color{black}\line(1,0){5}}
\put(10,10){\color{black}\line(1,0){5}}
\put(15,10){\color{black}\line(0,-1){5}}

\put(70,77){\color{black}\line(0,-1){5}}
\put(70,72){\color{black}\line(1,0){5}}
\put(70,77){\color{black}\line(1,0){5}}
\put(75,77){\color{black}\line(0,-1){5}}

\put(70,10){\color{black}\line(0,-1){5}}
\put(70,5){\color{black}\line(1,0){5}}
\put(70,10){\color{black}\line(1,0){5}}
\put(75,10){\color{black}\line(0,-1){5}}

\end{overpic}
\vspace{2.5cm}
\caption{{\bf Transition to a clustered rainfall state in the simplified model.}
{\bf a}, Surface rainfall average during the first day ($t=1\;d$, spin up) for $\Delta T=2\;K$.
{\bf b}, Similar to (a), but for $t=6\;d$.
{\bf c}, Similar to (a), but for $t=7\;d$.
{\bf d---f}, Similar to (a)---(c), but for large amplitude.
To highlight the spatial and temporal variation, boxes of side length $l=180\;km$ are shown at equal positions in panels b,c,e,f.
Pixels in white, blue and green correspond to zero, one, or two rain events, occurring within the respective model day.
}
\label{fig:daily_mean_simplified_model}
\end{figure*}

\end{document}